\documentclass[preprint,12pt,authoryear]{elsarticle} 

\usepackage{color}
\usepackage{fullpage}
\usepackage{amsmath}
\usepackage{mathrsfs}
\usepackage[table]{xcolor}
\usepackage{newfloat}
\usepackage{graphicx}
\usepackage{subfig}
\usepackage{multirow}
\usepackage{caption}
\usepackage{arydshln}
\usepackage{tikz}
\usetikzlibrary{arrows,positioning,calc}
\usepackage{pstool}
\usepackage{textcomp}
\usepackage[hidelinks]{hyperref}
\hypersetup{colorlinks,bookmarksopen,linkcolor=blue,citecolor=blue,urlcolor=blue}

\newcommand{\bs}[1]{{\boldsymbol{#1}}}

\newcommand{\corr}[1]{\mathrm{corr\,{#1}}}

\newcommand{\tr}[1]{\mathrm{tr\,{#1}}}

\newcommand{\ignore}[1]{}
\DeclareFloatingEnvironment{algorithm}

\usepackage{amssymb}

\journal{Int. J. Solids. Struct.}

\begin{document}

\begin{frontmatter}



\title{On Micromechanical Parameter Identification With Integrated DIC and the Role of Accuracy in Kinematic Boundary Conditions\tnoteref{titlefoot}}


\author[TUe]{O. Roko\v{s}\corref{mycorrespondingauthor}}
\cortext[mycorrespondingauthor]{Corresponding author.}
\ead{o.rokos@tue.nl}

\author[TUe]{J.P.M. Hoefnagels}

\author[TUe]{R.H.J. Peerlings}

\author[TUe]{M.G.D. Geers}

\address[TUe]{Eindhoven University of Technology, Department of Mechanical Engineering, P.O. Box~513, 5600~MB Eindhoven, The Netherlands}

\tnotetext[titlefoot]{The post-print version of this article is published in \emph{Int. J. Solids. Struct.}, \href{https://www.sciencedirect.com/science/article/pii/S0020768318301422}{10.1016/j.ijsolstr.2018.04.004}.}

\begin{abstract}
Integrated Digital Image Correlation~(IDIC) is nowadays a well established full-field experimental procedure for reliable and accurate identification of material parameters. It is based on the correlation of a series of images captured during a mechanical experiment, that are matched by displacement fields derived from an underlying mechanical model. In recent studies, it has been shown that when the applied boundary conditions lie outside the employed field of view, IDIC suffers from inaccuracies. A typical example is a micromechanical parameter identification inside a Microstructural Volume Element~(MVE), whereby images are usually obtained by electron microscopy or other microscopy techniques but the loads are applied at a much larger scale. For any IDIC model, MVE boundary conditions still need to be specified, and any deviation or fluctuation in these boundary conditions may significantly influence the quality of identification. Prescribing proper boundary conditions is generally a challenging task, because the MVE has no free boundary, and the boundary displacements are typically highly heterogeneous due to the underlying microstructure. The aim of this paper is therefore first to quantify the effects of errors in the prescribed boundary conditions on the accuracy of the identification in a systematic way. To this end, three kinds of mechanical tests, each for various levels of material contrast ratios and levels of image noise, are carried out by means of virtual experiments. For simplicity, an elastic compressible Neo-Hookean constitutive model under plane strain assumption is adopted. It is shown that a high level of detail is required in the applied boundary conditions. This motivates an improved boundary condition application approach, which considers constitutive material parameters as well as kinematic variables at the boundary of the entire MVE as degrees of freedom in the IDIC procedure, assuring that both are identified with equal precision and importance. This problem has been studied in the literature with a different method, i.e. Finite Element Method Updating framework.
\end{abstract}


\end{frontmatter}


%
%
\section{Introduction}
\label{Sect:Introduction}
Accurate identification of micromechanical parameters is important in numerous areas of science and engineering. On the one hand, parameters are required for (complex) constitutive laws that help to predict, e.g., mechanical response, performance, or lifespan of electronic, micro-electro-mechanical, or other mechanical devices. On the other hand, they help to better understand complex physical processes in materials occurring across the scales, such as plasticity, failure, ductile damage, or delamination and crack growth~\cite{Hoc:2003,Rupil:2011,Blaysat:2015,Buljac:2017}.

Due to their intrinsically small dimensions, micro- or nanoscale mechanical tests are challenging and necessitate advanced experimental methods. One such method is Digital Image Correlation~(DIC), which is a non-intrusive full-field measurement technique with high accuracy and reliability that emerged from the recent progress in computer technology and digital imaging. In particular, its integrated variant called Integrated Digital Image Correlation (IDIC) has proven to be a reliable and accurate technique for the identification of material parameters, see e.g.~\cite{Roux:2006,Leclerc:2009,Rethore:2009,Rethore:2013,Neggers:2015,Ruybalid:2016}. It relies on the minimization of the difference between two images captured during an experiment (corresponding to the reference and a deformed configuration) inside the Region Of Interest~(ROI). The deformed image is back-deformed using a displacement field that is obtained from an underlying mechanical model with assumed constitutive laws and Boundary Conditions~(BCs). The required basics of IDIC together with geometry, constitutive model, and mechanical tests employed throughout this paper are specified in more detail in Section~\ref{Sect:Method}.

\begin{figure}
	\centering
	\includegraphics[scale=1]{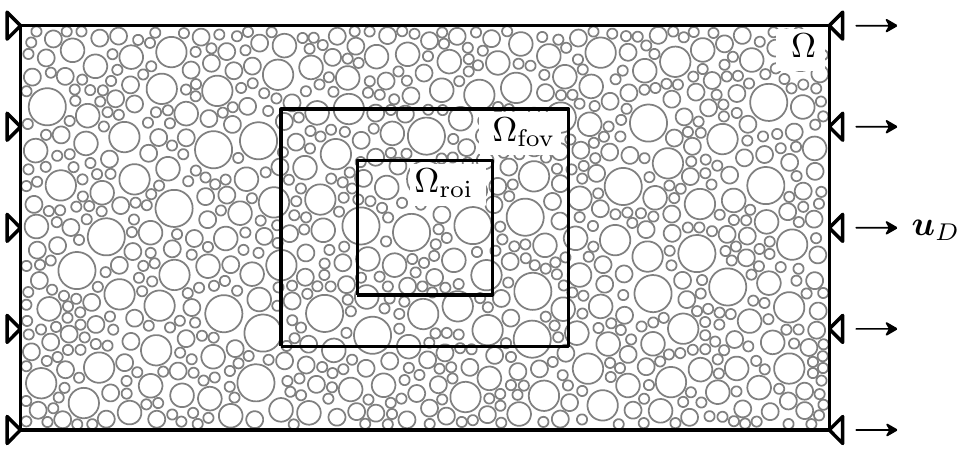}
	\caption{Sketch of a typical experimental set-up. Mechanical test carried out on a specimen with a domain~$\Omega$, field of view~$\Omega_\mathrm{fov}$, and a region of interest~$\Omega_\mathrm{roi}$.}
	\label{Sect:Introduction:Fig1}
\end{figure}

If the BCs applied to a tested specimen lie outside the Field Of View~(FOV), IDIC suffers from inaccuracies~\citep{Andre:2017}. This problem typically applies to micromechanical parameter identification, see Fig.~\ref{Sect:Introduction:Fig1}, whereby images are obtained by electron microscopy or other microscopy techniques and the loads are applied at a much larger scale. Prescribing proper boundary conditions to a given Microstructural Volume Element~(MVE) is a challenging task, as the MVE has no free boundary, and the displacements along its boundary are highly heterogeneous due to the presence of microstructural constituents with (highly) contrasting mechanical behavior at or near the boundary. This renders any kind of idealized boundary conditions inappropriate. Several approaches have been proposed and tested in the literature to resolve this issue, based on Virtual Fields Method~(VFM), as reported e.g. in~\citep{VFM:2006,Rahmani:2014}, or Finite Element Method Updating~(FEMU) with virtual boundaries, as proposed by~\cite{Fedele:2015}. In this contribution, the IDIC methodology will be adopted, which has been reported e.g. by~\cite{Tian:2010}, \cite{Hild:2016}, or~\cite{Shakoor:2017}. According to~\cite{Shakoor:2017}, so far the most accurate methodology employs Global Digital Image Correlation~(GDIC) in order to identify displacements that are subsequently applied as BCs to the MVE associated with IDIC; this approach will be referred to as GDIC-IDIC in the sequel.

As well-known from the literature, cf. e.g.~\cite{Bornert:2009,leclerc:2012,Hild:2016}, in general (G)DIC on the one hand lacks sufficient kinematic freedom when large elements or globally supported polynomials are used (kinematic smoothing), while on the other hand it suffers from random errors when relatively small elements or locally supported interpolation functions are employed. This indicates a possible pitfall for the GDIC-IDIC approach because, as the BCs are kept fixed during the IDIC parameter identification procedure, any errors introduced through the BCs remain locked. The only way in which the MVE model can compensate for erroneous BCs is by adjusting the material parameters---hence resulting in an inaccurate identification of these parameters.
\begin{figure}
 	\flushleft
 	\mbox{}\hspace{4.5em}\includegraphics[scale=1]{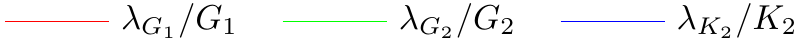}\vspace{-0.5em}\\
 	\centering
	\subfloat[identified parameter]{\includegraphics[scale=1]{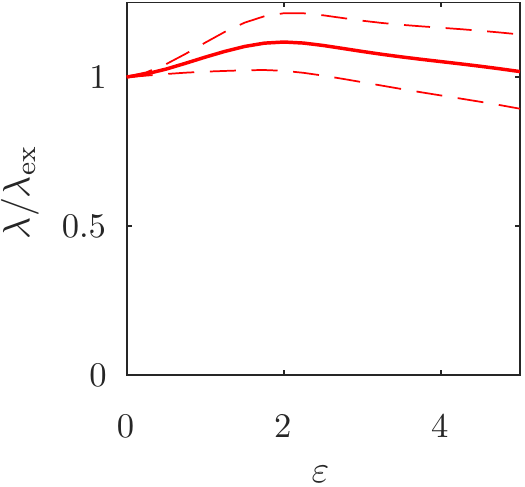}\label{SubSect:BCMA:Fig1a}}
	\subfloat[identified parameters]{\includegraphics[scale=1]{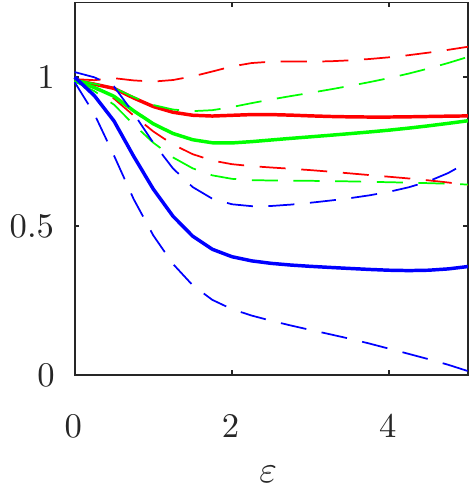}\label{SubSect:BCMA:Fig1b}}\hspace{0.5em}
	\subfloat[$\bs{\mathsf{u}}(\partial\Omega_\mathrm{mve})$ for~$\varepsilon = 5$]{\includegraphics[scale=1]{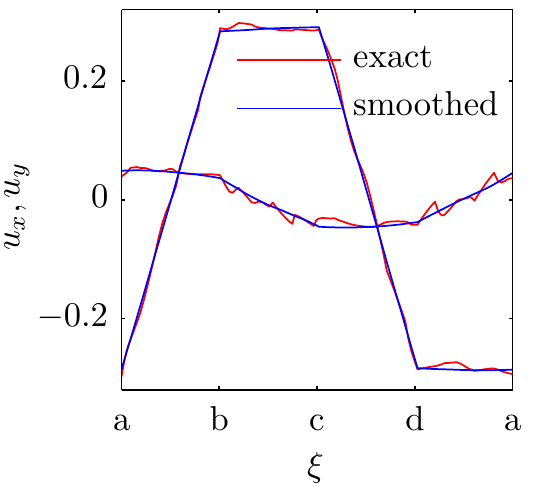}\label{SubSect:BCMA:Fig1c}}
	\caption{An example of identified results for a sheared specimen consisting of randomly distributed stiff inclusions embedded in a soft matrix, corresponding to~$50$ Monte Carlo realizations. Before identification, the exact boundary displacements are smoothed using the pillbox-shaped kernel with a dimensionless (normalized by inclusion's size) diameter~$\varepsilon \in [0, 5]$; the exact and smoothed displacements are compared in~(c). For identification, $\Omega_\mathrm{mve} = \Omega_\mathrm{roi}$ and zero image noise are used. A single parameter identification of the matrix shear modulus ($\lambda = G_1$) is shown in~(a); multiple parameter identification of the matrix and inclusion's shear moduli together with inclusion's bulk modulus ($\bs{\lambda} = [G_1,G_2,K_2]^\mathsf{T}$) are shown in~(b). The thick lines correspond to the mean values whereas shaded areas delimit the standard deviations over all realizations.}
	\label{SubSect:BCMA:Fig1}
\end{figure}

The first aim of this paper is therefore to systematically quantify the effects of inaccuracies in prescribed BCs on the accuracy of the identification by means of virtual experiments. Some of the obtained results can already be inferred from Fig.~\ref{SubSect:BCMA:Fig1}, where effects of kinematic smoothing are demonstrated. Without going into many details, about which the interested reader is invited to learn more in Sections~\ref{Sect:Method} and~\ref{Sect:BCErrors}, we note that the identified parameters rapidly deviate from their exact value with increasing smoothing kernel size~$\varepsilon$. The exact and smoothed BCs, for the worst case considered ($\varepsilon = 5$), are shown in Fig.~\ref{SubSect:BCMA:Fig1c}, indicating that small deviations are at the root of relatively poor identification. This kind of behaviour is typical and can be explained by extensive constraints of the MVE system by Dirichlet BCs applied along the entire boundary, and by associated sensitivity fields of low magnitudes, as we will detail in Section~\ref{Sect:BCErrors}.

The second objective of this paper is to provide a methodology ensuring the desired high accuracy in identifying material parameters and boundary data. The proposed approach essentially incorporates all Degrees Of Freedom~(DOFs) associated with boundary nodes of the MVE model as DOFs in the IDIC procedure, and will be referred to as Boundary-Enriched Integrated Digital Image Correlation~(BE-IDIC) in what follows. The method significantly improves the accuracy of the identified parameters while being robust with respect to image noise and material contrast ratio. Although this methodology may resemble the one proposed by~\cite{Fedele:2015}, in which kinematic BCs are also introduced as DOFs of the micromechanical parameter identification routine, important differences exist. These differences will be discussed in detail in Section~\ref{Sect:OurIDIC}, along with a detailed description of the BE-IDIC. The paper finally closes with a summary and conclusions in Section~\ref{Sect:Conclusion}.

%
%
\section{Theory and Problem Statement}
\label{Sect:Method}
The basics of DIC, needed for subsequent developments, are first recalled in this section. Next, three mechanical tests are described that serve to demonstrate the sensitivity of the IDIC technology to Dirichlet BCs. In Section~\ref{Sect:OurIDIC}, the same mechanical tests will be used to assess the BE-IDIC approach. Next, the constitutive model employed throughout this work is specified, and sensitivity fields are shown. Finally, the speckle pattern and creation of deformed images are briefly described.
%
%
\subsection{Digital Image Correlation}
\label{SubSect:DIC}
In its simplest form, DIC correlates two images captured during an experiment, one in the reference configuration and one deformed. These images are in essence scalar fields supported in the FOV, storing usually gray level values (e.g. integer numbers ranging~$[0, 255]$ when $8$-bit digitization is used). Upon defining a ROI, one aims to find a vector~$\bs{\lambda}$ of~$n_\lambda$ IDIC DOFs that minimizes in the least-square sense the difference between grey values in the reference image and in the corresponding material points in the deformed image as predicted by a displacement field~$\bs{u}$, i.e.
\begin{equation}
\begin{aligned}
\bs{\lambda} &\in \underset{\widehat{\bs{\lambda}}\in\mathbb{R}^{n_\lambda}}{\text{arg min}}\ \mathcal{R}(\widehat{\bs{\lambda}}), \\
\mathcal{R}(\widehat{\bs{\lambda}}) &= \frac{1}{2}\int_{\Omega_\mathrm{roi}}[f(\bs{X}) - g(\bs{X}+\bs{u}(\bs{X},\widehat{\bs{\lambda}}))]^2\,\mathrm{d}\bs{X}.
\end{aligned}
\label{SubSect:DIC:Eq1}
\end{equation}
In~\eqref{SubSect:DIC:Eq1}, $\bs{X} = [X_1,X_2]^\mathsf{T} \in \mathbb{R}^2$ stores the material coordinates in the reference configuration, and~$\bs{u}(\bs{X},\widehat{\bs{\lambda}}) = [u_1(\bs{X},\widehat{\bs{\lambda}}), u_2(\bs{X},\widehat{\bs{\lambda}})]^\mathsf{T}$ is an approximate displacement field that is required in order to regularize the otherwise ill-posed problem; for more details see e.g.~\cite{HORN:1981}. Throughout this work, the hatted variables~$\widehat{\bullet}$ relate to arbitrary admissible values, whereas the absence of hats indicates minimizers of the corresponding cost functional. As indicated by the inclusion sign~$\in$, the cost functional~$\mathcal{R}$ may be non-convex with multiple minima; in such a case, the global minimum is sought.

If the approximate field~$\bs{u}(\bs{X},\bs{\widehat{\lambda}})$ is chosen such that
\begin{equation}
\bs{u}(\bs{X},\bs{\widehat{\lambda}}) = \sum_{i = 1}^{n_\lambda}\bs{\psi}_i(\bs{X})\widehat{\lambda}_i,
\label{SubSect:DIC:Eq2}
\end{equation}
one recovers GDIC, where~$\bs{\psi}_i(\bs{X})$ are user-selected vector interpolation (or basis) functions, usually expressed in terms of globally- or locally-supported polynomials. The variable~$\widehat{\bs{\lambda}} = [\widehat{\lambda}_1,\dots,\widehat{\lambda}_{n_\lambda}]^\mathsf{T} \in \mathbb{R}^{n_\lambda}$ then constitutes an admissible vector of generalized displacements.

On the contrary, if
\begin{equation}
\bs{u}(\bs{X},\widehat{\bs{\lambda}}) \in \underset{\widehat{\bs{u}}(\bs{X},\widehat{\bs{\lambda}}) \in \mathscr{U}(\widehat{\bs{\lambda}})}{\text{arg min}}\ \mathcal{E}(\widehat{\bs{u}}(\bs{X},\widehat{\bs{\lambda}}),\widehat{\bs{\lambda}}),
\label{SubSect:DIC:Eq3}
\end{equation}
is a solution to an underlying (elastic for simplicity) mechanical system specified by its stored energy~$\mathcal{E}$ and a proper function space~$\mathscr{U}$ (see e.g.~\citealt{EvansPDE}), the IDIC method results. In practice, a Finite Element~(FE) discretization of~$\widehat{\bs{u}}(\bs{X},\widehat{\bs{\lambda}})$ is used (see e.g.~\citealt{zienkiewicz:Vol1,Ciarlet:FEM}), typically given by
\begin{equation}
\widehat{\bs{u}}(\bs{X},\widehat{\bs{\lambda}}) = \sum_{i=1}^{n_\mathrm{u}/2} N_i(\bs{X})\widehat{\bs{\mathsf{u}}}_i(\widehat{\bs{\lambda}}),
\label{SubSect:DIC:Eq4}
\end{equation}
where~$\widehat{\bs{\mathsf{u}}} = [\widehat{\bs{\mathsf{u}}}_1^\mathsf{T},\dots,\widehat{\bs{\mathsf{u}}}_{n_\mathrm{u}/2}^\mathsf{T}]^\mathsf{T} \in \mathbb{R}^{n_\mathrm{u}}$, $\widehat{\bs{\mathsf{u}}}_i = [\widehat{\mathsf{u}}_1^i, \widehat{\mathsf{u}}_2^i]^\mathsf{T} \in \mathbb{R}^2$, stores displacements of the $i$-th node associated with a FE mesh in~$X_1$ and~$X_2$ directions, and~$N_i(\bs{X})$ are standard FE shape functions. In IDIC, $\widehat{\bs{\lambda}}$ can store kinematic variables such as prescribed BCs, or material constants---hence the dependence of~$\mathscr{U}$ as well as~$\mathcal{E}$ on~$\widehat{\bs{\lambda}}$. Similarly to the DIC cost functional~$\mathcal{R}$ specified in Eq.~\eqref{SubSect:DIC:Eq1}, $\mathcal{E}$ may be non-convex, allowing, e.g., for structural buckling and bifurcation.

In order to minimize~\eqref{SubSect:DIC:Eq1}, various approaches are being used. Although working only in the proximity of a local minimum, the most frequently employed one is a standard Newton, or more precisely a Gauss-Newton algorithm, that iteratively solves the following system of linear equations (obtained by a Taylor expansion of the first-order optimality conditions in~$\widehat{\bs{\lambda}}$):
\begin{equation}
\bs{H}^l(\widehat{\bs{\lambda}}^{l+1}-\widehat{\bs{\lambda}}^l) = -\bs{g}^l.
\label{SubSect:DIC:Eq5}
\end{equation}
The individual components of the gradient~$\bs{g}$ and Hessian~$\bs{H}$, derived by differentiating~\eqref{SubSect:DIC:Eq1}, read
\begin{equation}
\begin{aligned}
(g^l)_i &= (g(\widehat{\bs{\lambda}}^l))_i = \left. -\int_{\Omega_\mathrm{roi}} \bs{\varphi}_i(\bs{X},\widehat{\bs{\lambda}}) \cdot \nabla f(\bs{X}) \left[f(\bs{X})-g(\bs{X}+\bs{u}(\bs{X},\bs{\widehat{\bs{\lambda}}}))\right]\,\mathrm{d}\bs{X}\right|_{\widehat{\bs{\lambda}}=\widehat{\bs{\lambda}}^l}, \\
(H^l)_{ij} &= (H(\widehat{\bs{\lambda}}^l))_{ij} = \left. \int_{\Omega_\mathrm{roi}} \bs{\varphi}_i(\bs{X},\widehat{\bs{\lambda}}) \cdot \nabla f(\bs{X}) \nabla f(\bs{X}) \cdot \bs{\varphi}_j(\bs{X},\widehat{\bs{\lambda}})\,\mathrm{d}\bs{X}\right|_{\widehat{\bs{\lambda}}=\widehat{\bs{\lambda}}^l}. \\
\end{aligned}
\label{SubSect:DIC:Eq6}
\end{equation}
Note that~$\nabla(\bullet) = \partial(\bullet)/\partial \bs{X}$, and that a simplified version of the Hessian is used here, see~\cite{Neggers:NME:2016} for further details. In Eqs.~\eqref{SubSect:DIC:Eq6}, the so-called sensitivity fields, defined as
\begin{equation}
\bs{\varphi}_i(\bs{X},\widehat{\bs{\lambda}}) = \frac{\partial\bs{u}(\bs{X},\widehat{\bs{\lambda}})}{\partial\widehat{\lambda}_i}, \quad i = 1, \dots, n_\lambda,
\label{SubSect:DIC:Eq7}
\end{equation}
are required. In the case of GDIC, $\bs{\varphi}_{i}(\bs{X}) = \bs{\psi}_i(\bs{X})$, $i = 1,\dots,n_\lambda$, whereas in the case of IDIC, $\bs{\varphi}_i(\bs{X},\widehat{\bs{\lambda}})$ are obtained usually by numerical perturbations of the FE solution, i.e.

\begin{equation}
\bs{\varphi}_i(\bs{X},\widehat{\bs{\lambda}}) = \frac{\bs{u}(\bs{X},\widehat{\bs{\lambda}}+\epsilon\widehat{\lambda}_i\bs{e}_i) - \bs{u}(\bs{X},\widehat{\bs{\lambda}})}{\epsilon\widehat{\lambda}_i}, \quad i = 1, \dots, n_\lambda,
\label{SubSect:DIC:Eq8}
\end{equation}
as explicit forms of the partial derivatives in Eq.~\eqref{SubSect:DIC:Eq7} are rarely available. In Eq.~\eqref{SubSect:DIC:Eq8}, $\epsilon > 0$ is a sufficiently small scalar perturbation factor (set to~$\epsilon =1 \cdot 10^{-3}$ in all examples below), $\widehat{\lambda}_i$ is the $i$-th component of~$\widehat{\bs{\lambda}}$, and~$\bs{e}_i$ denotes the $i$-th standard basis vector in~$\mathbb{R}^{n_\lambda}$. 

In order to solve the elastic mechanical minimization problem specified in Eq.~\eqref{SubSect:DIC:Eq3}, standard solution techniques can be used, see e.g.~\cite{zienkiewicz:Vol2,Crisfield:vol1,Jirasek:2002,BonnansOptim,Nocedal:Optim}.
%
%
\subsection{Considered Virtual Laboratory Tests}
\label{SubSect:Tests}
\begin{figure}
	\centering
	\subfloat[specimen geometry]{\includegraphics[scale=1]{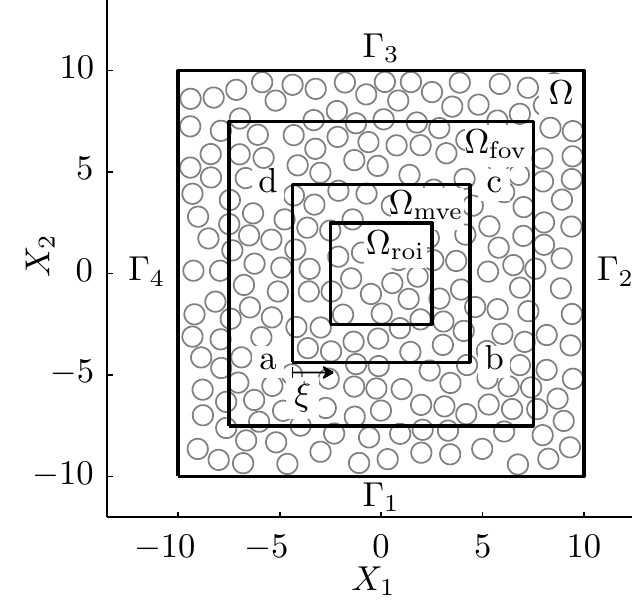}\label{SubSect:Tests:Fig1a}}\hspace{3em}
	\subfloat[pure bending test]{\includegraphics[scale=1]{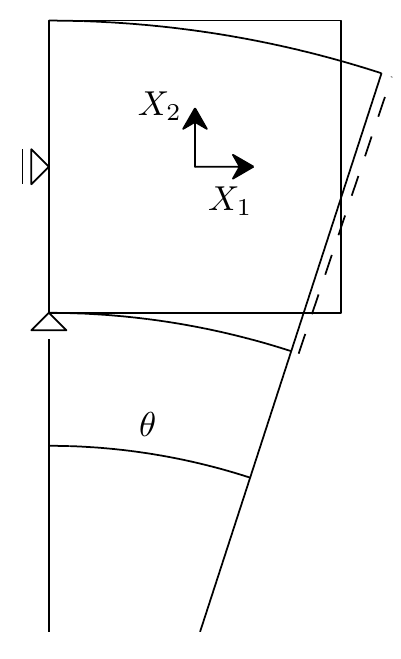}\label{SubSect:Tests:Fig1b}}
	\caption{Sketch of the considered mechanical tests---tension, shear, and bending. (a)~Specimen geometry, (b)~pure bending. $\Omega$ denotes specimen domain, $\Omega_\mathrm{fov}$~corresponds to the field of view,  $\Omega_\mathrm{roi}$ to the region of interest, and~$\Omega_\mathrm{mve}$ is the microstructural volume element representing the mechanical system in IDIC.}
	\label{SubSect:Tests:Fig1}
\end{figure}

Three virtual mechanical tests will be considered, which predominantly introduce tension, shear, and bending, respectively. They reflect different mechanical behaviour, and most importantly yield different sensitivity fields with respect to individual material parameters. This is important especially when for instance a shear test is performed and the bulk modulus is to be identified. Because the sensitivity of the bulk modulus is in this particular case low (cf. Section~\ref{SubSect:Sfields}), one can expect large errors in the identified values. In order to identify all parameters accurately and reliably, multiple tests may be carried out. All specimen geometries, BCs, ROI, FOV, and MVE are sketched in Fig.~\ref{SubSect:Tests:Fig1}. Here, one particular realization of randomly distributed inclusions with a fixed diameter~$d = 1$ in a homogeneous matrix is depicted as well. All geometric properties are dimensionless, but can be thought of as~$[\mu\mathrm{m}]$ for microscale images. This is done for compactness, as the material models used are size insensitive.

The displacements prescribed at the specimen's boundary, $\partial\Omega = \bigcup\limits_{i=1}^{4}\Gamma_i$, in the case of tension and shear read
\begin{equation}
\begin{aligned}
\bs{u}(\bs{X}) &= (\overline{\bs{F}}-\bs{I})\cdot\bs{X}, && \bs{X} \in \Gamma_2 \cup \Gamma_4, \\
\overline{\bs{F}} &= \bs{I}+0.1\,\bs{e}_1\otimes\bs{e}_1, && \text{for tension}, \\
\overline{\bs{F}} &= \bs{I}+0.1\,\bs{e}_2\otimes\bs{e}_1, && \text{for shear},
\end{aligned}
\label{Sect:Tests:Eq1}
\end{equation}
whereas~$\Gamma_1$ and~$\Gamma_3$ are free edges. In the case of bending, prescribed boundary conditions read
\begin{equation}
\begin{aligned}
u_1(\bs{X}) &= 0, && \bs{X} \in \Gamma_4, \\
\bs{u}(\bs{X}) &= \bs{0}, && \bs{X} = \Gamma_1 \cap \Gamma_4, \\
\bs{n}(\theta) \cdot ( \bs{X} + \bs{u}(\bs{X}) - \bs{X}_0 - \bs{u}(\bs{X}_0) ) &= 0, && \bs{X} \in \Gamma_2, \bs{X}_0 \in \Gamma_2 \mbox{ arbitrary but fixed}, 
\end{aligned}
\label{Sect:Tests:Eq2}
\end{equation}
where~$\bs{n}(\theta) = [\cos{\theta}, -\sin{\theta}]^\mathsf{T}$, $\theta \in [0, \pi/24]$, is the outer unit normal to the rotated boundary edge~$\Gamma_2$ inducing the bending effect, $\bs{e}_1 = (1, 0)^\mathsf{T}$, $\bs{e}_2 = (0, 1)^\mathsf{T}$, $(\bs{A}\cdot\bs{b})_i = A_{ij}b_j$ and~$\bs{a} \cdot \bs{b} = a_ib_i$ denote the single contraction with implicitly implied summation rule, and~$\bs{u}(\bs{X})$, $\bs{X} \in \Gamma$, is to be interpreted as displacements located on~$\Gamma$. The two horizontal edges, $\Gamma_1$ and~$\Gamma_3$, are left free again. After discretization, Eq.~\eqref{Sect:Tests:Eq2}$_3$ is enforced for all~$n_{\Gamma_2}$ nodes situated on~$\Gamma_2$ part of the boundary. This yields a system of~$n_{\Gamma_2}-1$ equations that can be enforced as a set of linear constraints
\begin{equation}
\bs{\mathsf{C}}(\theta)\bs{\mathsf{u}} = \bs{\mathsf{d}}.
\label{Sect:Tests:Eq3}
\end{equation}
The mechanical problem in Eq.~\eqref{SubSect:DIC:Eq3} then transforms to an equality constrained minimization, which can be solved using, e.g., the primal-dual formulation; for further details see~\cite{BonnansOptim} or~\cite{Nocedal:Optim}.
%
%
\subsection{Constitutive Model}
\label{SubSect:ConstModel}
\begin{figure}
	\centering
	\subfloat[tension, $F_{11}-1$]{\includegraphics[scale=1]{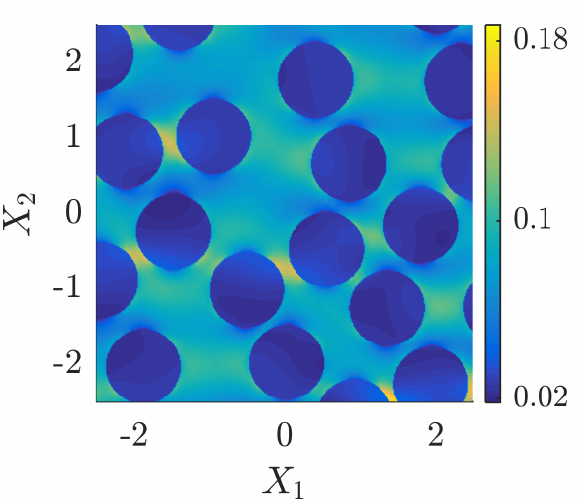}\label{Sect:ConstModel:Fig0a}}
	\subfloat[shear, $F_{21}$]{\includegraphics[scale=1]{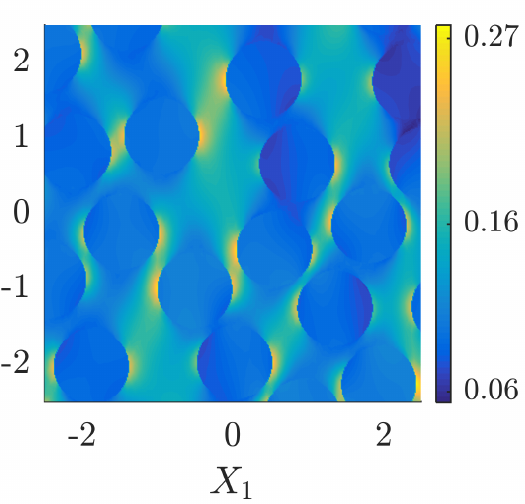}\label{Sect:ConstModel:Fig0b}}
	\subfloat[bending, $F_{11}-1$]{\includegraphics[scale=1]{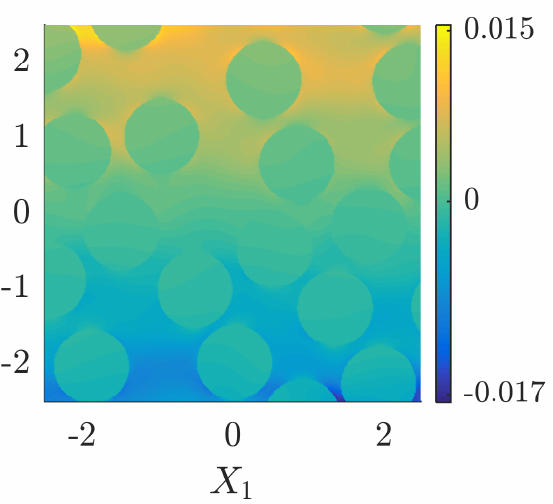}\label{Sect:ConstModel:Fig0c}}
	\caption{Typical realizations of resulting DNS strain fields corresponding to individual mechanical tests. (a)~$F_{11}(\bs{X}) - 1$ for the tension test, (b)~$F_{21}(\bs{X})$ for the shear test, and~(c) $F_{11}(\bs{X}) - 1$ for the bending test. In all cases, $\bs{X} \in \Omega_\mathrm{roi}$.}
	\label{Sect:ConstModel:Fig0}
\end{figure}

A compressible Neo-Hookean hyperelastic material is adopted, specified by the following elastic energy density
\begin{equation}
W_\alpha(\bs{F}) = \frac{1}{2}G_\alpha(\overline{I}_1-3)+\frac{1}{2}K_\alpha(\ln(J))^2,
\label{Sect:ConstModel:Eq1}
\end{equation}
where~$\alpha = 1$ corresponds to the matrix and~$\alpha = 2$ to the inclusions. In Eq.~\eqref{Sect:ConstModel:Eq1}, $\bs{F}(\bs{u}(\bs{X})) = \bs{I} + \nabla\bs{u}(\bs{X})$ denotes the deformation gradient tensor (recall that~$\bs{X}$ relates to the reference configuration), $J = \det{\bs{F}}$, and~$\overline{I}_1 = J^{-2/3}\,\tr(\bs{C})$ is the first modified invariant of the right Cauchy--Green deformation tensor~$\bs{C} = \bs{F}^\mathsf{T} \cdot \bs{F}$. The reference values of material parameters are summarized in Tab.~\ref{Sect:ConstModel:Tab1} as functions of the material contrast ratio~$\rho > 1$. The underlying mechanical system, occupying domain~$\Omega$, is then specified by its stored energy
\begin{equation}
\mathcal{E}(\bs{u}(\bs{X})) = \int_\Omega \chi_1(\bs{X})W_1(\bs{F}(\bs{u}(\bs{X})))+\chi_2(\bs{X})W_2(\bs{F}(\bs{u}(\bs{X})))\,\mathrm{d}\bs{X},
\label{Sect:ConstModel:Eq2}
\end{equation}
and by Dirichlet BCs reflected by the space of admissible solutions~$\mathscr{U}$; Neumann BCs are omitted, as these are typically not experimentally available. In Eq.~\eqref{Sect:ConstModel:Eq2}, $\chi_1(\bs{X})$ and~$\chi_2(\bs{X})$ are indicator functions associated with the matrix and inclusions. For the solution of the mechanical system, recall Eq.~\eqref{SubSect:DIC:Eq3}, the Total Lagrangian formulation is used, see e.g.~\cite{tadmor:2012:Continuum}. Spatial discretization relies on the Gmsh mesh generator, presented by~\cite{gmsh}, employing quadratic iso-parametric triangular elements and the three-point Gaussian quadrature rule. For the Direct Numerical Simulation~(DNS), the fine mesh shown in Fig.~\ref{SubSect:BCGDIC:Fig1a} is used, whereas three typical MVE triangulations can be found in~Fig.~\ref{SubSect:BEIDIC:Fig2}. Because both Poisson's ratios are significantly smaller than~$0.5$ and because deformations in all simulations remain moderate, no incompressibility issues arise. Typical DNS results are presented in terms of strain fields in Fig.~\ref{Sect:ConstModel:Fig0}. The results show that, in accordance with Eq.~\eqref{Sect:Tests:Eq1}, the overall strain for the tension and shear test corresponds to~$10\,\%$, whereas peak strains achieve values as high as~$27\,\%$. In the case of bending, the overall strain is zero, whereas peak values achieve approximately~$2\,\%$ of strain.
\begin{table}
	\centering
	\caption{Material parameters for all employed examples.}
	\renewcommand{\arraystretch}{1.5}
	\begin{tabular}{l|r@{}lr@{}l}
		Physical parameters                                                     & 	\multicolumn{2}{c}{\renewcommand{\arraystretch}{0.8}\begin{tabular}{@{}c@{}}
			    matrix     \\
			{\footnotesize($\alpha = 1$)}
		\end{tabular}} & \multicolumn{2}{c}{\renewcommand{\arraystretch}{0.8}\begin{tabular}{@{}c@{}} inclusions \\ {\footnotesize($\alpha = 2$)} \end{tabular}}  \\\hline
		Shear modulus, \hfill $G_\alpha$                                                  & 1 &                        &  $\rho$ &                        \\
		Bulk modulus, \hfill $K_\alpha$                                                   & 3 &                        & $3\rho$ &                        \\
		Poisson's ratio, $\nu_\alpha = \frac{3K_\alpha-2G_\alpha}{2(3K_\alpha+G_\alpha)}$ & 0 & .35                    &       0 & .35                  
	\end{tabular}
	\label{Sect:ConstModel:Tab1}
\end{table}  

In the context of IDIC, the parameters to be identified are
\begin{itemize}
\item the matrix shear and bulk moduli~$G_1$ and~$K_1$ 
\item the inclusions' shear and bulk moduli~$G_2$ and~$K_2$.
\end{itemize}
As Dirichlet BCs are applied on the entire boundary of the MVE, $\partial\Omega_\mathrm{mve}$, only material parameter ratios can be extracted from the IDIC procedure. This holds true unless additional measurements, such as the applied load, are included in the objective function defined in Eq.~\eqref{SubSect:DIC:Eq1}, which is not done here as such data are not readily accessible in micro-mechanical testing of a microstructure; recall the discussion in the introduction. As a consequence, in order to induce normalization one needs to fix one of the parameters to an arbitrary value (exact in our case of virtual experiments), and identify the remaining parameters relative to that reference. The fixed material parameter can be estimated by an independent force-based mechanical test or from reliable experimental sources for one of the phases.
%
%
\subsection{Sensitivity Fields}
\label{SubSect:Sfields}
\begin{figure}
	\centering
	\mbox{}\hspace{1.5em}\includegraphics[scale=1]{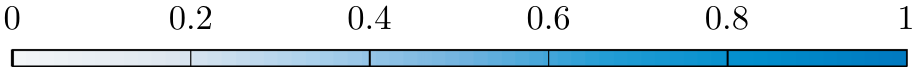}\vspace{-1.0em}\\
	\subfloat[$\max{\widetilde{\varphi}_{G_1}} = 5.404 \cdot 10^{-2}$]{\includegraphics[scale=1]{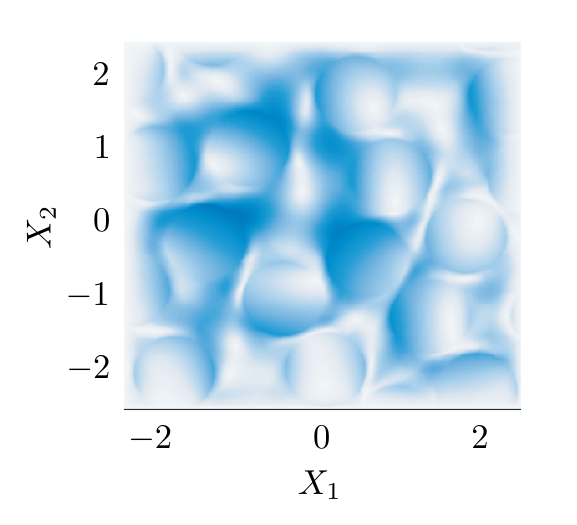}\label{SubSect:ConstModel:Fig1a}}\hspace{1em}
	\subfloat[$\max{\widetilde{\varphi}_{K_1}} = 1.693 \cdot 10^{-2}$]{\includegraphics[scale=1]{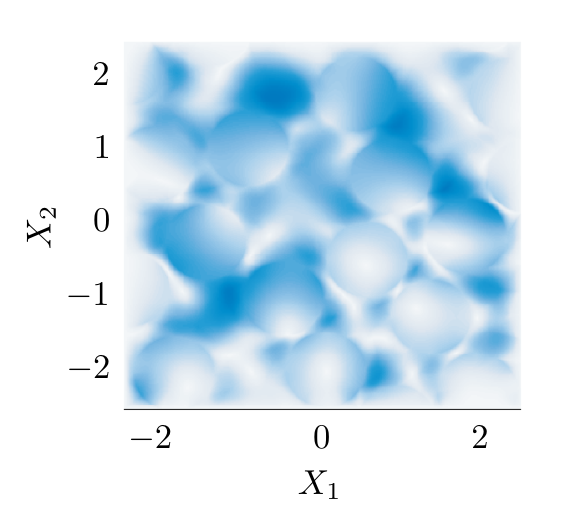}\label{SubSect:ConstModel:Fig1b}}\\
	\subfloat[$\max{\widetilde{\varphi}_{G_2}} = 5.710 \cdot 10^{-2}$]{\includegraphics[scale=1]{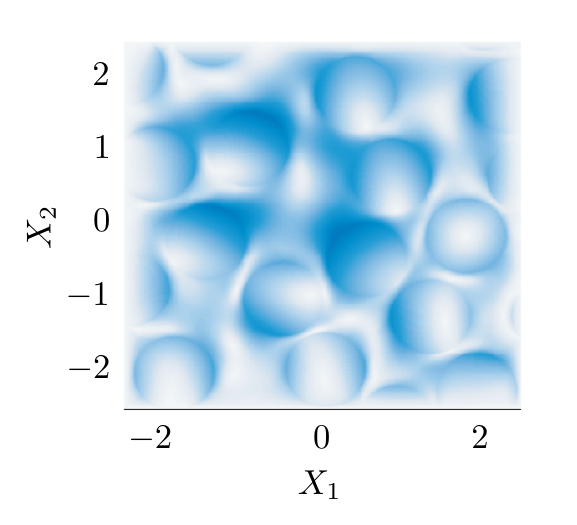}\label{SubSect:ConstModel:Fig1c}}\hspace{1em}
	\subfloat[$\max{\widetilde{\varphi}_{K_2}} = 5.083 \cdot 10^{-3}$]{\includegraphics[scale=1]{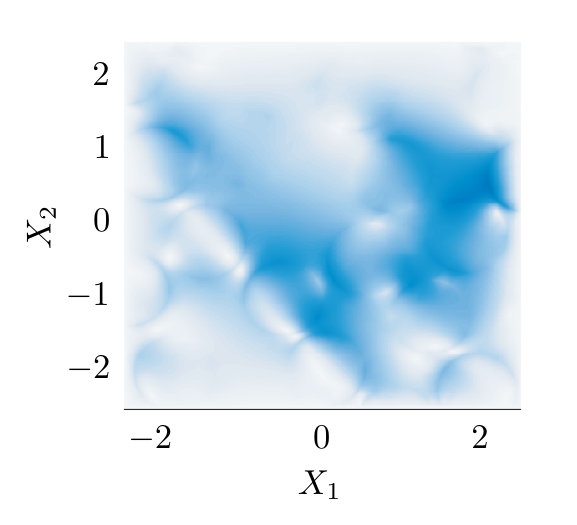}\label{SubSect:ConstModel:Fig1d}}\\
	\caption{Normalized sensitivity fields~$\widetilde{\varphi}_i$, recall Eq.~\eqref{Sect:ConstModel:Eq3}, evaluated for the shear test, exact material and kinematic data, $\rho = 4$, and for~$\Omega_\mathrm{mve} = \Omega_\mathrm{roi}$. For clarity, presented plots are normalized to one whereas corresponding magnitudes are mentioned in individual captions. Sensitivities correspond to~(a) shear modulus of the matrix~$G_1$, (b)~bulk modulus of the matrix~$K_1$, (c)~shear modulus of the inclusions~$G_2$, and~(d) bulk modulus of the inclusions~$K_2$.}
	\label{SubSect:ConstModel:Fig1}
\end{figure}

The normalized sensitivity fields corresponding to the shear test, exact Dirichlet BCs applied to~$\partial\Omega_\mathrm{mve}$, and all material parameters for~$\rho = 4$, are shown in Fig.~\ref{SubSect:ConstModel:Fig1} inside the ROI ($\Omega_\mathrm{mve} = \Omega_\mathrm{roi}$). The adopted normalization reads
\begin{equation}
\widetilde{\varphi}_i(\bs{X},\widehat{\bs{\lambda}}) = \frac{|\lambda_i| \, \| \bs{\varphi}_i(\bs{X},\widehat{\bs{\lambda}})\|_2 }{\max_{\bs{Y}\in\Omega_\mathrm{roi}} \|\bs{u}(\bs{Y},\widehat{\bs{\lambda}})\|_2  },
\label{Sect:ConstModel:Eq3}
\end{equation}
i.e. the magnitude of the sensitivity field is normalized by the peak displacement measured inside ROI over the value of the IDIC DOF. Fig.~\ref{SubSect:ConstModel:Fig1} shows that the sensitivity field corresponding to the inclusion's bulk modulus~$K_2$ (Fig.~\ref{SubSect:ConstModel:Fig1d}) is one order of magnitude smaller compared to the remaining sensitivity fields. This implies that lower accuracy in identified parameter~$K_2$ compared to~$G_1$, $G_2$, and~$K_1$ should be expected. Furthermore, patterns corresponding to the two shear moduli~$G_1$ and~$G_2$ (shown in Figs.~\ref{SubSect:ConstModel:Fig1a} and~\ref{SubSect:ConstModel:Fig1c}) are surprisingly similar, meaning that accurate identification of associated material parameters may be compromised because a change in one parameter has almost the same (or the opposite) mechanical effect as a change in the other parameter. Similarity of two sensitivity fields is quantified by their cross-correlation, attaining the value~$\corr (\bs{\varphi}_{G_1},\bs{\varphi}_{G_2}) \approx -0.945$ in the case of~$G_1$ and~$G_2$, whereas cross-correlations of the remaining combinations are smaller than~$0.35$ in their absolute values.

\begin{figure}
	\centering
	\includegraphics[scale=1]{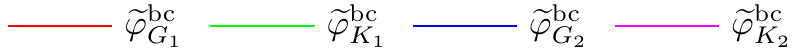}\vspace{-0.75em}\\
	\subfloat[tension]{\includegraphics[scale=1]{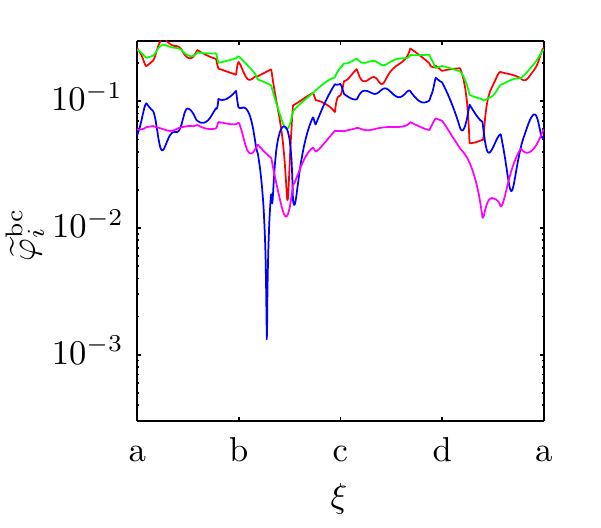}\label{SubSect:ConstModel:Fig2a}}
	\subfloat[shear]{\includegraphics[scale=1]{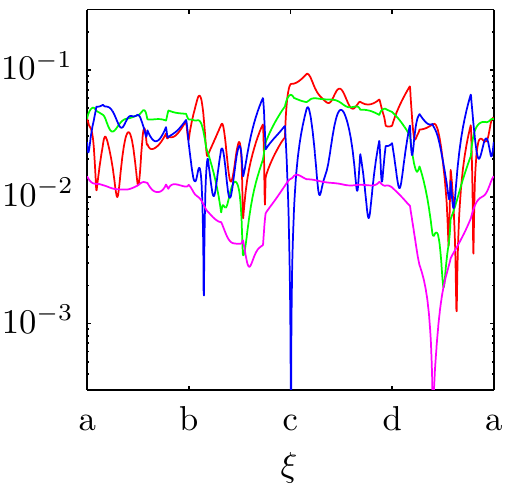}\label{SubSect:ConstModel:Fig2b}}
	\subfloat[bending]{\includegraphics[scale=1]{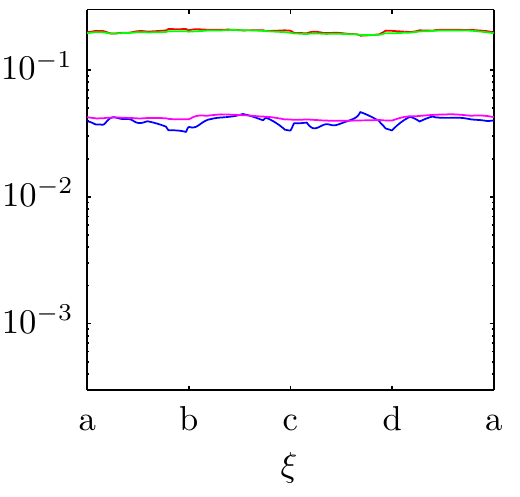}\label{SubSect:ConstModel:Fig2c}}
	\caption{Boundary sensitivity functions~$\widetilde{\varphi}_i^\mathrm{bc}(\xi,\widehat{\bs{\lambda}})$ on the MVE boundary for~(a) tension, (b)~shear, and~(c) bending tests corresponding to material contrast ratio~$\rho = 4$ and~$\Omega_\mathrm{mve} = \Omega_\mathrm{roi}$.}
	\label{SubSect:ConstModel:Fig2}
\end{figure}

Further, we introduce boundary sensitivity functions~$\bs{\varphi}_i^\mathrm{bc}(\xi,\widehat{\bs{\lambda}})$, defined as traces (on~$\partial\Omega_\mathrm{mve}$) of material sensitivity fields associated with the DNS. They are obtained according to the definition of Eq.~\eqref{SubSect:DIC:Eq8} with the only difference that they are computed over the entire domain~$\Omega$, evaluated at~$\partial\Omega_\mathrm{mve}$, and expressed as functions of~$\xi$, which is a parametric coordinate along~$\partial\Omega_\mathrm{mve}$ (see Fig.~\ref{SubSect:Tests:Fig1}). The boundary sensitivity functions normalized according to Eq.~\eqref{Sect:ConstModel:Eq3} are denoted~$\widetilde{\varphi}_i^\mathrm{bc}(\xi,\widehat{\bs{\lambda}})$ and presented in Fig.~\ref{SubSect:ConstModel:Fig2}. By definition, $\widetilde{\varphi}_i^\mathrm{bc}$ measure how the DNS displacements at the MVE boundary change under perturbations of the material parameters.\footnote{It is important to realize that when MVE boundary conditions are fixed during an IDIC minimization procedure (GDIC-IDIC approach), boundary sensitivity functions are not part of the optimization problem. Hence, $\widetilde{\varphi}_i^\mathrm{bc}$ measure how rapidly solutions to approximate optimization problems (with erroneous boundary conditions) deviate from the solutions corresponding to the correct optimization problems (with the exact boundary data). Although derived conclusions hold only in the vicinity of a given configuration of the system due to linearization (providing thus only qualitative information), low absolute values of boundary sensitivity functions confirm the importance of the accuracy in the prescribed MVE BCs.} These curves reveal that, in the case of shear for instance, when the material parameters change in the order of~$100\,\%$, the boundary displacements change on average in the order of~$3\,\%$ relative to their peak values. Notice also that various parts of the boundary react differently: whereas for shear and tension the vertical MVE boundaries change less under perturbations in material parameters than the horizontal boundaries, in the case of bending the boundary sensitivity functions are almost constant. Moreover, the tension test is approximately one order of magnitude more robust compared to the shear and bending tests; this observation may be useful in real experiments, or may serve for the design of experiments that are optimal with respect to boundary sensitivity functions. It is worth mentioning that although containing essential information, the boundary sensitivity functions require considerable computational effort for virtual experiments or DNSs.
%
%
%
%
\subsection{Speckle Pattern, Reference and Deformed Images}
\label{SubSect:Speckle}
\begin{figure}
	\centering
	\subfloat[speckle pattern]{\includegraphics[scale=1]{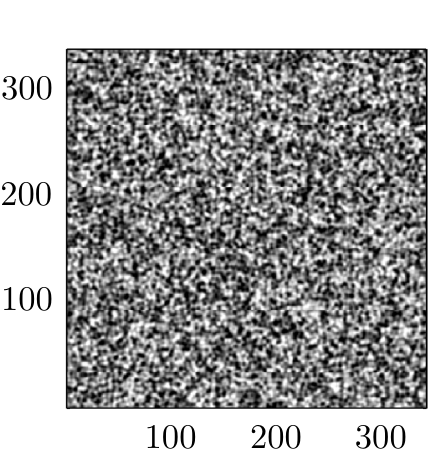}\label{SubSect:Speckle:Fig1a}}\hspace{1em}
	\subfloat[brightness histogram]{\includegraphics[scale=1]{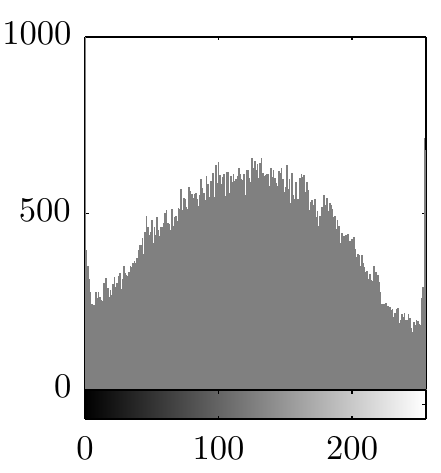}\label{SubSect:Speckle:Fig1b}}\hspace{1em}
	\subfloat[autocorrelation function, $\ell_\mathrm{c} = 2.18$ pixels]{\includegraphics[scale=1]{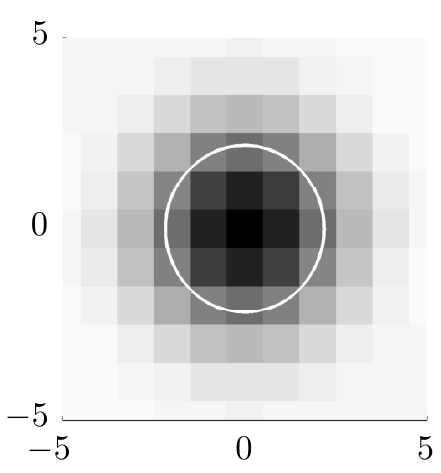}\label{SubSect:Speckle:Fig1c}}
	\caption{Initial image~$f$; (a)~speckle pattern inside ROI, (b)~corresponding histogram, and~(c) autocorrelation function.}
	\label{SubSect:Speckle:Fig1}
\end{figure}

The reference image~$f$, employed to represent the applied speckle pattern has been adopted from~\cite["medium pattern size"]{Bornert:2009} and is partly shown in Fig.~\ref{SubSect:Speckle:Fig1}. Its resolution is~$512 \times 512$ pixels inside FOV, which corresponds approximately to~$340 \times 340$ pixels inside ROI. For completeness, the corresponding histogram and autocorrelation function are shown as well. Additional image quality descriptors are summarized in Tab.~\ref{Sect:Speckle:Tab1}, where the correlation length~$\ell_\mathrm{c}$ is defined as the radial distance at which the autocorrelation function equals~$1/2$. More details about the mean intensity gradient~$\delta_f$ can be found e.g. in~\cite{Pan:2010}.

In order to produce deformed images~$g$ resulting from all mechanical tests, the DNS results (recall Fig.~\ref{Sect:ConstModel:Fig0}) are used. The computed displacement fields are used to map the initial image~$f$ into the deformed configuration in~$10$ time increments. Subsequently, the deformed images are interpolated at pixel positions using bi-cubic polynomial interpolation. Note that the peak displacements inside the ROI measure approximately to~$21$ (tension), $20$ (shear), and~$16$ (bending) pixels, i.e. relatively large displacements compared to the typical correlation length~$\ell_\mathrm{c} = 2.18$ reported in Tab.~\ref{Sect:Speckle:Tab1}.
\begin{table}
	\centering
	\caption{Speckle pattern parameters.}
	\renewcommand{\arraystretch}{1.5}
	\begin{tabular}{l@{}l|r@{}l}
		\multicolumn{2}{l|}{Pattern quality parameters}           & \multicolumn{2}{c}{image~$f$} \\ \hline
		Root-mean-square value,  & ~$\mathrm{RMS}$                 & 138 & .769                    \\
		Mean intensity gradient, & ~$\delta_f$                     &  38 & .940                    \\
		Correlation length,      & ~$\ell_\mathrm{c}$              &   2 & .179  pixels                  \\
		Quality factor,          & ~$Q = \delta_f/\ell_\mathrm{c}$ &  17 & .870 
	\end{tabular}
	\label{Sect:Speckle:Tab1}
\end{table}
%
%
\section{Influence of Inaccuracy in Kinematic Boundary Conditions}
\label{Sect:BCErrors}
Using the proposed methodology, models, test examples, and data presented in Section~\ref{Sect:Method}, the influence of two kinds of errors in BCs prescribed to the MVE model are next examined. First, the effects of uncorrelated random noise, followed by smoothing of kinematic fields, and finally the combined effect of both error sources stemming directly from the GDIC method itself are studied. In all cases, and throughout this paper, IDIC is always carried out for two images only (the reference and deformed ones at the beginning and at the end of all time increments), whereas GDIC is carried out as an evolutionary process at all increments due to its lower robustness with respect to large displacement changes. This has no practical implications except that multiple time steps help GDIC to locate the proper minimum.
%
%
\subsection{Influence of Random Noise}
\label{SubSect:BCNoise}
To quantify the effect of random noise, the following test is performed. $\Omega_\mathrm{mve} = \Omega_\mathrm{roi}$ is adopted and Dirichlet BCs are sampled by interpolating the DNS displacements directly at the nodal positions of the MVE boundary (i.e. without the use of GDIC). Note that below, all interpolations at nodal or pixel positions are carried out by inverting the iso-parametric mappings of the underlying FE approximations, unless explicitly stated otherwise. Subsequently, random uncorrelated noise is superimposed on the boundary displacement, i.e.
\begin{equation}
\bs{\mathsf{u}}_\mathrm{mve}(\bs{X}) = \bs{\mathsf{u}}_\mathrm{dns}(\bs{X}) + \sigma \max_{\bs{Y}\in\Omega_\mathrm{mve}}(\|\bs{u}_\mathrm{dns}(\bs{Y})\|_2)\,\bs{\mathcal{U}}, \quad \bs{X} \in \partial\Omega_\mathrm{mve},
\label{Sect:BCNoise:Eq1}
\end{equation}
where~$\bs{\mathsf{u}}_\mathrm{mve}(\bs{X})$, $\bs{X} \in \partial\Omega_\mathrm{mve}$, denotes a column storing the nodal displacements at the boundary nodes of the MVE, $\bs{u}_\mathrm{dns}(\bs{Y})$, $\bs{Y} \in \Omega_\mathrm{mve}$, denotes a vector of DNS displacements restricted on~$\Omega_\mathrm{mve}$, $\bs{\mathsf{u}}_\mathrm{dns}(\bs{X})$, $\bs{X} \in \partial\Omega_\mathrm{mve}$, is a column of DNS displacements~$\bs{u}_\mathrm{dns}$ evaluated at the MVE boundary nodes, $\bs{\mathcal{U}}$ is the corresponding column of Independent and Identically Distributed~(iid) random variables with uniform distribution over~$[-0.5, 0.5]$, and~$\sigma$ reflects the standard deviation of the random noise in the prescribed BCs. The iid variables can be used because of the rather homogeneous triangulations, see Fig.~\ref{SubSect:BEIDIC:Fig2}. In general, the noise in prescribed BCs has an experimental origin in image noise.

\begin{figure}
	\flushleft
	\mbox{}\hspace{4.5em}\includegraphics[scale=1]{BCNoiseLsigLegend.pdf}\vspace{-0.5em}\\
	\centering
	\subfloat[identified parameter]{\includegraphics[scale=1]{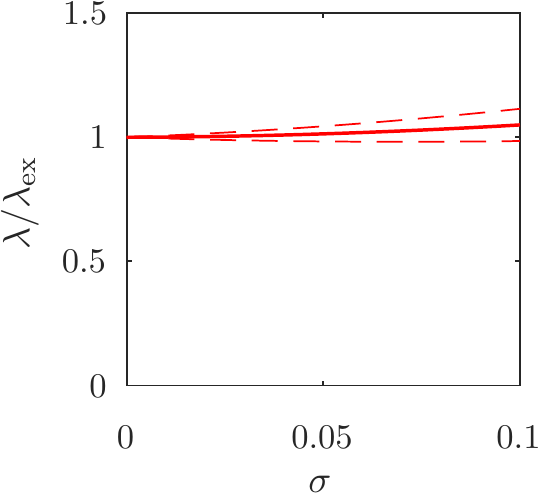}\label{SubSect:BCNoise:Fig1a}}
	\subfloat[identified parameters]{\includegraphics[scale=1]{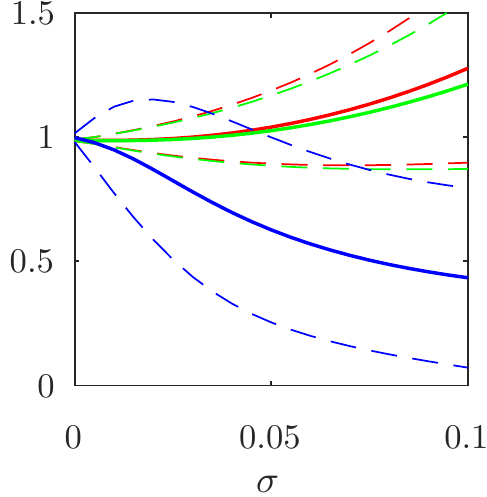}\label{SubSect:BCNoise:Fig1b}}\hspace{0.5em}
	\subfloat[$\bs{\mathsf{u}}(\partial\Omega_\mathrm{mve})$ for~$\sigma = 0.1$]{\includegraphics[scale=1]{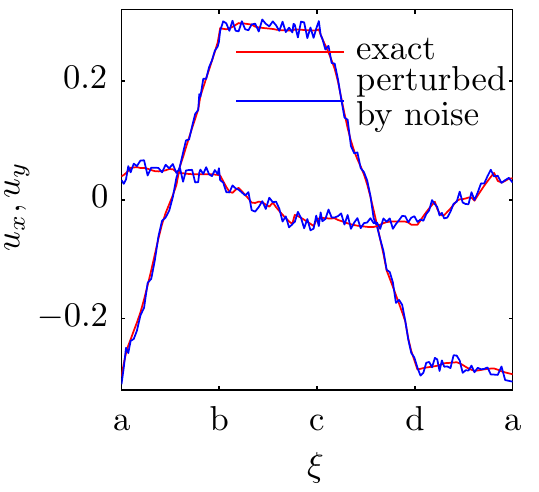}\label{SubSect:BCNoise:Fig1c}}
	\caption{Identified results in the case of the shear test, random noise in BCs with increasing~$\sigma \in [0, 0.1]$, cf. Eq.~\eqref{Sect:BCNoise:Eq1}, $\Omega_\mathrm{mve} = \Omega_\mathrm{roi}$, and zero image noise; (a)~$\lambda = G_1$, (b)~$\bs{\lambda} = [G_1,G_2,K_2]^\mathsf{T}$, and~(c) an example of boundary data for~$\sigma = 0.1$.}
	\label{SubSect:BCNoise:Fig1}
\end{figure}

\begin{figure}
	\flushleft
	\mbox{}\hspace{4.5em}\includegraphics[scale=1]{BCNoiseLsigLegend.pdf}\vspace{-0.5em}\\
	\centering
	\subfloat[tension]{\includegraphics[scale=1]{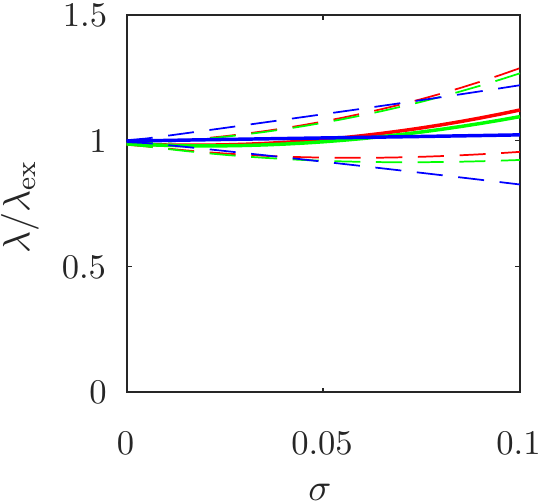}\label{SubSect:BCNoise:Fig2a}}
	\subfloat[bending]{\includegraphics[scale=1]{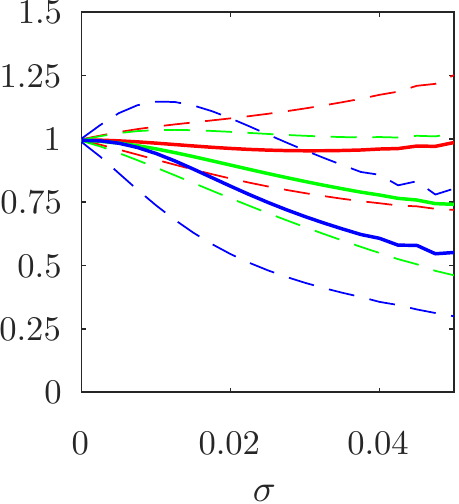}\label{SubSect:BCNoise:Fig2b}}\hspace{0.5em}
	\subfloat[displacement error at~$\partial\Omega_\mathrm{mve}$]{\includegraphics[scale=1]{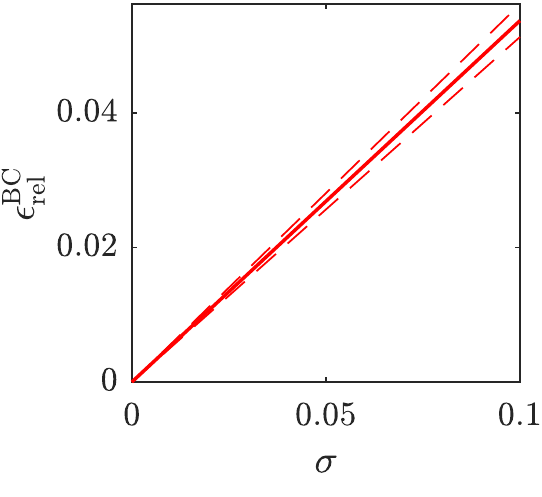}\label{SubSect:BCNoise:Fig2c}}
	\caption{Identified results in the case of random noise in BCs with increasing~$\sigma$, cf. Eq.~\eqref{Sect:BCNoise:Eq1}, $\Omega_\mathrm{mve} = \Omega_\mathrm{roi}$, and zero image noise; (a)~tension and~(b)~bending test for material parameters~$\bs{\lambda} = [G_1,G_2,K_2]^\mathsf{T}$. (c)~Typical dependence of the relative error in BCs, cf. Eq.~\eqref{Sect:BCNoise:Eq1a}, on~$\sigma$ for the case of tension.}
	\label{SubSect:BCNoise:Fig2}
\end{figure}

The results for the shear test, zero image noise, medium MVE mesh (shown in Fig.~\ref{SubSect:BEIDIC:Fig2b}), $\rho = 4$, $\sigma \in [0, 0.1]$, and~$50$ Monte Carlo~(MC) realizations for each value of~$\sigma$ with random noise in boundary data are presented in Fig.~\ref{SubSect:BCNoise:Fig1}. The peak noise displacement deviations (corresponding to~$\sigma = 0.1$) attain values of approximately~$0.5 \times 0.1 \times 20 = 1$ pixel (recall Section~\ref{SubSect:Speckle}). Note also that for each MC realization, a different microstructure with random spatial distribution of circular inclusions is generated in order to avoid any bias due to morphology. In Fig.~\ref{SubSect:BCNoise:Fig1}, the thick lines denote the mean values over all realizations, whereas dashed lines delimit the mean values~$\pm$ corresponding standard deviations. The results indicate that even though the identification of a single material parameter~$\lambda = G_1$ may be rather satisfactory (Fig.~\ref{SubSect:BCNoise:Fig1a}), the accuracy is compromised by the random noise in the case of multiple parameters~$\bs{\lambda} = [G_1, G_2, K_2]^\mathsf{T}$, as the curves start to deviate from~$1$ for values of~$\sigma$ as low as~$0.025$ (Fig.~\ref{SubSect:BCNoise:Fig1b}). The typical relative error in the prescribed boundary conditions, defined as
\begin{equation}
\epsilon_\mathrm{rel}^\mathrm{BC} = \frac{\|\bs{\mathsf{u}}_\mathrm{mve}(\bs{X}) - \bs{\mathsf{u}}_\mathrm{dns}(\bs{X})\|_2}{\|\bs{\mathsf{u}}_\mathrm{dns}(\bs{X})\|_2}, \quad \bs{X} \in \partial\Omega_\mathrm{mve},
\label{Sect:BCNoise:Eq1a}
\end{equation}
can be inspected in Fig.~\ref{SubSect:BCNoise:Fig2c}. Although not all presented, the remaining mechanical tests, material contrast ratios, and material parameter combinations display similar trends, cf. Fig.~\ref{SubSect:BCNoise:Fig2}, except for the tension test, which is more robust as already remarked in Section~\ref{SubSect:Sfields}, recall also Fig.~\ref{SubSect:ConstModel:Fig2}.

\begin{figure}
 	\centering
	\includegraphics[scale=1]{BCNoiseLsigLegend.pdf}\vspace{-0.5em}\\
	\subfloat[$\sigma_\mathrm{px} = 0.01$~px]{\includegraphics[scale=1]{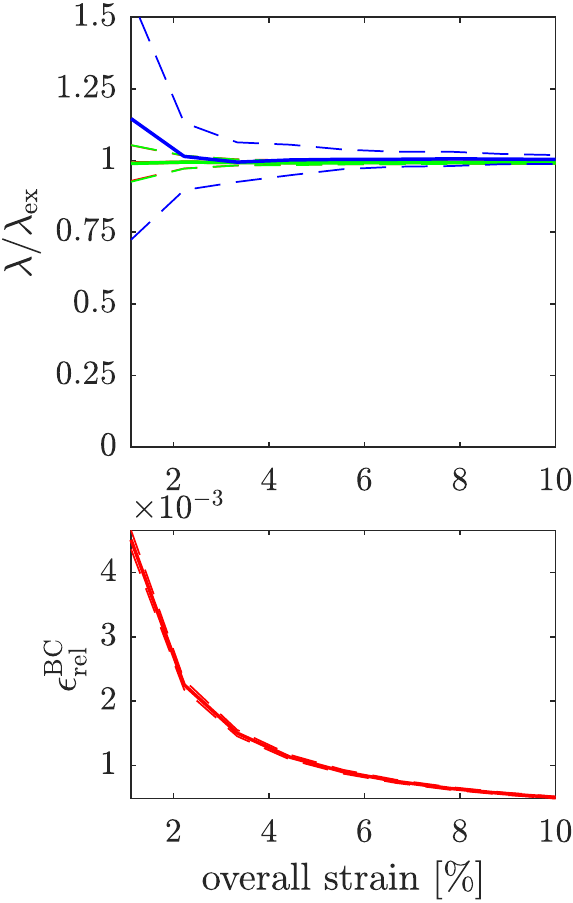}\label{SubSect:BCNoise:Fig3a}}
	\subfloat[$\sigma_\mathrm{px} = 0.10$~px]{\includegraphics[scale=1]{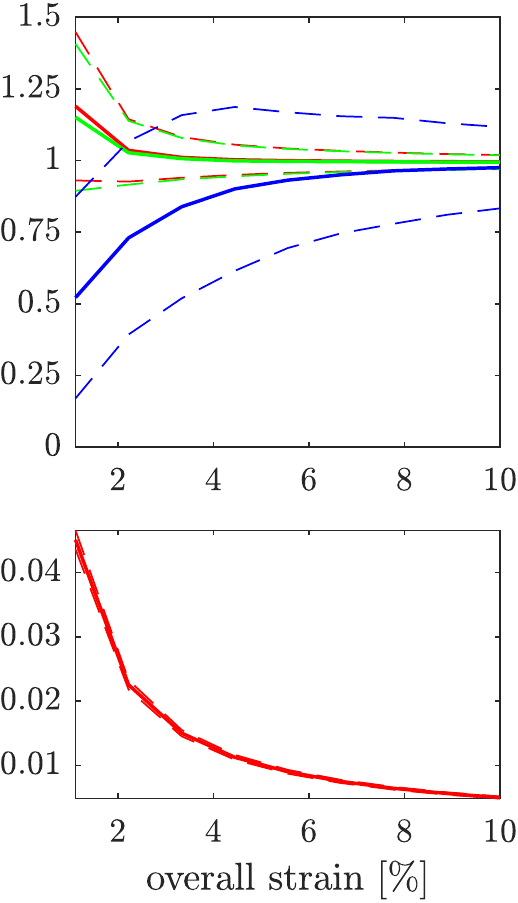}\label{SubSect:BCNoise:Fig3b}}
	\subfloat[$\sigma_\mathrm{px} = 0.25$~px]{\includegraphics[scale=1]{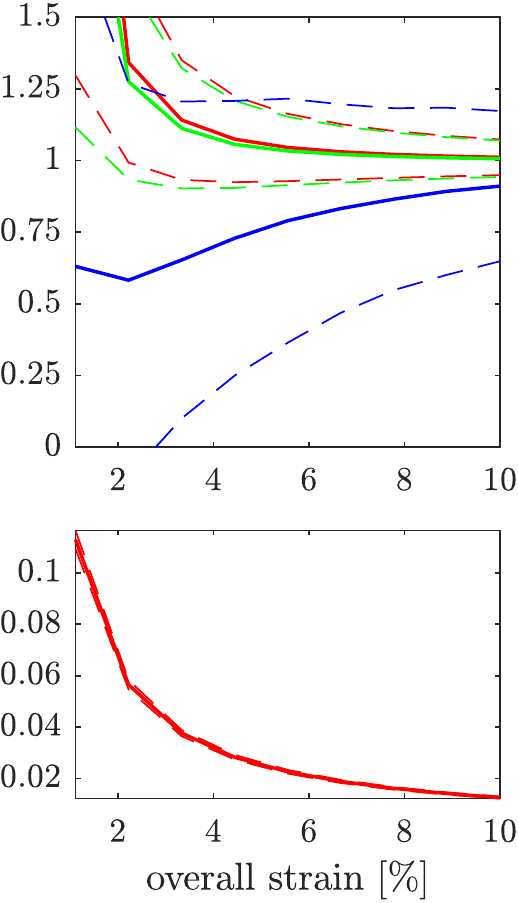}\label{SubSect:BCNoise:Fig3c}}
	\caption{Identified material parameters~$\bs{\lambda} = [G_1,G_2,K_2]^\mathsf{T}$ for the case of random noise in BCs with fixed magnitude~$\sigma_\mathrm{px}$, cf. Eq.~\eqref{Sect:BCNoise:Eq1b}, expressed as a function of overall applied strain. (a) $\sigma_\mathrm{px} = 0.01$, (b) $\sigma_\mathrm{px} = 0.1$, and (c) $\sigma_\mathrm{px} = 0.25$~px. In all cases, the shear test with~$\Omega_\mathrm{mve} = \Omega_\mathrm{roi}$ and zero image noise has been used.}
	\label{SubSect:BCNoise:Fig3}
\end{figure}

Because random errors in DIC are usually expressed relative to the given pixel size, we next present a noise study in which the magnitude of random noise added to exact DNS boundary displacements is kept constant. At the same time, the level of overall applied strain is monotonically increased according to Eqs.~\eqref{Sect:Tests:Eq1} and~\eqref{Sect:Tests:Eq2}. Analogously to Eq.~\eqref{Sect:BCNoise:Eq1}, applied boundary displacements are expressed as
\begin{equation}
\bs{\mathsf{u}}_\mathrm{mve}(\bs{X}) = \bs{\mathsf{u}}_\mathrm{dns}(\bs{X}) + 2\,\sigma_\mathrm{px}\,\bs{\mathcal{U}}, \quad \bs{X} \in \partial\Omega_\mathrm{mve},
\label{Sect:BCNoise:Eq1b}
\end{equation}
where~$\sigma_\mathrm{px}$ is the fixed level of the displacement noise magnitude in pixels, while the remaining quantities have the same meaning as in Eq.~\eqref{Sect:BCNoise:Eq1}. Obtained results for the shear test, zero image noise, medium MVE mesh (cf. Fig.~\ref{SubSect:BEIDIC:Fig2b}), $\rho = 4$, $\sigma_\mathrm{px} \in \mathrm{px} \cdot \{ 0.01, 0.1, 0.25 \}$, and~$50$ MC realizations are shown along with relative errors in prescribed BCs in Fig.~\ref{SubSect:BCNoise:Fig3}. Here we notice that although the error in prescribed BCs is rather small, and naturally decreases with the applied level of overall strain, the corresponding deviations in the material parameters from the exact values are significant, especially in cases with~$\sigma_\mathrm{px} = 0.1$ and~$0.25$~px. For the case of lower DIC error bound, i.e.~$\sigma_\mathrm{px} = 0.01$~px, the results seem to rapidly reach accurate values. Note, however, that in the case of highly heterogeneous displacement fields, such a level of accuracy may be challenging to reach, cf. also Section~\ref{SubSect:BCGDIC} where actual DIC data is used. Although not presented, we note that the results corresponding to the tensile test display less sensitivity to errors in prescribed BCs, and hence the achieved accuracy is higher. The bending test on the other hand shows error levels that are comparable to those of the shear test.

It is important to note that in practice only a limited number of experiments or observations is carried out (e.g. two or three), meaning that standard deviation is of more importance than the mean value of the identified parameter. Therefore, in situations in which the mean value is accurate and the standard deviation is large, erroneous identification may be expected as not enough statistical data is usually available.
%
%
\subsection{Influence of Smoothing}
\label{SubSect:BCMA}
As a next step, the effect of smoothing is examined. To this end, the exact DNS displacement field is smoothed according to
\begin{equation}
\widetilde{\bs{u}}_\mathrm{dns}(\bs{X}) = \int_\Omega \bs{u}_\mathrm{dns}(\bs{Y})h_\varepsilon(\bs{Y}-\bs{X})\,\mathrm{d}\bs{Y},
\label{Sect:BCNoise:Eq2}
\end{equation}
where~$h_\varepsilon$ denotes the pillbox-shaped kernel with a dimensionless diameter~$\varepsilon \geq 0$ (normalized by the inclusion's diameter~$d$). The smoothed data are subsequently prescribed as nodal displacements to the discretized MVE model:
\begin{equation}
\bs{\mathsf{u}}_\mathrm{mve}(\bs{X}) = \widetilde{\bs{\mathsf{u}}}_\mathrm{dns}(\bs{X}), \quad \bs{X} \in \partial\Omega_\mathrm{mve},
\label{Sect:BCNoise:Eq3}
\end{equation}
In Eq.~\eqref{Sect:BCNoise:Eq3}, $\widetilde{\bs{\mathsf{u}}}_\mathrm{dns}(\bs{X})$, $\bs{X} \in \partial\Omega_\mathrm{mve}$, again represents a column of displacement evaluations of~$\widetilde{\bs{u}}_\mathrm{dns}$ at the MVE boundary nodes. For ease of implementation, the integral in Eq.~\eqref{Sect:BCNoise:Eq2} has been carried out at discrete pixel positions numerically, while the corresponding displacements have been interpolated using a linear interpolation scheme.

\begin{figure}
	\flushleft
 	\mbox{}\hspace{4.5em}\includegraphics[scale=1]{BCNoiseLsigLegend.pdf}\vspace{-0.5em}\\
 	\centering
	\subfloat[tension]{\includegraphics[scale=1]{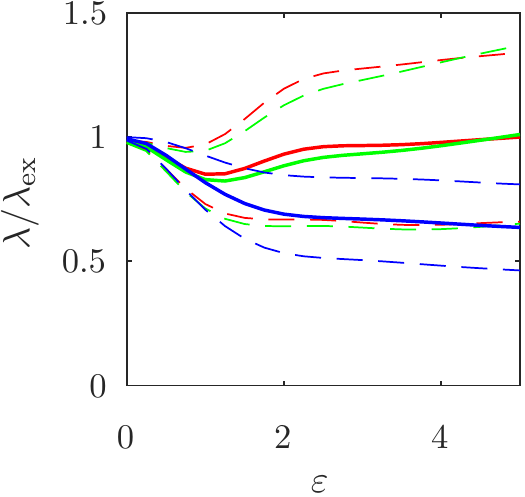}\label{SubSect:BCMA:Fig2a}}
	\subfloat[bending]{\includegraphics[scale=1]{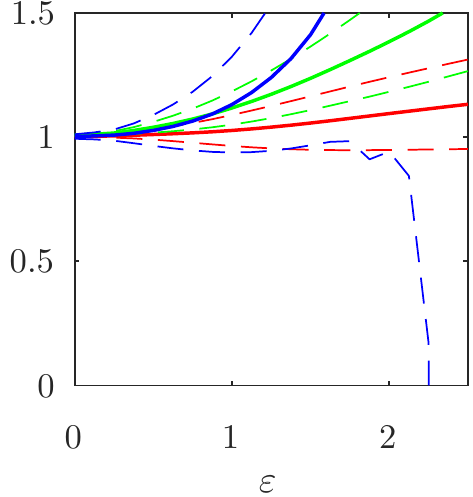}\label{SubSect:BCMA:Fig2b}}\hspace{0.5em}
	\subfloat[displacement error at~$\partial\Omega_\mathrm{mve}$]{\includegraphics[scale=1]{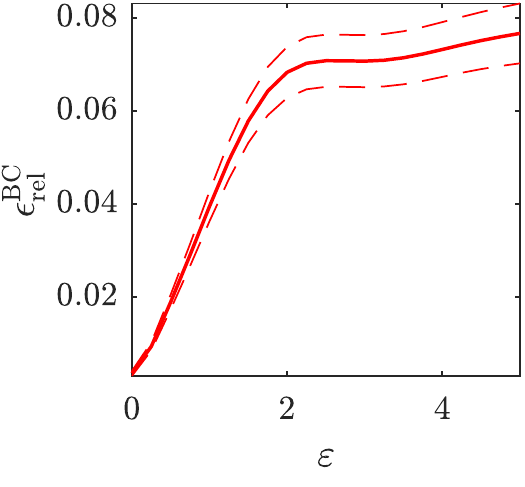}\label{SubSect:BCMA:Fig2c}}
	\caption{Identified results as a function of increasing dimensionless kernel size~$\varepsilon$ of the moving average for: (a)~tension and~(b) bending test, $\Omega_\mathrm{mve} = \Omega_\mathrm{roi}$, zero image noise, $\bs{\lambda} = [G_1,G_2,K_2]^\mathsf{T}$. (c)~Typical dependence of the relative error in BCs, cf. Eq.~\eqref{Sect:BCNoise:Eq1a}, on~$\varepsilon$ for the case of tension.}
	\label{SubSect:BCMA:Fig2}
\end{figure}

Fig.~\ref{SubSect:BCMA:Fig1} shows the obtained results for the case of shear, medium MVE mesh (shown in Fig.~\ref{SubSect:BEIDIC:Fig2b}), $\rho = 4$, $\Omega_\mathrm{mve} = \Omega_\mathrm{roi}$, and zero image noise, which once again confirms the need for accurate boundary data. Similarly to the random errors presented in Fig.~\ref{SubSect:BCNoise:Fig1}, it is clear that not only the standard deviations, but also the mean values rapidly deviate from~$1$ for erroneous BCs. Note that the smoothing effect for the applied maximum kernel size ($\varepsilon = 5$), shown in Fig.~\ref{SubSect:BCMA:Fig1c}, is not excessively large (see also Fig.~\ref{SubSect:BCMA:Fig2c}), yet the mean values start to deviate from~$1$ already at~$\varepsilon = 0.5$. Eliminating boundary fluctuations by smoothing therefore has a significant erroneous influence. Results for the other two mechanical tests, three material contrast ratios, and all other parameter combinations exhibit similar trends to those of Fig.~\ref{SubSect:BCMA:Fig1}, and can be inspected in Fig.~\ref{SubSect:BCMA:Fig2}.

The non-zero, but extremely small, values of the standard deviations observed in Figs.~\ref{SubSect:BCMA:Fig1} and~\ref{SubSect:BCNoise:Fig1} for~$\varepsilon = 0$ and~$\sigma = 0$ originate from the image and displacement interpolations.
%
%
\subsection{Influence of Global Digital Image Correlation}
\label{SubSect:BCGDIC}
In the light of the results obtained from the two previous sections, a question arises how important the effects of random noise and smoothing induced by GDIC are in the GDIC-IDIC approach, recalled for completeness in Algorithm~\ref{SubSect:BCGDIC:Alg1}. As is known from the literature, cf. e.g.~\cite{leclerc:2012}, and indicated in the introduction, a delicate balance between smoothing and random errors has to be found. A limit in terms of displacement accuracy that cannot be overcome by GDIC will therefore always exist for a given pixel resolution. For vanishing error in the GDIC data, however, one can expect accurate identification.

\begin{figure}
 	\centering
 	\scalebox{0.75}{
	\begin{tikzpicture}[node distance=1em, auto]  
	\linespread{1}
	\tikzset{
    	mynode/.style={rectangle,rounded corners,draw=black, top color=white, thick, inner sep=0.5em, minimum size=3em, text centered},
	    myarrow/.style={->, >=latex', shorten >=1pt},
	}  
	
	\node[mynode] (head) {\begin{tabular}{c}Computed\\ results
		\end{tabular}};

	\node[mynode,below=2em of head] (rho8) {$\rho = 8$};
	\node[mynode,left=11.5em of rho8] (rho4) {$\rho = 4$};
	\node[mynode,right=2.75em of rho8] (rho16) {$\rho = 16$};
	
	\node[mynode,below=2em of rho4] (Shear) {Shear};
	\node[mynode,left=11.5em of Shear] (Tension) {Tension};
	\node[mynode,right=1em of Shear] (Bending) {Bending};
	
	\node[mynode,below=2em of Tension] (MVEmedium) {\begin{tabular}{c}MVE\\ medium mesh
	\end{tabular}};
	\node[mynode,left=0.5em of MVEmedium] (MVEfine) {\begin{tabular}{c}MVE\\ fine mesh
	\end{tabular}};
	\node[mynode,right=0.5em of MVEmedium] (MVEcoarse) {\begin{tabular}{c}MVE\\ coarse mesh
	\end{tabular}};	

	\node[below=2em of MVEfine,anchor=west] (Fmat1) {\footnotesize $G_1, K_1, G_2$};
	\node[below=1.5em of Fmat1.west,anchor=west] (Fmat2) {\footnotesize $G_1, K_1, K_2$};
	\node[below=1.5em of Fmat2.west,anchor=west] (Fmat3) {\footnotesize $G_1, G_2, K_2$};
	\node[below=1.5em of Fmat3.west,anchor=west] (Fmat4) {\footnotesize $K_1, G_2, K_2$};
	\node[below=2em of MVEmedium,anchor=west] (Mmat1) {\footnotesize $G_1, K_1, G_2$};
	\node[below=1.5em of Mmat1.west,anchor=west] (Mmat2) {\footnotesize $G_1, K_1, K_2$};
	\node[below=1.5em of Mmat2.west,anchor=west] (Mmat3) {\footnotesize $G_1, G_2, K_2$};
	\node[below=1.5em of Mmat3.west,anchor=west] (Mmat4) {\footnotesize $K_1, G_2, K_2$};	
	\node[below=2em of MVEcoarse,anchor=west] (Cmat1) {\footnotesize $G_1, K_1, G_2$};
	\node[below=1.5em of Cmat1.west,anchor=west] (Cmat2) {\footnotesize $G_1, K_1, K_2$};
	\node[below=1.5em of Cmat2.west,anchor=west] (Cmat3) {\footnotesize $G_1, G_2, K_2$};
	\node[below=1.5em of Cmat3.west,anchor=west] (Cmat4) {\footnotesize $K_1, G_2, K_2$};	

	\draw[myarrow] (head.south) -- ++(0,-0.30) -| (rho4.north);	
	\draw[myarrow] (head.south) -- ++(0,-0.30) -| (rho8.north);
	\draw[myarrow] (head.south) -- ++(0,-0.30) -| (rho16.north);
	\draw[myarrow] (rho4.south) -- ++(0,-0.30) -| (Tension.north);	
	\draw[myarrow] (rho4.south) -- ++(0,-0.30) -| (Shear.north);
	\draw[myarrow] (rho4.south) -- ++(0,-0.30) -| (Bending.north);	
	\draw[myarrow] (Tension.south) -- ++(0,-0.30) -| (MVEfine.north);	
	\draw[myarrow] (Tension.south) -- ++(0,-0.30) -| (MVEmedium.north);
	\draw[myarrow] (Tension.south) -- ++(0,-0.30) -| (MVEcoarse.north);
	\draw[] (MVEfine.south) -- ++(0,-0.30) -| (Fmat1.west) -- ++(0.1,0);
	\draw[] (MVEfine.south) -- ++(0,-0.30) -| (Fmat2.west) -- ++(0.1,0);
	\draw[] (MVEfine.south) -- ++(0,-0.30) -| (Fmat3.west) -- ++(0.1,0);
	\draw[] (MVEfine.south) -- ++(0,-0.30) -| (Fmat4.west) -- ++(0.1,0);
	\draw[] (MVEmedium.south) -- ++(0,-0.30) -| (Mmat1.west) -- ++(0.1,0);
	\draw[] (MVEmedium.south) -- ++(0,-0.30) -| (Mmat2.west) -- ++(0.1,0);
	\draw[] (MVEmedium.south) -- ++(0,-0.30) -| (Mmat3.west) -- ++(0.1,0);
	\draw[] (MVEmedium.south) -- ++(0,-0.30) -| (Mmat4.west) -- ++(0.1,0);
	\draw[] (MVEcoarse.south) -- ++(0,-0.30) -| (Cmat1.west) -- ++(0.1,0);
	\draw[] (MVEcoarse.south) -- ++(0,-0.30) -| (Cmat2.west) -- ++(0.1,0);
	\draw[] (MVEcoarse.south) -- ++(0,-0.30) -| (Cmat3.west) -- ++(0.1,0);
	\draw[] (MVEcoarse.south) -- ++(0,-0.30) -| (Cmat4.west) -- ++(0.1,0);
	
	\node[below=1.25em of Shear,anchor=north,rectangle,inner sep=0em,outer sep=0em] (ZoomShear) {
	\begin{tikzpicture}[node distance=1em, auto] 
	\begin{scope}[transform canvas={scale=0.25}]

		\node[inner sep=0em,outer sep=0em] (Tension) {};
	
		\node[mynode,below=2em of Tension] (MVEmedium) {\begin{tabular}{c}MVE\\ medium mesh
		\end{tabular}};
		\node[mynode,left=0.5em of MVEmedium] (MVEfine) {\begin{tabular}{c}MVE\\ fine mesh
		\end{tabular}};
		\node[mynode,right=0.5em of MVEmedium] (MVEcoarse) {\begin{tabular}{c}MVE\\ coarse mesh
		\end{tabular}};	

		\node[below=2em of MVEfine,anchor=west] (Fmat1) {\footnotesize ~$G_1, K_1, G_2$};
		\node[below=1.5em of Fmat1.west,anchor=west] (Fmat2) {\footnotesize ~$G_1, K_1, K_2$};
		\node[below=1.5em of Fmat2.west,anchor=west] (Fmat3) {\footnotesize ~$G_1, G_2, K_2$};
		\node[below=1.5em of Fmat3.west,anchor=west] (Fmat4) {\footnotesize ~$K_1, G_2, K_2$};
		\node[below=2em of MVEmedium,anchor=west] (Mmat1) {\footnotesize ~$G_1, K_1, G_2$};
		\node[below=1.5em of Mmat1.west,anchor=west] (Mmat2) {\footnotesize ~$G_1, K_1, K_2$};
		\node[below=1.5em of Mmat2.west,anchor=west] (Mmat3) {\footnotesize ~$G_1, G_2, K_2$};
		\node[below=1.5em of Mmat3.west,anchor=west] (Mmat4) {\footnotesize ~$K_1, G_2, K_2$};	
		\node[below=2em of MVEcoarse,anchor=west] (Cmat1) {\footnotesize ~$G_1, K_1, G_2$};
		\node[below=1.5em of Cmat1.west,anchor=west] (Cmat2) {\footnotesize ~$G_1, K_1, K_2$};
		\node[below=1.5em of Cmat2.west,anchor=west] (Cmat3) {\footnotesize ~$G_1, G_2, K_2$};
		\node[below=1.5em of Cmat3.west,anchor=west] (Cmat4) {\footnotesize ~$K_1, G_2, K_2$};	

		\draw[myarrow] (Tension.south) -- ++(0,-0.30) -| (MVEfine.north);	
		\draw[myarrow] (Tension.south) -- ++(0,-0.30) -| (MVEmedium.north);
		\draw[myarrow] (Tension.south) -- ++(0,-0.30) -| (MVEcoarse.north);
		\draw[] (MVEfine.south) -- ++(0,-0.30) -| (Fmat1.west) -- ++(0.1,0);
		\draw[] (MVEfine.south) -- ++(0,-0.30) -| (Fmat2.west) -- ++(0.1,0);
		\draw[] (MVEfine.south) -- ++(0,-0.30) -| (Fmat3.west) -- ++(0.1,0);
		\draw[] (MVEfine.south) -- ++(0,-0.30) -| (Fmat4.west) -- ++(0.1,0);
		\draw[] (MVEmedium.south) -- ++(0,-0.30) -| (Mmat1.west) -- ++(0.1,0);
		\draw[] (MVEmedium.south) -- ++(0,-0.30) -| (Mmat2.west) -- ++(0.1,0);
		\draw[] (MVEmedium.south) -- ++(0,-0.30) -| (Mmat3.west) -- ++(0.1,0);
		\draw[] (MVEmedium.south) -- ++(0,-0.30) -| (Mmat4.west) -- ++(0.1,0);
		\draw[] (MVEcoarse.south) -- ++(0,-0.30) -| (Cmat1.west) -- ++(0.1,0);
		\draw[] (MVEcoarse.south) -- ++(0,-0.30) -| (Cmat2.west) -- ++(0.1,0);
		\draw[] (MVEcoarse.south) -- ++(0,-0.30) -| (Cmat3.west) -- ++(0.1,0);
		\draw[] (MVEcoarse.south) -- ++(0,-0.30) -| (Cmat4.west) -- ++(0.1,0);

	\end{scope}
	\end{tikzpicture}
	};
	
	\draw[] (Shear.south) -- ++(0,-0.6175) -| (ZoomShear.north);	

	\node[below=0.75em of Bending,anchor=north,rectangle,inner sep=0em,outer sep=0em] (ZoomBending) {
	\begin{tikzpicture}[node distance=1em, auto] 
	\begin{scope}[transform canvas={scale=0.15}]

		\node[inner sep=0em,outer sep=0em] (Tension) {};
	
		\node[mynode,below=2em of Tension] (MVEmedium) {\begin{tabular}{c}MVE\\ medium mesh
		\end{tabular}};
		\node[mynode,left=0.5em of MVEmedium] (MVEfine) {\begin{tabular}{c}MVE\\ fine mesh
		\end{tabular}};
		\node[mynode,right=0.5em of MVEmedium] (MVEcoarse) {\begin{tabular}{c}MVE\\ coarse mesh
		\end{tabular}};	

		\node[below=2em of MVEfine,anchor=west] (Fmat1) {\footnotesize ~$G_1, K_1, G_2$};
		\node[below=1.5em of Fmat1.west,anchor=west] (Fmat2) {\footnotesize ~$G_1, K_1, K_2$};
		\node[below=1.5em of Fmat2.west,anchor=west] (Fmat3) {\footnotesize ~$G_1, G_2, K_2$};
		\node[below=1.5em of Fmat3.west,anchor=west] (Fmat4) {\footnotesize ~$K_1, G_2, K_2$};
		\node[below=2em of MVEmedium,anchor=west] (Mmat1) {\footnotesize ~$G_1, K_1, G_2$};
		\node[below=1.5em of Mmat1.west,anchor=west] (Mmat2) {\footnotesize ~$G_1, K_1, K_2$};
		\node[below=1.5em of Mmat2.west,anchor=west] (Mmat3) {\footnotesize ~$G_1, G_2, K_2$};
		\node[below=1.5em of Mmat3.west,anchor=west] (Mmat4) {\footnotesize ~$K_1, G_2, K_2$};	
		\node[below=2em of MVEcoarse,anchor=west] (Cmat1) {\footnotesize ~$G_1, K_1, G_2$};
		\node[below=1.5em of Cmat1.west,anchor=west] (Cmat2) {\footnotesize ~$G_1, K_1, K_2$};
		\node[below=1.5em of Cmat2.west,anchor=west] (Cmat3) {\footnotesize ~$G_1, G_2, K_2$};
		\node[below=1.5em of Cmat3.west,anchor=west] (Cmat4) {\footnotesize ~$K_1, G_2, K_2$};	

		\draw[myarrow] (Tension.south) -- ++(0,-0.30) -| (MVEfine.north);	
		\draw[myarrow] (Tension.south) -- ++(0,-0.30) -| (MVEmedium.north);
		\draw[myarrow] (Tension.south) -- ++(0,-0.30) -| (MVEcoarse.north);
		\draw[] (MVEfine.south) -- ++(0,-0.30) -| (Fmat1.west) -- ++(0.1,0);
		\draw[] (MVEfine.south) -- ++(0,-0.30) -| (Fmat2.west) -- ++(0.1,0);
		\draw[] (MVEfine.south) -- ++(0,-0.30) -| (Fmat3.west) -- ++(0.1,0);
		\draw[] (MVEfine.south) -- ++(0,-0.30) -| (Fmat4.west) -- ++(0.1,0);
		\draw[] (MVEmedium.south) -- ++(0,-0.30) -| (Mmat1.west) -- ++(0.1,0);
		\draw[] (MVEmedium.south) -- ++(0,-0.30) -| (Mmat2.west) -- ++(0.1,0);
		\draw[] (MVEmedium.south) -- ++(0,-0.30) -| (Mmat3.west) -- ++(0.1,0);
		\draw[] (MVEmedium.south) -- ++(0,-0.30) -| (Mmat4.west) -- ++(0.1,0);
		\draw[] (MVEcoarse.south) -- ++(0,-0.30) -| (Cmat1.west) -- ++(0.1,0);
		\draw[] (MVEcoarse.south) -- ++(0,-0.30) -| (Cmat2.west) -- ++(0.1,0);
		\draw[] (MVEcoarse.south) -- ++(0,-0.30) -| (Cmat3.west) -- ++(0.1,0);
		\draw[] (MVEcoarse.south) -- ++(0,-0.30) -| (Cmat4.west) -- ++(0.1,0);

	\end{scope}
	\end{tikzpicture}
	};
	
	\draw[] (Bending.south) -- ++(0,-0.3785) -| (ZoomBending.north);	
	
	\node[below=1.25em of rho8,anchor=north,rectangle,inner sep=0em,outer sep=0em] (ZoomRho8) {
	\begin{tikzpicture}[node distance=1em, auto] 
	\begin{scope}[transform canvas={scale=0.25}]

	\node[inner sep=0em,outer sep=0em] (rho4) {};
	
	\node[mynode,below=2em of rho4] (Shear) {Shear};
	\node[mynode,left=11.5em of Shear] (Tension) {Tension};
	\node[mynode,right=1em of Shear] (Bending) {Bending};
	
	\node[mynode,below=2em of Tension] (MVEmedium) {\begin{tabular}{c}MVE\\ medium mesh
	\end{tabular}};
	\node[mynode,left=0.5em of MVEmedium] (MVEfine) {\begin{tabular}{c}MVE\\ fine mesh
	\end{tabular}};
	\node[mynode,right=0.5em of MVEmedium] (MVEcoarse) {\begin{tabular}{c}MVE\\ coarse mesh
	\end{tabular}};	

	\node[below=2em of MVEfine,anchor=west] (Fmat1) {\footnotesize ~$G_1, K_1, G_2$};
	\node[below=1.5em of Fmat1.west,anchor=west] (Fmat2) {\footnotesize ~$G_1, K_1, K_2$};
	\node[below=1.5em of Fmat2.west,anchor=west] (Fmat3) {\footnotesize ~$G_1, G_2, K_2$};
	\node[below=1.5em of Fmat3.west,anchor=west] (Fmat4) {\footnotesize ~$K_1, G_2, K_2$};
	\node[below=2em of MVEmedium,anchor=west] (Mmat1) {\footnotesize ~$G_1, K_1, G_2$};
	\node[below=1.5em of Mmat1.west,anchor=west] (Mmat2) {\footnotesize ~$G_1, K_1, K_2$};
	\node[below=1.5em of Mmat2.west,anchor=west] (Mmat3) {\footnotesize ~$G_1, G_2, K_2$};
	\node[below=1.5em of Mmat3.west,anchor=west] (Mmat4) {\footnotesize ~$K_1, G_2, K_2$};	
	\node[below=2em of MVEcoarse,anchor=west] (Cmat1) {\footnotesize ~$G_1, K_1, G_2$};
	\node[below=1.5em of Cmat1.west,anchor=west] (Cmat2) {\footnotesize ~$G_1, K_1, K_2$};
	\node[below=1.5em of Cmat2.west,anchor=west] (Cmat3) {\footnotesize ~$G_1, G_2, K_2$};
	\node[below=1.5em of Cmat3.west,anchor=west] (Cmat4) {\footnotesize ~$K_1, G_2, K_2$};	

	\draw[myarrow] (rho4.south) -- ++(0,-0.30) -| (Tension.north);	
	\draw[myarrow] (rho4.south) -- ++(0,-0.30) -| (Shear.north);
	\draw[myarrow] (rho4.south) -- ++(0,-0.30) -| (Bending.north);	
	\draw[myarrow] (Tension.south) -- ++(0,-0.30) -| (MVEfine.north);	
	\draw[myarrow] (Tension.south) -- ++(0,-0.30) -| (MVEmedium.north);
	\draw[myarrow] (Tension.south) -- ++(0,-0.30) -| (MVEcoarse.north);
	\draw[] (MVEfine.south) -- ++(0,-0.30) -| (Fmat1.west) -- ++(0.1,0);
	\draw[] (MVEfine.south) -- ++(0,-0.30) -| (Fmat2.west) -- ++(0.1,0);
	\draw[] (MVEfine.south) -- ++(0,-0.30) -| (Fmat3.west) -- ++(0.1,0);
	\draw[] (MVEfine.south) -- ++(0,-0.30) -| (Fmat4.west) -- ++(0.1,0);
	\draw[] (MVEmedium.south) -- ++(0,-0.30) -| (Mmat1.west) -- ++(0.1,0);
	\draw[] (MVEmedium.south) -- ++(0,-0.30) -| (Mmat2.west) -- ++(0.1,0);
	\draw[] (MVEmedium.south) -- ++(0,-0.30) -| (Mmat3.west) -- ++(0.1,0);
	\draw[] (MVEmedium.south) -- ++(0,-0.30) -| (Mmat4.west) -- ++(0.1,0);
	\draw[] (MVEcoarse.south) -- ++(0,-0.30) -| (Cmat1.west) -- ++(0.1,0);
	\draw[] (MVEcoarse.south) -- ++(0,-0.30) -| (Cmat2.west) -- ++(0.1,0);
	\draw[] (MVEcoarse.south) -- ++(0,-0.30) -| (Cmat3.west) -- ++(0.1,0);
	\draw[] (MVEcoarse.south) -- ++(0,-0.30) -| (Cmat4.west) -- ++(0.1,0);
	
	\node[below=1.25em of Shear,anchor=north,rectangle,inner sep=0em,outer sep=0em] (ZoomShear) {
	\begin{tikzpicture}[node distance=1em, auto] 
	\begin{scope}[transform canvas={scale=0.25}]

		\node[inner sep=0em,outer sep=0em] (Tension) {};
	
		\node[mynode,below=2em of Tension] (MVEmedium) {\begin{tabular}{c}MVE\\ medium mesh
		\end{tabular}};
		\node[mynode,left=0.5em of MVEmedium] (MVEfine) {\begin{tabular}{c}MVE\\ fine mesh
		\end{tabular}};
		\node[mynode,right=0.5em of MVEmedium] (MVEcoarse) {\begin{tabular}{c}MVE\\ coarse mesh
		\end{tabular}};	

		\node[below=2em of MVEfine,anchor=west] (Fmat1) {\footnotesize ~$G_1, K_1, G_2$};
		\node[below=1.5em of Fmat1.west,anchor=west] (Fmat2) {\footnotesize ~$G_1, K_1, K_2$};
		\node[below=1.5em of Fmat2.west,anchor=west] (Fmat3) {\footnotesize ~$G_1, G_2, K_2$};
		\node[below=1.5em of Fmat3.west,anchor=west] (Fmat4) {\footnotesize ~$K_1, G_2, K_2$};
		\node[below=2em of MVEmedium,anchor=west] (Mmat1) {\footnotesize ~$G_1, K_1, G_2$};
		\node[below=1.5em of Mmat1.west,anchor=west] (Mmat2) {\footnotesize ~$G_1, K_1, K_2$};
		\node[below=1.5em of Mmat2.west,anchor=west] (Mmat3) {\footnotesize ~$G_1, G_2, K_2$};
		\node[below=1.5em of Mmat3.west,anchor=west] (Mmat4) {\footnotesize ~$K_1, G_2, K_2$};	
		\node[below=2em of MVEcoarse,anchor=west] (Cmat1) {\footnotesize ~$G_1, K_1, G_2$};
		\node[below=1.5em of Cmat1.west,anchor=west] (Cmat2) {\footnotesize ~$G_1, K_1, K_2$};
		\node[below=1.5em of Cmat2.west,anchor=west] (Cmat3) {\footnotesize ~$G_1, G_2, K_2$};
		\node[below=1.5em of Cmat3.west,anchor=west] (Cmat4) {\footnotesize ~$K_1, G_2, K_2$};	

		\draw[myarrow] (Tension.south) -- ++(0,-0.30) -| (MVEfine.north);	
		\draw[myarrow] (Tension.south) -- ++(0,-0.30) -| (MVEmedium.north);
		\draw[myarrow] (Tension.south) -- ++(0,-0.30) -| (MVEcoarse.north);
		\draw[] (MVEfine.south) -- ++(0,-0.30) -| (Fmat1.west) -- ++(0.1,0);
		\draw[] (MVEfine.south) -- ++(0,-0.30) -| (Fmat2.west) -- ++(0.1,0);
		\draw[] (MVEfine.south) -- ++(0,-0.30) -| (Fmat3.west) -- ++(0.1,0);
		\draw[] (MVEfine.south) -- ++(0,-0.30) -| (Fmat4.west) -- ++(0.1,0);
		\draw[] (MVEmedium.south) -- ++(0,-0.30) -| (Mmat1.west) -- ++(0.1,0);
		\draw[] (MVEmedium.south) -- ++(0,-0.30) -| (Mmat2.west) -- ++(0.1,0);
		\draw[] (MVEmedium.south) -- ++(0,-0.30) -| (Mmat3.west) -- ++(0.1,0);
		\draw[] (MVEmedium.south) -- ++(0,-0.30) -| (Mmat4.west) -- ++(0.1,0);
		\draw[] (MVEcoarse.south) -- ++(0,-0.30) -| (Cmat1.west) -- ++(0.1,0);
		\draw[] (MVEcoarse.south) -- ++(0,-0.30) -| (Cmat2.west) -- ++(0.1,0);
		\draw[] (MVEcoarse.south) -- ++(0,-0.30) -| (Cmat3.west) -- ++(0.1,0);
		\draw[] (MVEcoarse.south) -- ++(0,-0.30) -| (Cmat4.west) -- ++(0.1,0);

	\end{scope}
	\end{tikzpicture}
	};
	
	\draw[] (Shear.south) -- ++(0,-0.6175) -| (ZoomShear.north);	

	\node[below=0.75em of Bending,anchor=north,rectangle,inner sep=0em,outer sep=0em] (ZoomBending) {
	\begin{tikzpicture}[node distance=1em, auto] 
	\begin{scope}[transform canvas={scale=0.15}]

		\node[inner sep=0em,outer sep=0em] (Tension) {};
	
		\node[mynode,below=2em of Tension] (MVEmedium) {\begin{tabular}{c}MVE\\ medium mesh
		\end{tabular}};
		\node[mynode,left=0.5em of MVEmedium] (MVEfine) {\begin{tabular}{c}MVE\\ fine mesh
		\end{tabular}};
		\node[mynode,right=0.5em of MVEmedium] (MVEcoarse) {\begin{tabular}{c}MVE\\ coarse mesh
		\end{tabular}};	

		\node[below=2em of MVEfine,anchor=west] (Fmat1) {\footnotesize ~$G_1, K_1, G_2$};
		\node[below=1.5em of Fmat1.west,anchor=west] (Fmat2) {\footnotesize ~$G_1, K_1, K_2$};
		\node[below=1.5em of Fmat2.west,anchor=west] (Fmat3) {\footnotesize ~$G_1, G_2, K_2$};
		\node[below=1.5em of Fmat3.west,anchor=west] (Fmat4) {\footnotesize ~$K_1, G_2, K_2$};
		\node[below=2em of MVEmedium,anchor=west] (Mmat1) {\footnotesize ~$G_1, K_1, G_2$};
		\node[below=1.5em of Mmat1.west,anchor=west] (Mmat2) {\footnotesize ~$G_1, K_1, K_2$};
		\node[below=1.5em of Mmat2.west,anchor=west] (Mmat3) {\footnotesize ~$G_1, G_2, K_2$};
		\node[below=1.5em of Mmat3.west,anchor=west] (Mmat4) {\footnotesize ~$K_1, G_2, K_2$};	
		\node[below=2em of MVEcoarse,anchor=west] (Cmat1) {\footnotesize ~$G_1, K_1, G_2$};
		\node[below=1.5em of Cmat1.west,anchor=west] (Cmat2) {\footnotesize ~$G_1, K_1, K_2$};
		\node[below=1.5em of Cmat2.west,anchor=west] (Cmat3) {\footnotesize ~$G_1, G_2, K_2$};
		\node[below=1.5em of Cmat3.west,anchor=west] (Cmat4) {\footnotesize ~$K_1, G_2, K_2$};	

		\draw[myarrow] (Tension.south) -- ++(0,-0.30) -| (MVEfine.north);	
		\draw[myarrow] (Tension.south) -- ++(0,-0.30) -| (MVEmedium.north);
		\draw[myarrow] (Tension.south) -- ++(0,-0.30) -| (MVEcoarse.north);
		\draw[] (MVEfine.south) -- ++(0,-0.30) -| (Fmat1.west) -- ++(0.1,0);
		\draw[] (MVEfine.south) -- ++(0,-0.30) -| (Fmat2.west) -- ++(0.1,0);
		\draw[] (MVEfine.south) -- ++(0,-0.30) -| (Fmat3.west) -- ++(0.1,0);
		\draw[] (MVEfine.south) -- ++(0,-0.30) -| (Fmat4.west) -- ++(0.1,0);
		\draw[] (MVEmedium.south) -- ++(0,-0.30) -| (Mmat1.west) -- ++(0.1,0);
		\draw[] (MVEmedium.south) -- ++(0,-0.30) -| (Mmat2.west) -- ++(0.1,0);
		\draw[] (MVEmedium.south) -- ++(0,-0.30) -| (Mmat3.west) -- ++(0.1,0);
		\draw[] (MVEmedium.south) -- ++(0,-0.30) -| (Mmat4.west) -- ++(0.1,0);
		\draw[] (MVEcoarse.south) -- ++(0,-0.30) -| (Cmat1.west) -- ++(0.1,0);
		\draw[] (MVEcoarse.south) -- ++(0,-0.30) -| (Cmat2.west) -- ++(0.1,0);
		\draw[] (MVEcoarse.south) -- ++(0,-0.30) -| (Cmat3.west) -- ++(0.1,0);
		\draw[] (MVEcoarse.south) -- ++(0,-0.30) -| (Cmat4.west) -- ++(0.1,0);

	\end{scope}
	\end{tikzpicture}
	};
	
	\draw[] (Bending.south) -- ++(0,-0.3785) -| (ZoomBending.north);	
	
	\end{scope}
	\end{tikzpicture}
	};	
	
	\draw[] (rho8.south) -- ++(0,-0.6175) -| (ZoomRho8.north);		

	\node[below=0.75em of rho16,anchor=north,rectangle,inner sep=0em,outer sep=0em] (ZoomRho16) {
	\begin{tikzpicture}[node distance=1em, auto] 
	\begin{scope}[transform canvas={scale=0.15}]

	\node[inner sep=0em,outer sep=0em] (rho4) {};
	
	\node[mynode,below=2em of rho4] (Shear) {Shear};
	\node[mynode,left=11.5em of Shear] (Tension) {Tension};
	\node[mynode,right=1em of Shear] (Bending) {Bending};
	
	\node[mynode,below=2em of Tension] (MVEmedium) {\begin{tabular}{c}MVE\\ medium mesh
	\end{tabular}};
	\node[mynode,left=0.5em of MVEmedium] (MVEfine) {\begin{tabular}{c}MVE\\ fine mesh
	\end{tabular}};
	\node[mynode,right=0.5em of MVEmedium] (MVEcoarse) {\begin{tabular}{c}MVE\\ coarse mesh
	\end{tabular}};	

	\node[below=2em of MVEfine,anchor=west] (Fmat1) {\footnotesize ~$G_1, K_1, G_2$};
	\node[below=1.5em of Fmat1.west,anchor=west] (Fmat2) {\footnotesize ~$G_1, K_1, K_2$};
	\node[below=1.5em of Fmat2.west,anchor=west] (Fmat3) {\footnotesize ~$G_1, G_2, K_2$};
	\node[below=1.5em of Fmat3.west,anchor=west] (Fmat4) {\footnotesize ~$K_1, G_2, K_2$};
	\node[below=2em of MVEmedium,anchor=west] (Mmat1) {\footnotesize ~$G_1, K_1, G_2$};
	\node[below=1.5em of Mmat1.west,anchor=west] (Mmat2) {\footnotesize ~$G_1, K_1, K_2$};
	\node[below=1.5em of Mmat2.west,anchor=west] (Mmat3) {\footnotesize ~$G_1, G_2, K_2$};
	\node[below=1.5em of Mmat3.west,anchor=west] (Mmat4) {\footnotesize ~$K_1, G_2, K_2$};	
	\node[below=2em of MVEcoarse,anchor=west] (Cmat1) {\footnotesize ~$G_1, K_1, G_2$};
	\node[below=1.5em of Cmat1.west,anchor=west] (Cmat2) {\footnotesize ~$G_1, K_1, K_2$};
	\node[below=1.5em of Cmat2.west,anchor=west] (Cmat3) {\footnotesize ~$G_1, G_2, K_2$};
	\node[below=1.5em of Cmat3.west,anchor=west] (Cmat4) {\footnotesize ~$K_1, G_2, K_2$};	

	\draw[myarrow] (rho4.south) -- ++(0,-0.30) -| (Tension.north);	
	\draw[myarrow] (rho4.south) -- ++(0,-0.30) -| (Shear.north);
	\draw[myarrow] (rho4.south) -- ++(0,-0.30) -| (Bending.north);	
	\draw[myarrow] (Tension.south) -- ++(0,-0.30) -| (MVEfine.north);	
	\draw[myarrow] (Tension.south) -- ++(0,-0.30) -| (MVEmedium.north);
	\draw[myarrow] (Tension.south) -- ++(0,-0.30) -| (MVEcoarse.north);
	\draw[] (MVEfine.south) -- ++(0,-0.30) -| (Fmat1.west) -- ++(0.1,0);
	\draw[] (MVEfine.south) -- ++(0,-0.30) -| (Fmat2.west) -- ++(0.1,0);
	\draw[] (MVEfine.south) -- ++(0,-0.30) -| (Fmat3.west) -- ++(0.1,0);
	\draw[] (MVEfine.south) -- ++(0,-0.30) -| (Fmat4.west) -- ++(0.1,0);
	\draw[] (MVEmedium.south) -- ++(0,-0.30) -| (Mmat1.west) -- ++(0.1,0);
	\draw[] (MVEmedium.south) -- ++(0,-0.30) -| (Mmat2.west) -- ++(0.1,0);
	\draw[] (MVEmedium.south) -- ++(0,-0.30) -| (Mmat3.west) -- ++(0.1,0);
	\draw[] (MVEmedium.south) -- ++(0,-0.30) -| (Mmat4.west) -- ++(0.1,0);
	\draw[] (MVEcoarse.south) -- ++(0,-0.30) -| (Cmat1.west) -- ++(0.1,0);
	\draw[] (MVEcoarse.south) -- ++(0,-0.30) -| (Cmat2.west) -- ++(0.1,0);
	\draw[] (MVEcoarse.south) -- ++(0,-0.30) -| (Cmat3.west) -- ++(0.1,0);
	\draw[] (MVEcoarse.south) -- ++(0,-0.30) -| (Cmat4.west) -- ++(0.1,0);
	
	\node[below=1.25em of Shear,anchor=north,rectangle,inner sep=0em,outer sep=0em] (ZoomShear) {
	\begin{tikzpicture}[node distance=1em, auto] 
	\begin{scope}[transform canvas={scale=0.25}]

		\node[inner sep=0em,outer sep=0em] (Tension) {};
	
		\node[mynode,below=2em of Tension] (MVEmedium) {\begin{tabular}{c}MVE\\ medium mesh
		\end{tabular}};
		\node[mynode,left=0.5em of MVEmedium] (MVEfine) {\begin{tabular}{c}MVE\\ fine mesh
		\end{tabular}};
		\node[mynode,right=0.5em of MVEmedium] (MVEcoarse) {\begin{tabular}{c}MVE\\ coarse mesh
		\end{tabular}};	

		\node[below=2em of MVEfine,anchor=west] (Fmat1) {\footnotesize ~$G_1, K_1, G_2$};
		\node[below=1.5em of Fmat1.west,anchor=west] (Fmat2) {\footnotesize ~$G_1, K_1, K_2$};
		\node[below=1.5em of Fmat2.west,anchor=west] (Fmat3) {\footnotesize ~$G_1, G_2, K_2$};
		\node[below=1.5em of Fmat3.west,anchor=west] (Fmat4) {\footnotesize ~$K_1, G_2, K_2$};
		\node[below=2em of MVEmedium,anchor=west] (Mmat1) {\footnotesize ~$G_1, K_1, G_2$};
		\node[below=1.5em of Mmat1.west,anchor=west] (Mmat2) {\footnotesize ~$G_1, K_1, K_2$};
		\node[below=1.5em of Mmat2.west,anchor=west] (Mmat3) {\footnotesize ~$G_1, G_2, K_2$};
		\node[below=1.5em of Mmat3.west,anchor=west] (Mmat4) {\footnotesize ~$K_1, G_2, K_2$};	
		\node[below=2em of MVEcoarse,anchor=west] (Cmat1) {\footnotesize ~$G_1, K_1, G_2$};
		\node[below=1.5em of Cmat1.west,anchor=west] (Cmat2) {\footnotesize ~$G_1, K_1, K_2$};
		\node[below=1.5em of Cmat2.west,anchor=west] (Cmat3) {\footnotesize ~$G_1, G_2, K_2$};
		\node[below=1.5em of Cmat3.west,anchor=west] (Cmat4) {\footnotesize ~$K_1, G_2, K_2$};	

		\draw[myarrow] (Tension.south) -- ++(0,-0.30) -| (MVEfine.north);	
		\draw[myarrow] (Tension.south) -- ++(0,-0.30) -| (MVEmedium.north);
		\draw[myarrow] (Tension.south) -- ++(0,-0.30) -| (MVEcoarse.north);
		\draw[] (MVEfine.south) -- ++(0,-0.30) -| (Fmat1.west) -- ++(0.1,0);
		\draw[] (MVEfine.south) -- ++(0,-0.30) -| (Fmat2.west) -- ++(0.1,0);
		\draw[] (MVEfine.south) -- ++(0,-0.30) -| (Fmat3.west) -- ++(0.1,0);
		\draw[] (MVEfine.south) -- ++(0,-0.30) -| (Fmat4.west) -- ++(0.1,0);
		\draw[] (MVEmedium.south) -- ++(0,-0.30) -| (Mmat1.west) -- ++(0.1,0);
		\draw[] (MVEmedium.south) -- ++(0,-0.30) -| (Mmat2.west) -- ++(0.1,0);
		\draw[] (MVEmedium.south) -- ++(0,-0.30) -| (Mmat3.west) -- ++(0.1,0);
		\draw[] (MVEmedium.south) -- ++(0,-0.30) -| (Mmat4.west) -- ++(0.1,0);
		\draw[] (MVEcoarse.south) -- ++(0,-0.30) -| (Cmat1.west) -- ++(0.1,0);
		\draw[] (MVEcoarse.south) -- ++(0,-0.30) -| (Cmat2.west) -- ++(0.1,0);
		\draw[] (MVEcoarse.south) -- ++(0,-0.30) -| (Cmat3.west) -- ++(0.1,0);
		\draw[] (MVEcoarse.south) -- ++(0,-0.30) -| (Cmat4.west) -- ++(0.1,0);

	\end{scope}
	\end{tikzpicture}
	};
	
	\draw[] (Shear.south) -- ++(0,-0.6175) -| (ZoomShear.north);	

	\node[below=0.75em of Bending,anchor=north,rectangle,inner sep=0em,outer sep=0em] (ZoomBending) {
	\begin{tikzpicture}[node distance=1em, auto] 
	\begin{scope}[transform canvas={scale=0.15}]

		\node[inner sep=0em,outer sep=0em] (Tension) {};
	
		\node[mynode,below=2em of Tension] (MVEmedium) {\begin{tabular}{c}MVE\\ medium mesh
		\end{tabular}};
		\node[mynode,left=0.5em of MVEmedium] (MVEfine) {\begin{tabular}{c}MVE\\ fine mesh
		\end{tabular}};
		\node[mynode,right=0.5em of MVEmedium] (MVEcoarse) {\begin{tabular}{c}MVE\\ coarse mesh
		\end{tabular}};	

		\node[below=2em of MVEfine,anchor=west] (Fmat1) {\footnotesize ~$G_1, K_1, G_2$};
		\node[below=1.5em of Fmat1.west,anchor=west] (Fmat2) {\footnotesize ~$G_1, K_1, K_2$};
		\node[below=1.5em of Fmat2.west,anchor=west] (Fmat3) {\footnotesize ~$G_1, G_2, K_2$};
		\node[below=1.5em of Fmat3.west,anchor=west] (Fmat4) {\footnotesize ~$K_1, G_2, K_2$};
		\node[below=2em of MVEmedium,anchor=west] (Mmat1) {\footnotesize ~$G_1, K_1, G_2$};
		\node[below=1.5em of Mmat1.west,anchor=west] (Mmat2) {\footnotesize ~$G_1, K_1, K_2$};
		\node[below=1.5em of Mmat2.west,anchor=west] (Mmat3) {\footnotesize ~$G_1, G_2, K_2$};
		\node[below=1.5em of Mmat3.west,anchor=west] (Mmat4) {\footnotesize ~$K_1, G_2, K_2$};	
		\node[below=2em of MVEcoarse,anchor=west] (Cmat1) {\footnotesize ~$G_1, K_1, G_2$};
		\node[below=1.5em of Cmat1.west,anchor=west] (Cmat2) {\footnotesize ~$G_1, K_1, K_2$};
		\node[below=1.5em of Cmat2.west,anchor=west] (Cmat3) {\footnotesize ~$G_1, G_2, K_2$};
		\node[below=1.5em of Cmat3.west,anchor=west] (Cmat4) {\footnotesize ~$K_1, G_2, K_2$};	

		\draw[myarrow] (Tension.south) -- ++(0,-0.30) -| (MVEfine.north);	
		\draw[myarrow] (Tension.south) -- ++(0,-0.30) -| (MVEmedium.north);
		\draw[myarrow] (Tension.south) -- ++(0,-0.30) -| (MVEcoarse.north);
		\draw[] (MVEfine.south) -- ++(0,-0.30) -| (Fmat1.west) -- ++(0.1,0);
		\draw[] (MVEfine.south) -- ++(0,-0.30) -| (Fmat2.west) -- ++(0.1,0);
		\draw[] (MVEfine.south) -- ++(0,-0.30) -| (Fmat3.west) -- ++(0.1,0);
		\draw[] (MVEfine.south) -- ++(0,-0.30) -| (Fmat4.west) -- ++(0.1,0);
		\draw[] (MVEmedium.south) -- ++(0,-0.30) -| (Mmat1.west) -- ++(0.1,0);
		\draw[] (MVEmedium.south) -- ++(0,-0.30) -| (Mmat2.west) -- ++(0.1,0);
		\draw[] (MVEmedium.south) -- ++(0,-0.30) -| (Mmat3.west) -- ++(0.1,0);
		\draw[] (MVEmedium.south) -- ++(0,-0.30) -| (Mmat4.west) -- ++(0.1,0);
		\draw[] (MVEcoarse.south) -- ++(0,-0.30) -| (Cmat1.west) -- ++(0.1,0);
		\draw[] (MVEcoarse.south) -- ++(0,-0.30) -| (Cmat2.west) -- ++(0.1,0);
		\draw[] (MVEcoarse.south) -- ++(0,-0.30) -| (Cmat3.west) -- ++(0.1,0);
		\draw[] (MVEcoarse.south) -- ++(0,-0.30) -| (Cmat4.west) -- ++(0.1,0);

	\end{scope}
	\end{tikzpicture}
	};
	
	\draw[] (Bending.south) -- ++(0,-0.3785) -| (ZoomBending.north);	
	
	\end{scope}
	\end{tikzpicture}
	};	
	
	\draw[] (rho16.south) -- ++(0,-0.3785) -| (ZoomRho16.north);		

	\end{tikzpicture}
	}
 	\caption{Scheme showing all combinations used for microstructural identification. For each of the~$108$ test cases shown, $50$ MC realizations with random microstructures have been computed.}
 	\label{SubSect:BCGDIC:Fig0}
\end{figure}
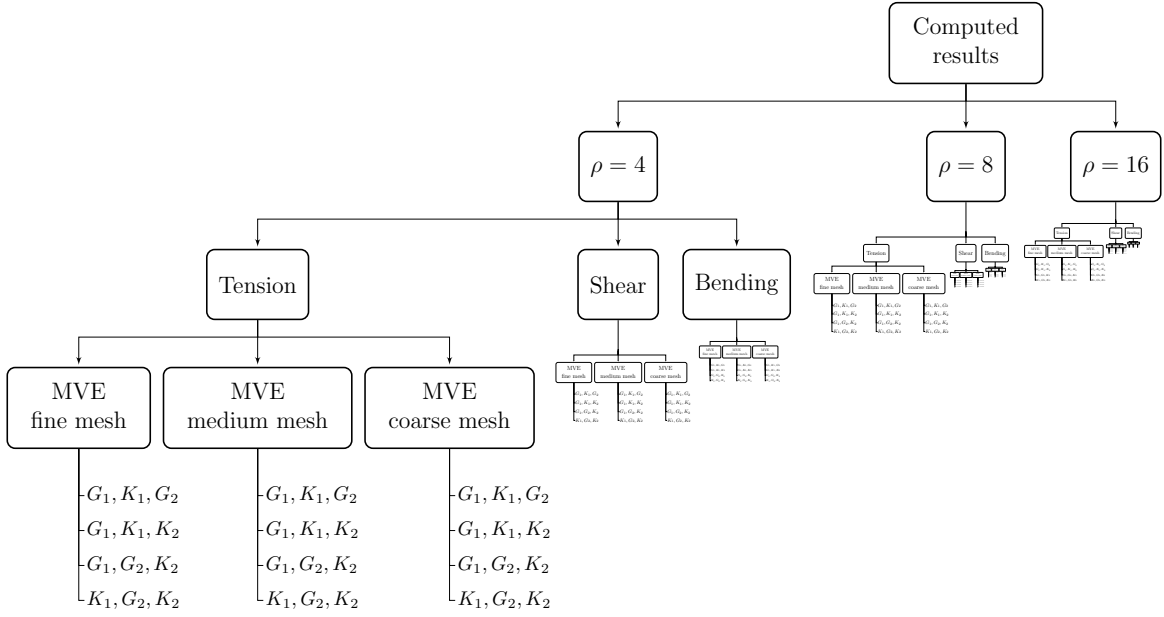

\begin{figure}
 	\centering
	\subfloat[DNS mesh]{\includegraphics[scale=1]{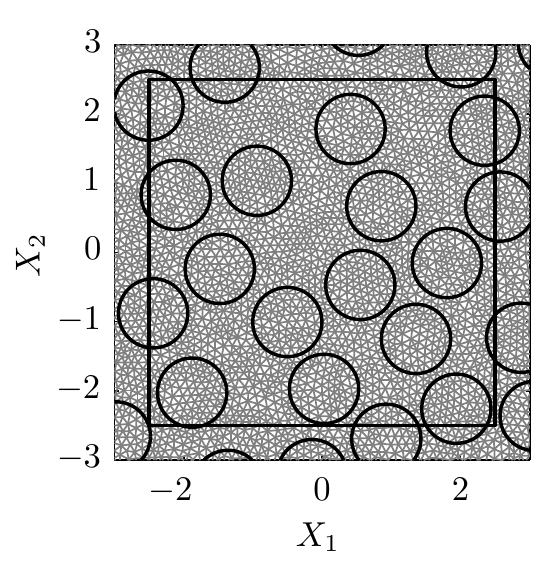}\label{SubSect:BCGDIC:Fig1a}}\hspace{0.5em}
	\subfloat[GDIC fine mesh]{\includegraphics[scale=1]{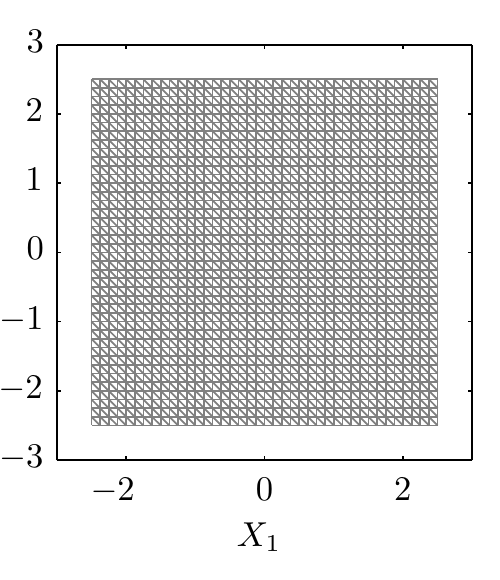}\label{SubSect:BCGDIC:Fig1b}}\hspace{0.5em}
	\subfloat[GDIC coarse mesh]{\includegraphics[scale=1]{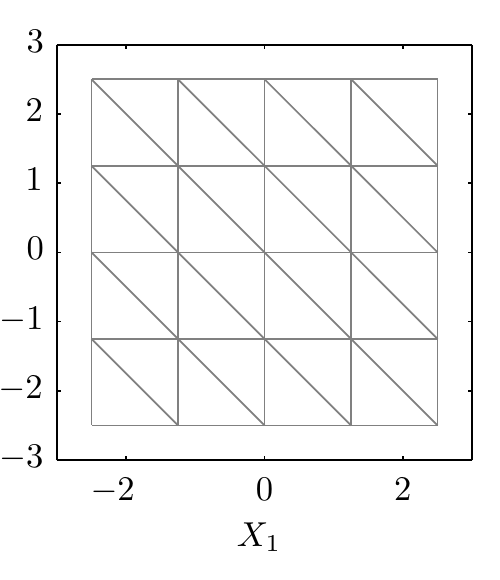}\label{SubSect:BCGDIC:Fig1c}}
	\caption{Typical discretizations employed in the simulations. (a)~DNS unstructured mesh, typical element size~$h \approx d/9$, approx.~25 pix/triangle, (b)~GDIC fine structured mesh, $h = d/8$, approx.~36 pix/triangle, and~(c) GDIC coarse structured mesh, $h =5d/4$, approx.~3600 pix/triangle. In all cases, quadratic iso-parametric elements were used.}
	\label{SubSect:BCGDIC:Fig1}
\end{figure}

\begin{figure}
 	\centering
	\subfloat[fine]{\includegraphics[scale=1]{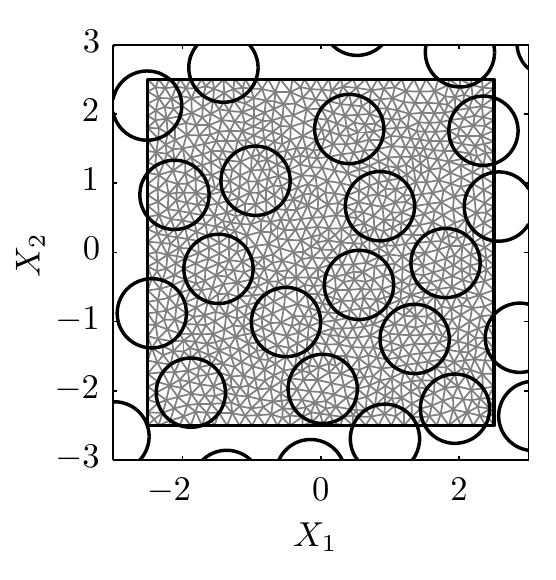}\label{SubSect:BEIDIC:Fig2a}}\hspace{0.5em}
	\subfloat[medium]{\includegraphics[scale=1]{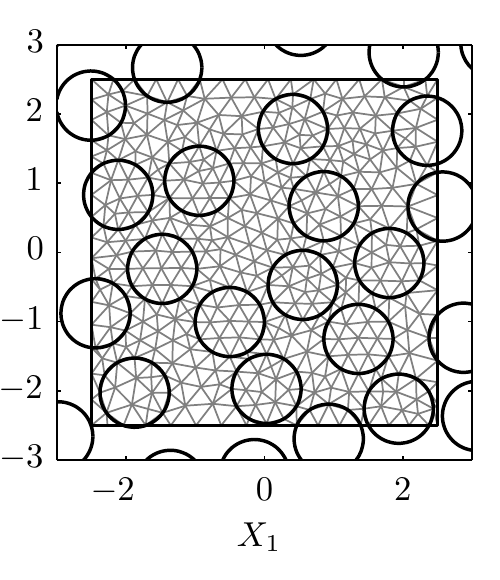}\label{SubSect:BEIDIC:Fig2b}}\hspace{0.5em}
	\subfloat[coarse]{\includegraphics[scale=1]{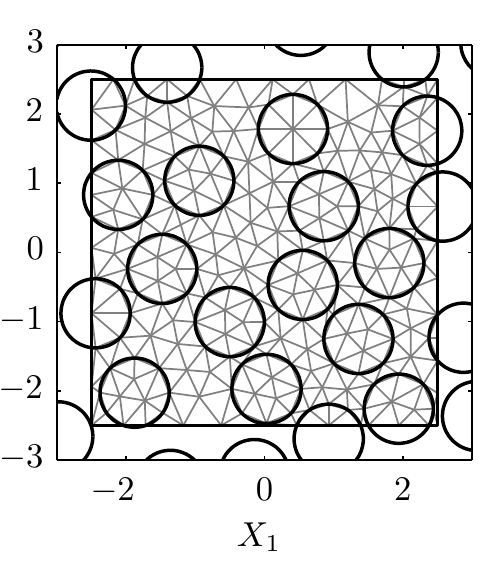}\label{SubSect:BEIDIC:Fig2c}}
	\caption{Three MVE meshes employed in the simulations. (a)~Fine mesh, typical element size~$h \approx d/6$, approx.~50 pix/triangle, (b)~medium mesh, $h \approx d/4$, approx.~140 pix/triangle, and~(c) coarse mesh, $h \approx d/3$, approx.~270 pix/triangle. In all cases, quadratic iso-parametric elements were used.}
	\label{SubSect:BEIDIC:Fig2}
\end{figure}

To systematically study the effects of the boundary conditions established by the GDIC on the subsequent IDIC identification, the following test is performed. GDIC with FE interpolation functions and quadratic iso-parametric triangular elements is employed. The element size of the structured GDIC mesh is increased from fine to coarse; the two extremes are shown in Figs.~\ref{SubSect:BCGDIC:Fig1b} and~\ref{SubSect:BCGDIC:Fig1c}. For each of these GDIC meshes, the IDIC identification is carried out for various combinations of material parameters, three MVE meshes (shown in Fig.~\ref{SubSect:BEIDIC:Fig2}), three material contrast ratios, and the three considered mechanical tests. All considered test cases are shown schematically in Fig.~\ref{SubSect:BCGDIC:Fig0}. As indicated in Algorithm~\ref{SubSect:BCGDIC:Alg1}, strict inclusion (i.e.~$\Omega_\mathrm{mve} \subset \Omega_\mathrm{roi}^\mathrm{gdic}$) is adopted to eliminate large errors close to the~$\Omega_\mathrm{roi}^\mathrm{gdic}$ boundary, cf. e.g.~\cite{Rethore:2008}. The margin is chosen to be one MVE mesh element size thick. Furthermore, because the GDIC minimization would fail in the case of fine triangulations (even for~$10$ time increments), a mechanical regularization based on the Equilibrium Gap method has been adopted; see~\cite{tomivcevc:2013} for further details. The weight associated with the elastic regularization potential is progressively decreased to zero throughout the iteration process, meaning that the employed regularization merely helps the GDIC algorithm to locate the proper minimum.

\begin{figure}
 	\centering
	\includegraphics[scale=1]{BCNoiseLsigLegend.pdf}\vspace{-0.5em}\\
	\subfloat[tension]{\includegraphics[scale=1]{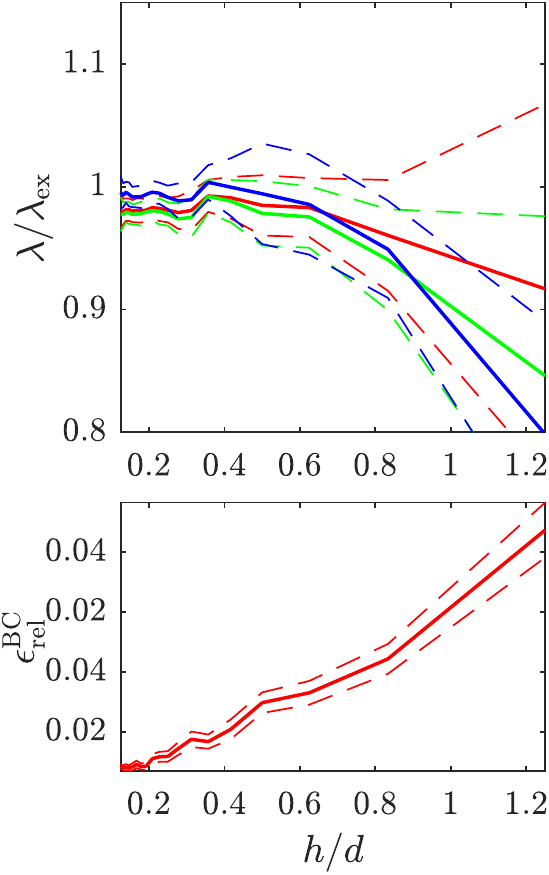}\label{SubSect:BCGDIC:Fig2a}}\hspace{0.1em}
	\subfloat[shear]{\includegraphics[scale=1]{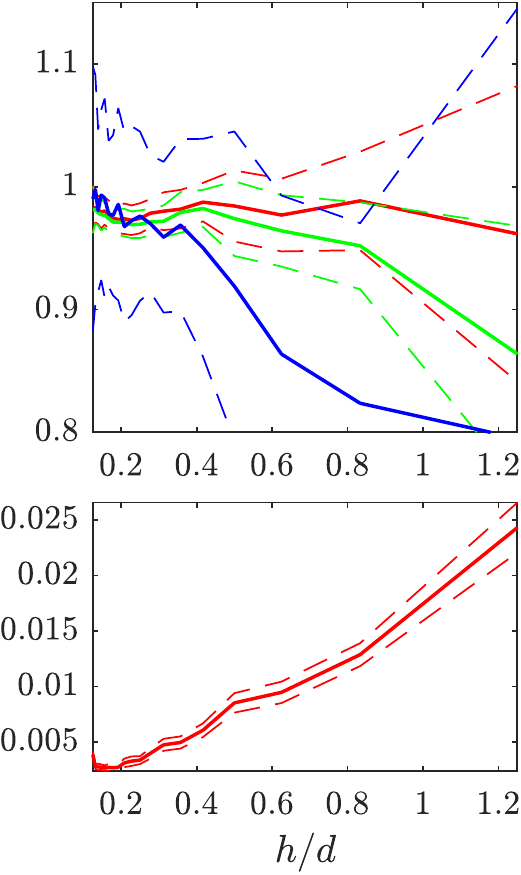}\label{SubSect:BCGDIC:Fig2b}}\hspace{0.1em}
	\subfloat[bending]{\includegraphics[scale=1]{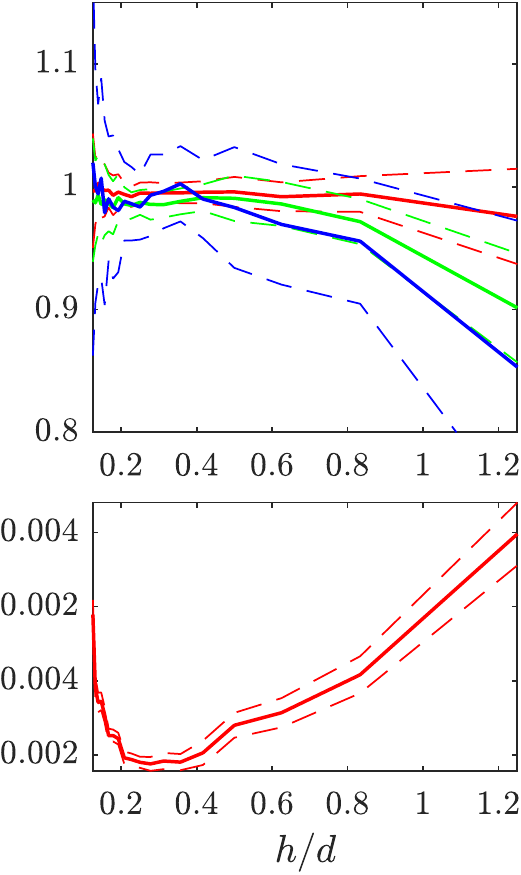}\label{SubSect:BCGDIC:Fig2c}}
\caption{Identified material parameters for the GDIC-IDIC approach as a function of the typical GDIC mesh element size~$h \in \frac{d}{8}\,[1, 10]$, recall Figs.~\ref{SubSect:BCGDIC:Fig1b} and~\ref{SubSect:BCGDIC:Fig1c}, material contrast ratio~$\rho = 4$, zero image noise, and~$\bs{\lambda} = [G_1,G_2,K_2]^\mathsf{T}$. The identification is shown for three mechanical tests: (a)~tension, (b)~shear, and~(c) bending.}
	\label{SubSect:BCGDIC:Fig2}
\end{figure}

Partial results obtained for the three mechanical tests, medium MVE meshes, zero image noise, and material contrast ratio~$\rho = 4$, are depicted in Fig.~\ref{SubSect:BCGDIC:Fig2}. Here, the effects of both random error (for small GDIC mesh element size~$h$) and smoothing (large~$h$) resulting from the GDIC can be observed. For large GDIC elements the effect of smoothing is highly pronounced, even significantly biasing the mean values, whereas the random error affects mainly the standard deviations and has a less extensive impact. This holds especially for the shear and bending tests, which are generally more sensitive to the accuracy of the prescribed BCs (recall Fig.~\ref{SubSect:ConstModel:Fig2}). Typical minimum values of the relative error (defined in Eq.~\eqref{Sect:BCNoise:Eq1a}) that were achieved by the GDIC are approximately~$1\,\%$, $0.5\,\%$, and $0.1\,\%$ for the tension, shear, and bending test. For a GDIC triangulation that may be considered reasonable ($h / d = 0.5$), the typical relative error increases approximately to~$1.5\,\%$, $0.9\,\%$, and~$0.2\,\%$. It is important to note that in practice no means are available to a priori determine the optimal GDIC mesh.

\begin{figure}
 	\centering
	\includegraphics[scale=1]{BCNoiseLsigLegend.pdf}\vspace{-0.5em}\\
	\subfloat[fine MVE mesh]{\includegraphics[scale=1]{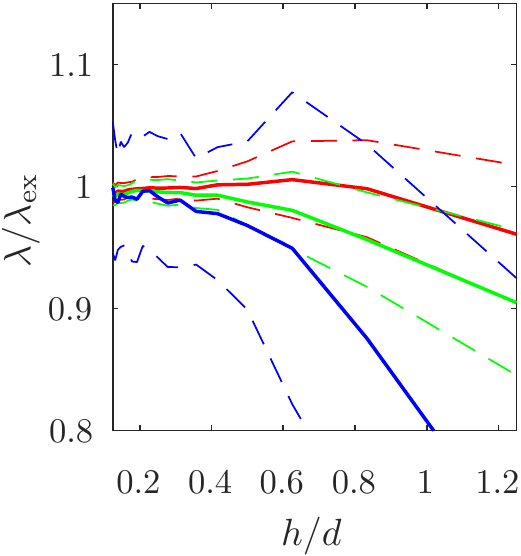}\label{SubSect:BCGDIC:Fig3a}}\hspace{0.5em}
	\subfloat[coarse MVE mesh]{\includegraphics[scale=1]{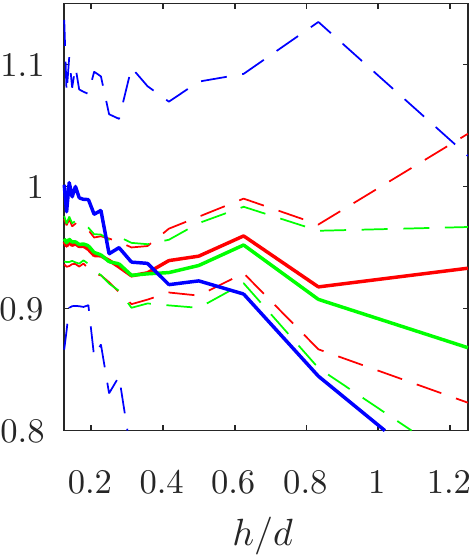}\label{SubSect:BCGDIC:Fig3b}}\hspace{0.5em}
	\subfloat[high materal contrast ratio]{\includegraphics[scale=1]{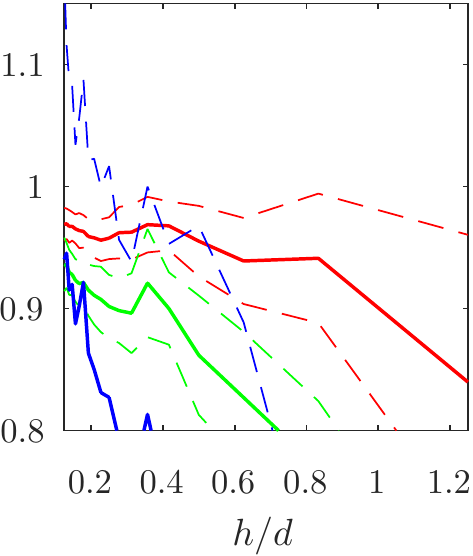}\label{SubSect:BCGDIC:Fig3c}}
\caption{Identified material parameters for the GDIC-IDIC approach as a function of the typical GDIC mesh element size~$h \in \frac{d}{8}\,[1, 10]$ for~$\bs{\lambda} = [G_1,G_2,K_2]^\mathsf{T}$ and zero image noise. (a)~Shear test for fine MVE meshes and~$\rho = 4$, (b)~shear test for coarse MVE meshes and~$\rho = 4$, and~(c) shear test for medium MVE meshes and~$\rho = 16$.}
	\label{SubSect:BCGDIC:Fig3}
\end{figure}

The remaining material combinations exhibit similar trends to those of Fig.~\ref{SubSect:BCGDIC:Fig2} and are therefore not all shown here. When fine MVE meshes instead of medium ones are used, the accuracy of the identified parameters increases, whereas for the coarse MVE meshes it decreases, see Figs.~\ref{SubSect:BCGDIC:Fig3a} and~\ref{SubSect:BCGDIC:Fig3b}. A decrease in accuracy is observed also for a higher material contrast ratio, cf. Fig.~\ref{SubSect:BCGDIC:Fig3c}. In general, the higher the contrast ratio, the more sensitive the IDIC is to the boundary data (and hence also the less accurate).

\begin{figure}
	\centering
	\includegraphics[scale=1]{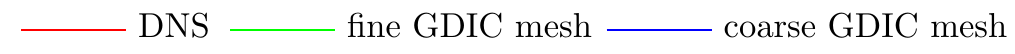}\vspace{0.1em}\\
	\begin{tikzpicture}
	\linespread{1}
	
	\tikzset{
    	mynode/.style={inner sep=0,outer sep=0},
	    myarrow/.style={black,thick,dashed},
	}	
	
	\node[mynode] (Tension) {
		\subfloat[tension]{\includegraphics[scale=1]{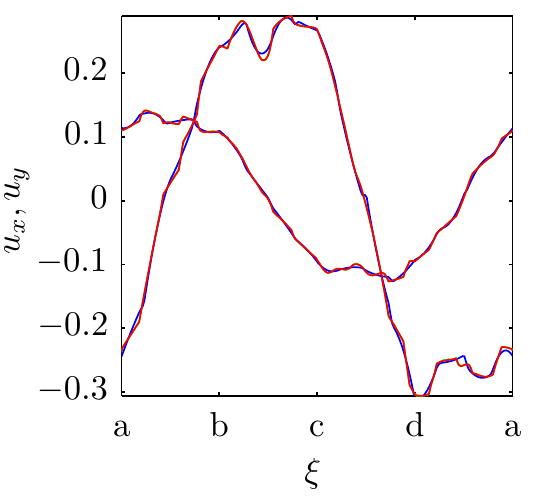}\label{SubSect:BCGDIC:Fig4a}}
		};
	\node[mynode,right=0.0em of Tension] (Shear) {
		\subfloat[shear]{\includegraphics[scale=1]{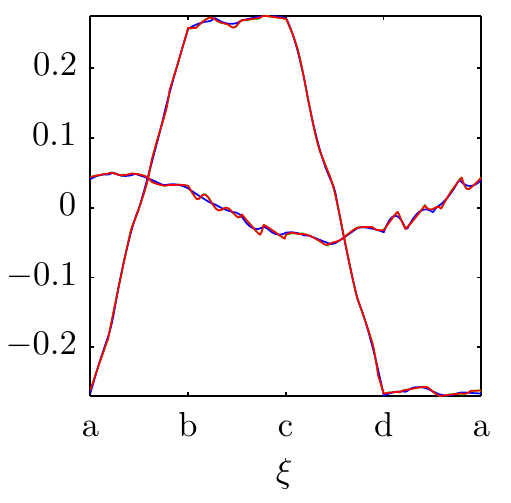}\label{SubSect:BCGDIC:Fig4b}}	
		};
	\node[mynode,right=0.0em of Shear] (Bending) {
		\subfloat[bending]{\includegraphics[scale=1]{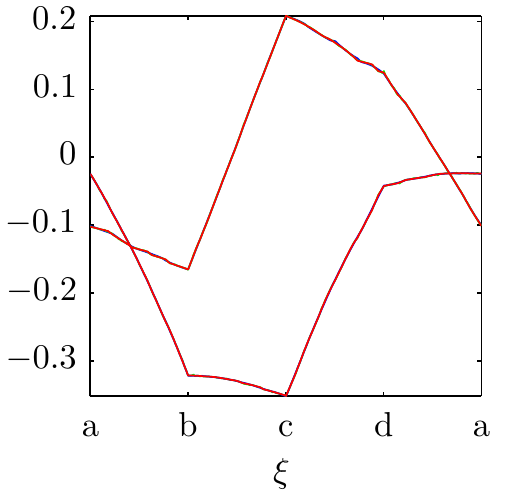}\label{SubSect:BCGDIC:Fig4c}}
	};

	\node[mynode,above=0.0em of Tension] (TensionZ) {
		\includegraphics[scale=1]{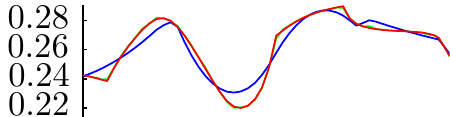}
		};
	\node[mynode,above=0.0em of Shear] (ShearZ) {
		\includegraphics[scale=1]{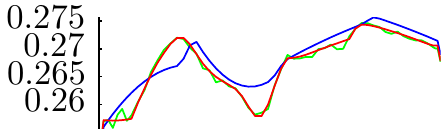}
		};
	\node[mynode,above=0.0em of Bending] (BendingZ) {
		\includegraphics[scale=1]{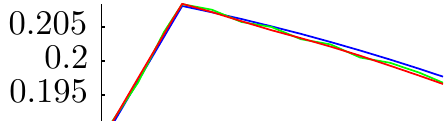}
	};	

	\draw[black, dashed, thick] (-0.7,2) rectangle (0.7,2.5);
	\draw[black, thick, dashed] (-0.7,2.25) -- ([shift={(0.85,0)}]TensionZ.south west);
	\draw[black, thick, dashed] (0.7,2.25) -- (TensionZ.south east);
	
	\draw[black, dashed, thick] (4.5,2) rectangle (5.8,2.5);
	\draw[black, thick, dashed] (4.5,2.25) -- ([shift={(1,0)}]ShearZ.south west);
	\draw[black, thick, dashed] (5.8,2.25) -- (ShearZ.south east);
	
	\draw[black, dashed, thick] (10.6,2) rectangle (11.6,2.5);
	\draw[black, thick, dashed] (10.6,2.25) -- ([shift={(1.05,0)}]BendingZ.south west);
	\draw[black, thick, dashed] (11.6,2.25) -- (BendingZ.south east);		
	
	
	\end{tikzpicture}
	
	\caption{GDIC boundary displacements corresponding to two extreme mesh element sizes, cf. Figs.~\ref{SubSect:BCGDIC:Fig1b} and~\ref{SubSect:BCGDIC:Fig1c}, compared to the DNS data for~$\rho = 4$, zero image noise, and~(a) tension, (b)~shear, and~(c) bending test.}
	\label{SubSect:BCGDIC:Fig4}
\end{figure}

Examples of boundary displacements obtained from the GDIC compared to the exact DNS solutions are presented in Fig.~\ref{SubSect:BCGDIC:Fig4}, which shows that the apparently accurate GDIC data, especially for the fine GDIC mesh, are in sharp contrast with the inaccurate identifications they induce, as shown in Figs.~\ref{SubSect:BCGDIC:Fig2} and~\ref{SubSect:BCGDIC:Fig3}. The errors of the GDIC data on the MVE boundary relative to the DNS solution are also indicated in the bottom part of Fig.~\ref{SubSect:BCGDIC:Fig2}.

Finally, let us note that the MVE mesh itself can be directly used for GDIC as well, removing thus one interpolation step. This option has also been tested, but has not brought any significant improvement of the statistical scatter in the data. The achieved accuracy improved only in some particular cases,
depending on the topology of employed MVE meshes.
\begin{algorithm}
\caption{GDIC-IDIC approach.}
\label{SubSect:BCGDIC:Alg1}
	\centering
	\vspace{-\topsep}
	\begin{enumerate}[1:]
	\item Construct a GDIC triangulation~$\mathcal{T}_\mathrm{gdic}$ of~$\Omega_\mathrm{roi}^\mathrm{gdic}$ and build~$\bs{\psi}_i$, cf. Eq.~\eqref{SubSect:DIC:Eq2}.
	\item Perform GDIC on~$\Omega_\mathrm{roi}^\mathrm{gdic}$.
	\item Triangulate~$\Omega_\mathrm{mve}$ and assemble MVE model.
	\item Sample the MVE BCs on~$\partial\Omega_\mathrm{mve}$ from GDIC data.
	\item Perform IDIC on~$\Omega_\mathrm{roi}^\mathrm{idic}$ ($\Omega_\mathrm{roi}^\mathrm{idic} \subseteq \Omega_\mathrm{mve} \subset \Omega_\mathrm{roi}^\mathrm{gdic}$).		
	\end{enumerate}
	\vspace{-\topsep}
\end{algorithm}
%
%
\section{Boundary-Enriched Integrated Digital Image Correlation}
\label{Sect:OurIDIC}
From Sections~\ref{SubSect:ConstModel} and mainly~\ref{Sect:BCErrors} it has become clear that slight inaccuracies in the BCs of the MVE model significantly deteriorate the accuracy of the identified parameters. One way of attenuating these adverse effects would be to decrease the overall sensitivity of the IDIC procedure to the prescribed MVE BCs. This can be achieved, for instance, by prescribing BCs in the weak sense, giving more freedom to the system to accommodate boundary fluctuations. Such an approach would, nevertheless, rely on the assumption that the given system spontaneously adopts a correct configuration, which is rather unlikely. Another strategy could rely on adopting a large MVE domain (while keeping the ROI relatively small) and letting physical effects smoothing out any errors in BCs through Saint-Venant's Principle, cf. e.g.~\cite{Toupin:1966}. This strategy would work, but presumably only for random and not for systematic errors. The last option is to provide as accurate boundary data as possible, relying on the continuous dependence of solutions of well-posed partial differential equations on the given data, cf. e.g.~\cite{EvansPDE}. Assuming a correct constitutive law, morphology of the MVE model, and omitting any instability or other softening effects, this means that the experimentally observed configuration can be reached only for unique boundary data. To this end, an approach that treats the displacements of all nodes on the boundary of the MVE model as DOFs of the IDIC procedure is introduced, referred to as Boundary-Enriched IDIC~(BE-IDIC) for short. This allows the MVE model to relax any inaccuracies in BCs which, when prescribed rigidly, lock errors that later propagate to the identified parameters. Although BE-IDIC may resemble the methodology proposed by~\cite{Fedele:2015}, the following important differences exist:
\begin{enumerate}[(i)]
	\item whereas the work of~\cite{Fedele:2015} is set within the Finite Element Method Updating~(FEMU) framework, BE-IDIC is defined within the realm of IDIC, with demonstrated advantages in terms of robustness and accuracy (see~\citealt{Ruybalid:2016}),

	\item as a consequence of~(i), the resulting IDIC problem is well-posed and hence solvable even for full kinematic resolution of the boundary; this is in contrast with the method by~\cite{Fedele:2015}, for which the author himself points its ill-posedness,

	\item because the proposed methodology addresses the general case of highly heterogeneous nonlinear materials, smooth regularization of boundary data is not possible (in contrast to the method of~\cite{Fedele:2015}),

	\item for cases slightly less heterogeneous, in which full resolution of the boundary kinematics is not required, an adaptive algorithm is proposed to automatically find the correct boundary kinematics regularization (with option to reach the full resolution case); the method by~\cite{Fedele:2015} requires, on the other hand, a prior choice of regularization (properly selected by the user).

\end{enumerate}
In order to demonstrate the advantages and robustness of the introduced method, the examples from Section~\ref{SubSect:Tests} are performed again and compared to the best results obtained from the GDIC-IDIC approach. Subsequently, a noise study is carried out to assess the robustness of both methods under more realistic measurement conditions.

Before proceeding, let us note that~\cite{Buljac:2017} mention that as long as the BCs capture the mesoscopic kinematic features, they are sufficient for identification of micromechanical properties of cast iron. Although their conclusion builds on a tension test, which is relatively robust (recall Figs.~\ref{SubSect:ConstModel:Fig2a} and~\ref{SubSect:BCGDIC:Fig2a}), the previous sections of this contribution indicate that such a statement should not be generalized for highly heterogeneous microstructures, because one cannot a priori conjecture on the kind of loading inside a chosen ROI due to heterogeneities, existing percolation paths, or other effects.
%
%
\subsection{Description of the Method}
\label{SubSect:BEIDIC}
The BE-IDIC is an IDIC methodology that considers material parameters as well as the vector of displacements associated with nodes on the MVE boundary as unknowns, i.e.
\begin{equation}
\widehat{\bs{\lambda}} = [\widehat{\bs{\lambda}}_\mathrm{mat}^\mathsf{T},\widehat{\bs{\lambda}}_\mathrm{kin}^\mathsf{T}]^\mathsf{T},
\label{SubSect:BEIDIC:Eq1}
\end{equation}
where
\begin{equation}
\begin{aligned}
\widehat{\bs{\lambda}}_\mathrm{mat} &= [G_1, K_1, \dots]^\mathsf{T}, \\
\widehat{\bs{\lambda}}_\mathrm{kin} &= \bs{\mathsf{u}}_\mathrm{mve}(\bs{X}), \quad \bs{X} \in \partial\Omega_\mathrm{mve}.
\end{aligned}
\label{SubSect:BEIDIC:Eq2}
\end{equation}
The brightness cost functional~$\mathcal{R}(\widehat{\bs{\lambda}})$, defined in Eq.~\eqref{SubSect:DIC:Eq1}, is subsequently minimized following the standard IDIC procedure detailed in Section~\ref{SubSect:DIC}, cf. also Algorithm~\ref{SubSect:BEIDIC:Alg1}. Compared to the GDIC-IDIC approach, the number of IDIC DOFs being optimized in the BE-IDIC method increases by~$n_{\lambda_\mathrm{kin}}$. Note also that when accurate kinematic initialization through GDIC is provided, Algorithm~\ref{SubSect:BEIDIC:Alg1} can be simplified by removing the refinement loop.

\begin{figure}
	\centering
	\mbox{}\hspace{1.5em}\includegraphics[scale=1]{SensitivityColorBar.pdf}\vspace{-0.5em}\\
	\subfloat[$\max{\widetilde{\varphi}_2} = 3.415 \cdot 10^{-1}$]{\includegraphics[scale=1]{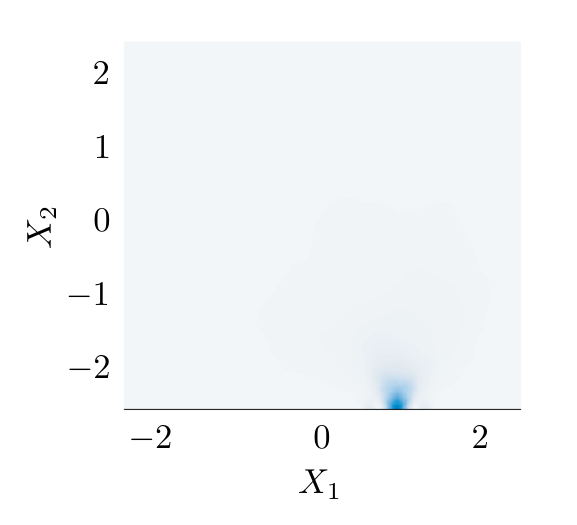}\label{SubSect:BEIDIC:Fig1a}}\hspace{1em}
	\subfloat[$\max{\widetilde{\varphi}_1} = 1.378 \cdot 10^{-2}$]{\includegraphics[scale=1]{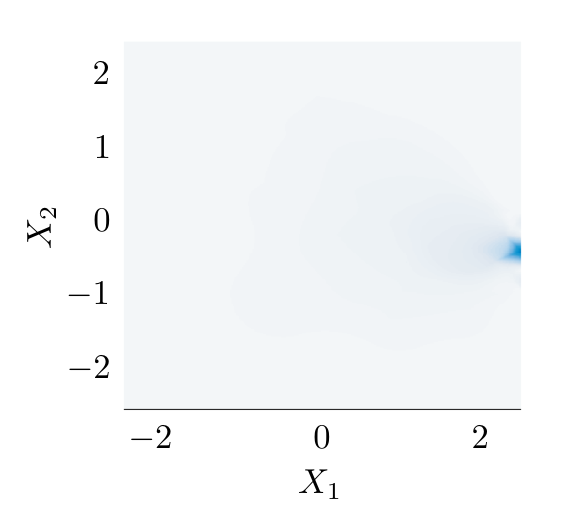}\label{SubSect:BEIDIC:Fig1b}}\\
	\caption{Normalized kinematic sensitivity fields~$\widetilde{\varphi}_i$, recall Eq.~\eqref{Sect:ConstModel:Eq3}, evaluated for the shear test, exact material and kinematic data, $\rho = 4$, and for~$\Omega_\mathrm{mve} = \Omega_\mathrm{roi}$. Sensitivities correspond to~(a) the vertical displacement of a node which is part of a stiff particle, and~(b) to the horizontal displacement of a node which is part of a compliant matrix. For clarity, presented plots are normalized to one whereas corresponding magnitudes are mentioned in individual captions.}
	\label{SubSect:BEIDIC:Fig1}
\end{figure}

Because GDIC is based purely on a geometric concept (in absence of a mechanical regularization), the mechanical significance and accuracy of the displacements relate to the shape and support size of individual interpolation functions~$\bs{\psi}_i$. On the contrary, in the BE-IDIC method the mechanical significance of kinematic boundary DOFs derives from the underlying mechanics through their sensitivity fields (this is in a sense true mechanical regularization). Recall for clarity Section~\ref{SubSect:Sfields} and remember that the sensitivity fields associated with~$\widehat{\bs{\lambda}}_\mathrm{kin}$ are computed through the MVE model, and are different from the boundary sensitivity functions defined as traces on~$\partial\Omega_\mathrm{mve}$ of the sensitivity fields computed through the DNS model of the entire specimen. Therefore, if a boundary node happens to be part of a stiff particle, cf. Fig.~\ref{SubSect:BEIDIC:Fig1a}, its sensitivity field has a larger magnitude compared to the sensitivity field corresponding to a node in a soft matrix, shown in Fig.~\ref{SubSect:BEIDIC:Fig1b}. The proposed method therefore automatically corrects for the displacements of all boundary nodes, while at the same time taking into account their mechanical importance. Fig.~\ref{SubSect:BEIDIC:Fig1} further shows that the kinematic sensitivity fields are supported only in a close vicinity of the boundary, whereas the material sensitivity fields are supported inside the full MVE but vanish on~$\partial\Omega_\mathrm{mve}$ (recall Fig.~\ref{SubSect:ConstModel:Fig1}). This means that no danger of high correlations between them exists. Finally, as already noted in Section~\ref{SubSect:BCGDIC}, for the GDIC-IDIC approach the employed~$\Omega_\mathrm{roi}^\mathrm{gdic}$ should be larger than~$\Omega_\mathrm{roi}^\mathrm{idic}$ in order to reduce the errors in the BCs. A certain portion of the micro-image is hence sacrificed for the identification of BCs, which is avoided in the BE-IDIC method.
\begin{algorithm}
\caption{Adaptive Boundary-Enriched Integrated Digital Image Correlation.}
\label{SubSect:BEIDIC:Alg1}
	\centering
	\vspace{-\topsep}
	\begin{enumerate}[1:]
	\item Construct triangulation~$\mathcal{T}_\mathrm{mve}$ of~$\Omega_\mathrm{mve}$ and assemble MVE micro-model.
	\item Initialize piecewise affine interpolation along the MVE boundary~$\partial\Omega_\mathrm{mve}$ such that only~4 MVE corner nodes result.
	\item \textbf{while} DOFs of all nodes at~$\partial\Omega_\mathrm{mve}$ are not included in~$\widehat{\bs{\lambda}}_\mathrm{kin}$ \textbf{and} given tolerance in~$\mathcal{R}(\widehat{\bs{\lambda}})$ is not met.
	\begin{enumerate}[(I):]
		\item Perform IDIC on~$\Omega_\mathrm{roi}^\mathrm{idic}$ for~$\widehat{\bs{\lambda}} = [\widehat{\bs{\lambda}}_\mathrm{mat}^\mathsf{T},\widehat{\bs{\lambda}}_\mathrm{kin}^\mathsf{T}]^\mathsf{T}$; iterate to convergence.
		\item Refine boundary interpolation: add mid-nodes between current boundary nodes, include their DOFs into~$\widehat{\bs{\lambda}}_\mathrm{kin}$ and initialize them through linear interpolation.
	\end{enumerate}
	\item \textbf{end}
	\end{enumerate}
	\vspace{-\topsep}
\end{algorithm}

Overall, the main assets of the BE-IDIC method can be summarized as follows:
\begin{itemize}
\item consistency; material parameters have the same influence in minimization of~$\mathcal{R}$ as BCs have, and are identified with an accuracy corresponding to their mechanical significance (reflected by their sensitivity fields);

\item boundary fluctuations are resolved automatically, weighted by their true mechanical significance; BCs do not lock errors;

\item the entire micro image is used for material identification;

\item simplicity; no direct need for a separate GDIC procedure.

\end{itemize}
However, some disadvantages should also be emphasized:
\begin{itemize}
\item computational intensity (a large number of IDIC DOFs, $\widehat{\bs{\lambda}}_\mathrm{kin}$);

\item high memory requirements (a large number of sensitivity fields);

\item sensitivity to initial guess due to high-dimensionality of~$\widehat{\bs{\lambda}}$;

\item for highly irregular meshes (when very short and long element edges at~$\partial\Omega_\mathrm{mve}$ occur) the IDIC Hessian~$\bs{H}$ may be poorly scaled.

\end{itemize}

All of the above-listed disadvantages can be partially remedied as follows. Although the high computational intensity may not be a real concern compared to the effort involved in performing an accurate micro-mechanical test under in-situ microscopic observation, it can be attenuated by computing sensitivity fields associated with~$\widehat{\bs{\lambda}}_\mathrm{kin}$ selectively, not in each iteration. As sensitivity fields are corrections from the current iterative state~$\widehat{\bs{\lambda}}$ to a perturbed state ($\widehat{\bs{\lambda}}+\epsilon\widehat{\lambda}_i\bs{e}_i$), recall Eq.~\eqref{SubSect:DIC:Eq8}, they can be resolved by a single Newton iteration, requiring only one factorization of the mechanical stiffness matrix solved for~$n_{\lambda_\mathrm{kin}}$ right hand sides. High memory requirements can be reduced by truncating all kinematic sensitivity fields in space, as they are locally supported in the close vicinity of the MVE boundary (recall Fig.~\ref{SubSect:BEIDIC:Fig1}), and by employing sparse data storage. The sensitivity to the initial guess values can be improved by adaptive refinement in the boundary, recall Algorithm~\ref{SubSect:BEIDIC:Alg1}, which systematically increases the number of IDIC DOFs. Adaptivity also addresses the last disadvantage, as too fine elements can be clustered to larger edge units.
%
%
\subsection{Examples}
\label{SubSect:Ex1}
\begin{figure}
 	\centering
	\includegraphics[scale=1]{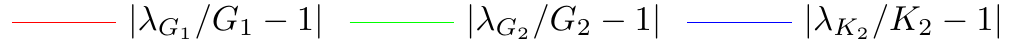}\\
	\includegraphics[scale=1]{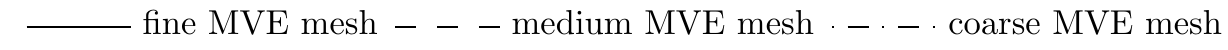}\vspace{-1.0em}\\
	\subfloat[tension]{\includegraphics[scale=1]{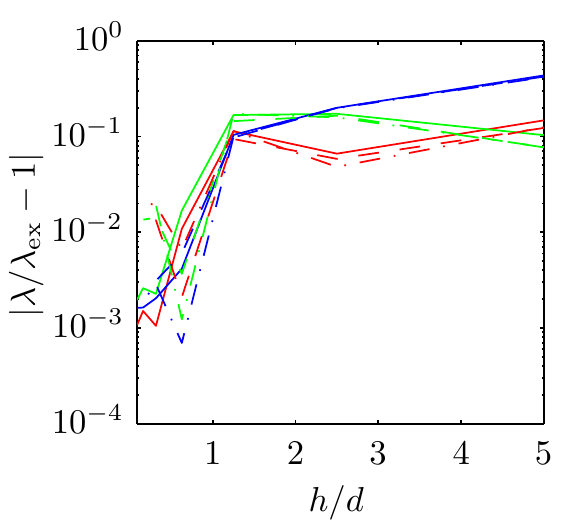}\label{SubSect:Ex1:Fig1a}}
	\subfloat[shear]{\includegraphics[scale=1]{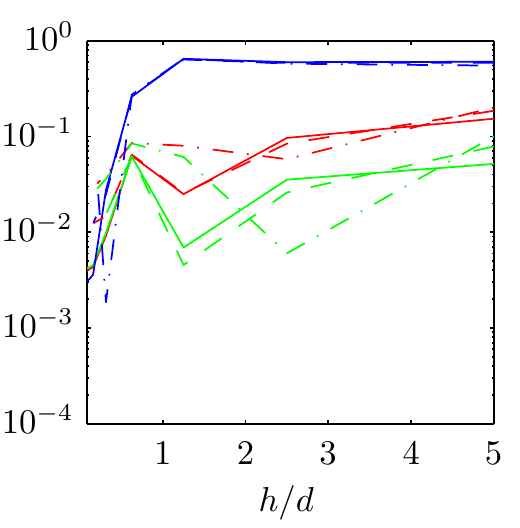}\label{SubSect:Ex1:Fig1b}}
	\subfloat[bending]{\includegraphics[scale=1]{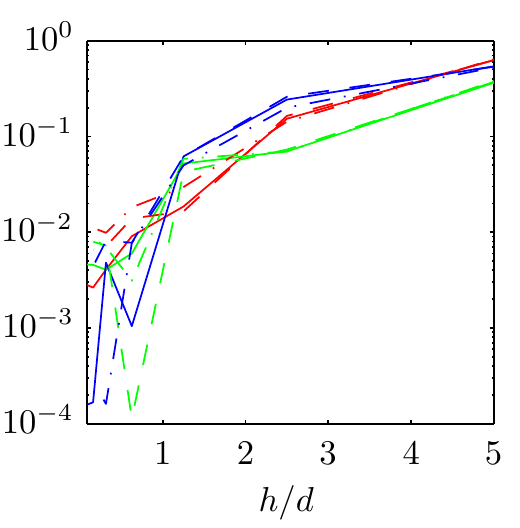}\label{SubSect:Ex1:Fig1c}}
	\caption{Errors in identified material parameters for the BE-IDIC approach as functions of average edge element size at the MVE boundary. The results correspond to one MC realization and~(a) tension, (b)~shear, and~(c) bending test; $\rho = 4$, $\Omega_\mathrm{mve} = \Omega_\mathrm{roi}$, images with zero noise, and all types of MVE meshes used.}
	\label{SubSect:Ex1:Fig1}
\end{figure}

First, convergence of the identified material parameters is demonstrated as a function of average element size on the MVE boundary. The obtained results are shown in Fig.~\ref{SubSect:Ex1:Fig1} for one realization, all three mechanical tests, material contrast ratio~$\rho = 4$, $\Omega_\mathrm{mve} = \Omega_\mathrm{roi}$, zero image noise, and all MVE meshes. The curves indicate that in all cases, a high level of detail is required (reflected by slow convergence). Identification starts to rapidly improve only for element sizes comparable to the microstructural geometric property~$d$, meaning that the BCs should capture microscopic features when accurate identification is required. Meso- or macroscopic features do not suffice.

\begin{figure}
 	\centering
	\subfloat[displacement component~$u_1$]{\includegraphics[scale=1]{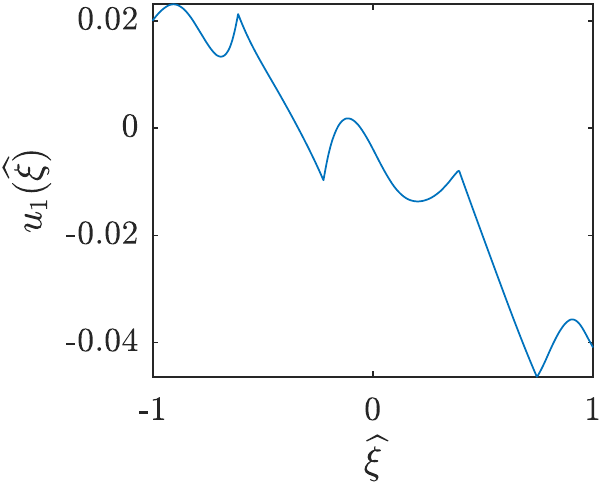}\label{SubSect:Ex1:Fig0a}}\hspace{0.5em}
	\subfloat[relative error in a least squares fit]{\includegraphics[scale=1]{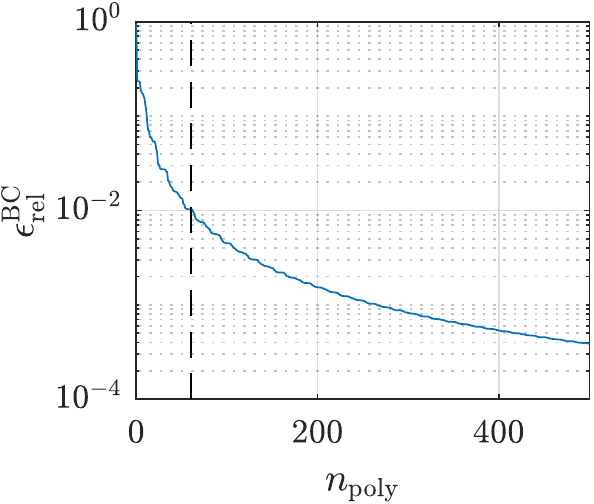}\label{SubSect:Ex1:Fig0b}}\hspace{0.1em}
	\caption{(a)~Typical heterogeneous displacement component~$u_1(\widehat{\xi})$ along the right vertical MVE edge, corresponding to the shear test. (b)~The relative displacement error (Eq.~\eqref{Sect:BCNoise:Eq1a}) as a function of the number of Chebyshev polynomials~$n_\mathrm{poly}$ used for the approximation of~$u_1$ by the least squares method.}
	\label{SubSect:Ex1:Fig0}
\end{figure}

Presented results also indicate that a straightforward regularization of boundary displacements may compromise accurate identification of material parameters if an insufficiently rich basis is used. This holds especially in the case of smooth functions such as Chebyshev polynomials, suggested by~\cite{Fedele:2015}. See for instance Fig.~\ref{SubSect:Ex1:Fig0a}, where a typical horizontal displacement component~$u_1(\widehat{\xi})$ is shown as a function of a normalized parametric coordinate~$\widehat{\xi}$ (spanning the right vertical MVE edge). The approximation quality of Chebyshev polynomials is measured by the relative displacement error in Fig.~\ref{SubSect:Ex1:Fig0b}, which quantifies the difference between the exact DNS results and a least squares fit; here, the dashed line corresponds to the number of FE nodes located on one MVE edge. The error is expressed as a function of the number of basis polynomials used, $n_\mathrm{poly}$. The resulting rate of convergence is rather slow due to sharp cusps and fluctuations.

\begin{figure}
 	\centering
	\includegraphics[scale=1]{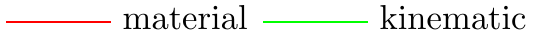}\vspace{-1.0em}\\ 	
	\subfloat[$\bs{\lambda}_\mathrm{kin}$ initialized with~$10\,\%$ error]{\includegraphics[scale=1]{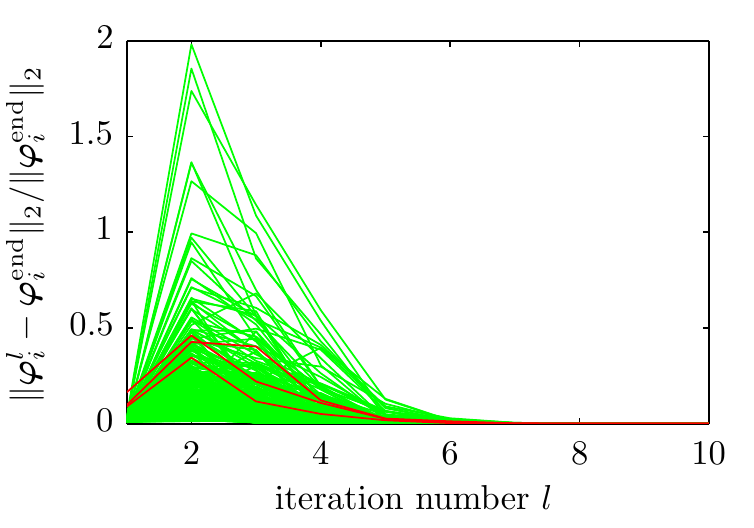}\label{SubSect:Ex1:Fig1.5a}}
	\subfloat[$\bs{\lambda}_\mathrm{kin}$ initialized with GDIC]{\includegraphics[scale=1]{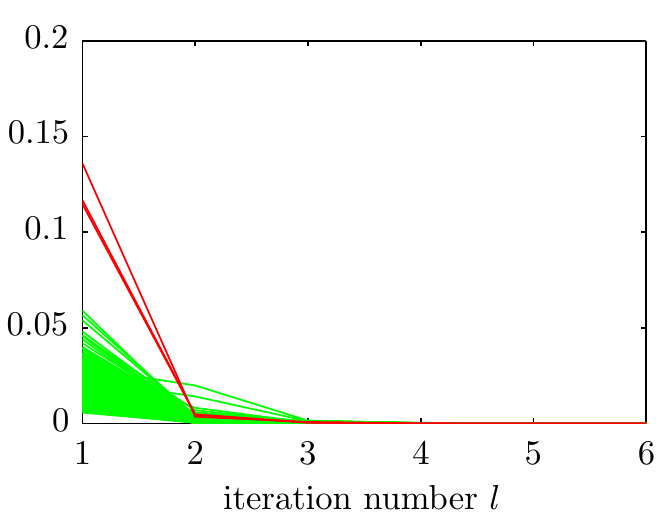}\label{SubSect:Ex1:Fig1.5b}}
	\caption{Typical convergence of relative errors in sensitivity fields corresponding to material and kinematic DOFs for fine MVE meshes. (a)~Both material and BCs are initialized with~$10\,\%$ systematic error, and~(b) only material parameters are initialized with~$10\,\%$ systematic error, whereas BCs are initialized through GDIC.}
	\label{SubSect:Ex1:Fig1.5}
\end{figure}

Typical convergence of relative errors in material and kinematic sensitivity fields~$\bs{\varphi}_i^l$ with respect to their converged values~$\bs{\varphi}_i^\mathrm{end}$ are plotted against the Newton iteration number~$l$ in Fig.~\ref{SubSect:Ex1:Fig1.5}. Here, two situations are depicted: first, material and kinematic DOFs are initialized with~$10\,\%$ systematic error (Fig.~\ref{SubSect:Ex1:Fig1.5a}); second, kinematic DOFs are initialized by GDIC and the material DOFs are initialized again with~$10\,\%$ systematic error. In both cases, fine MVE meshes and the fully resolved boundary are used. The curves show a fast convergence of the kinematic sensitivities when the relatively accurate initialization through GDIC is adopted, whereas they converge somewhat slower for inaccurate initialization. The observed behaviour thus suggests that updating kinematic sensitivity fields only selectively, or only once at the beginning of the iteration process, may suffice when displacements are initialized close to their correct values (recall Section~\ref{SubSect:BEIDIC}) as they are approximately within~$5\,\%$ accuracy already for the first Newton iteration.

\begin{figure}
 	\centering
	\includegraphics[scale=1]{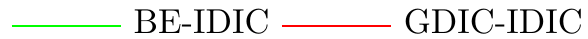}\vspace{-0.75em}\\ 	
	\subfloat[tension]{\includegraphics[scale=1]{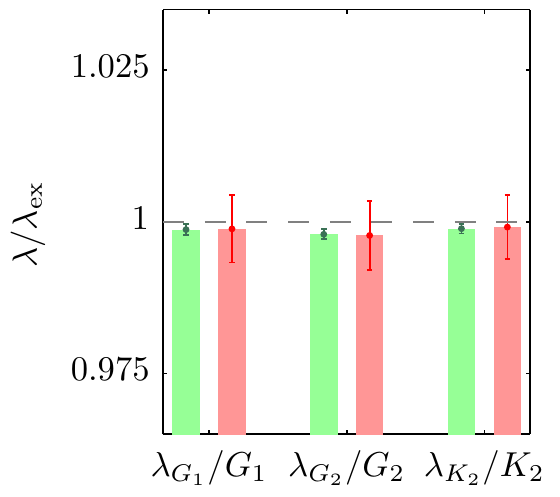}\label{SubSect:Ex1:Fig2a}}
	\subfloat[shear]{\includegraphics[scale=1]{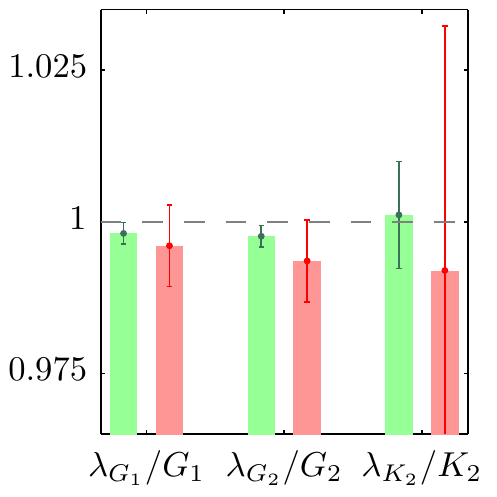}\label{SubSect:Ex1:Fig2b}}
	\subfloat[bending]{\includegraphics[scale=1]{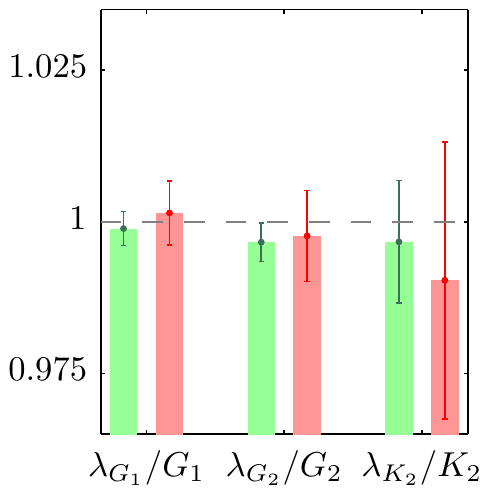}\label{SubSect:Ex1:Fig2c}}
	\caption{The mean values and standard deviations for the identified material parameters obtained for the BE-IDIC method and the best configuration of the GDIC-IDIC method. The results correspond to fine MVE meshes, $\rho = 4$, zero image noise, one set of identified material parameters~$\bs{\lambda} = [G_1,G_2,K_2]^\mathsf{T}$, and~$50$ MC realizations.}
	\label{SubSect:Ex1:Fig2}
\end{figure}

In order to assess the accuracy of the BE-IDIC procedure, all~$108$ test cases summarized in the diagram of Fig.~\ref{SubSect:BCGDIC:Fig0} have been repeated for the same~$50$ MC realizations of random microstructures. Typical results are presented in Fig.~\ref{SubSect:Ex1:Fig2} in terms of the mean values and standard deviations. These results are directly compared to the best identification of the GDIC-IDIC method, characterized by the optimal GDIC mesh element size~$h_\mathrm{opt}$. The optimal element size is established by minimizing the following Root-Mean-Square~(RMS) norm:
\begin{equation}
\begin{aligned}
h_\mathrm{opt} &= \underset{\widehat{h}\in\mathscr{H}}{\text{arg min}}\ \eta_\mathrm{rms}(\widehat{h}), \\
\eta_\mathrm{rms}(\widehat{h}) &= \sqrt{\sum_{i=1}^{n_{\lambda_\mathrm{mat}}} m_{2,\widetilde{\lambda}_i}(\widehat{h}) },
\end{aligned}
\label{SubSect:Ex1:Eq1}
\end{equation}
where~$m_{2,\widetilde{\lambda}_i}(\widehat{h}) = \frac{1}{n_\mathrm{mc}}\sum_{j=1}^{n_\mathrm{mc}} \widetilde{\lambda}_{i,j}^2(\widehat{h})$ is the second raw moment of relative error associated with $i$-th identified material parameter computed for~$j = 1, \dots, n_\mathrm{mc} = 50$ realizations, whereas~$\mathscr{H}$ is a set of all employed GDIC element mesh sizes. The adopted relative error reads
\begin{equation}
\widetilde{\lambda} = \frac{\lambda}{\lambda_\mathrm{ex}}-1.
\label{SubSect:Ex1:Eq2}
\end{equation}
The results in Fig.~\ref{SubSect:Ex1:Fig2} show an improved accuracy of the BE-IDIC method over the best results for the GDIC-IDIC approach, both in terms of the mean values as well as standard deviations. Quantified in terms of the RMS norm, $\eta_\mathrm{rms}$ decreases approximately~$3$, $4$, and~$2$ times for the tension, shear, and bending test.
\begin{figure}
 	\centering
	\includegraphics[scale=1]{BargraphLegend.pdf}\vspace{0.25em}\\
 	
	\begin{tikzpicture}[node distance=1em, auto]  
	\linespread{1}
	\tikzset{
    	mynode/.style={rectangle,rounded corners,draw=black, top color=white,inner sep=0.25em,outer sep=0.0em,minimum size=3em,text centered,scale=0.45,transform shape},
	    myarrow/.style={->, >=latex', shorten >=1pt},
	}  

	\node[inner sep=0em,outer sep=0em] (Tension) {
		\includegraphics[scale=1]{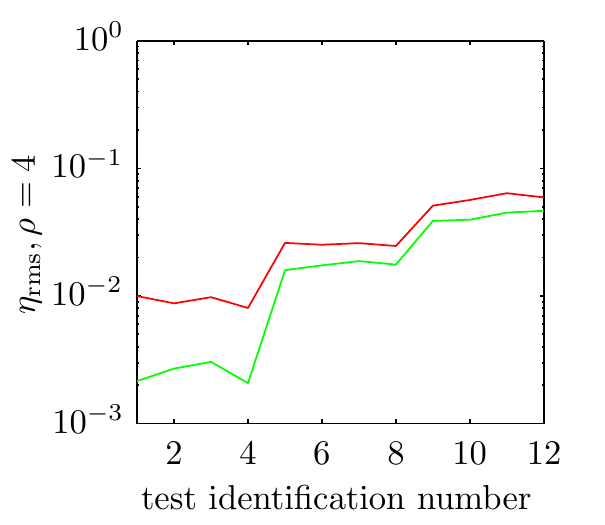}
		};
	\node[inner sep=0em,outer sep=0em,right=0.0em of Tension] (Shear) {
		\includegraphics[scale=1]{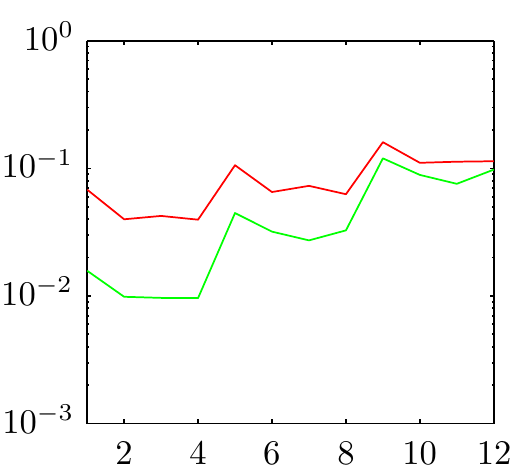}
		};
	\node[inner sep=0em,outer sep=0em,right=0.0em of Shear] (Bending) {
		\includegraphics[scale=1]{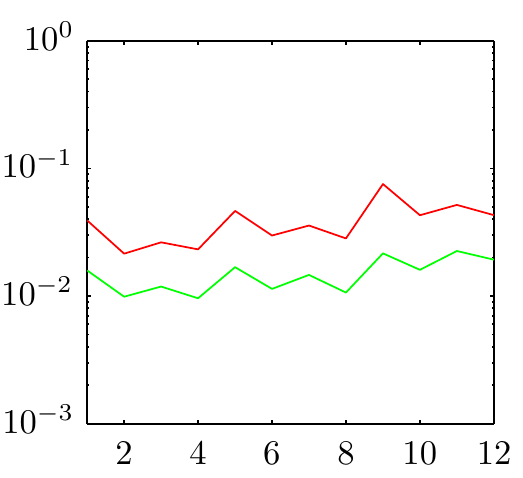}
	}; 

	\node[inner sep=0em,outer sep=0em,below=0.0em of Tension] (Tensionb) {
		\includegraphics[scale=1]{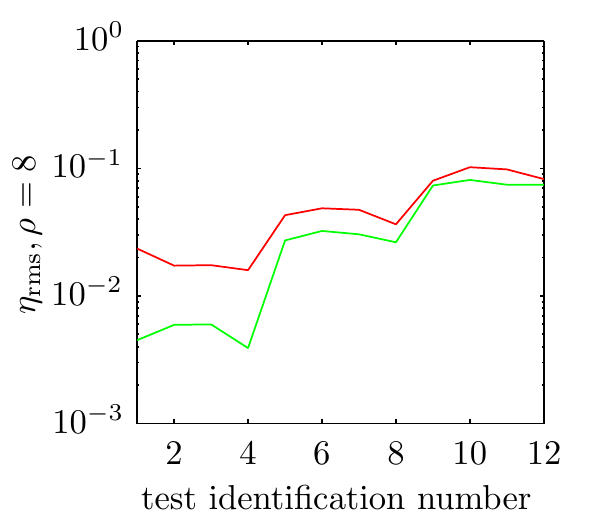}
		};
	\node[inner sep=0em,outer sep=0em,below=0.0em of Shear] (Shearb) {
		\includegraphics[scale=1]{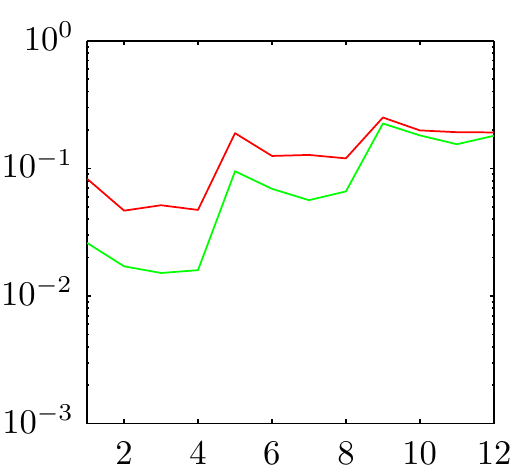}
		};
	\node[inner sep=0em,outer sep=0em,below=0.0em of Bending] (Bendingb) {
		\includegraphics[scale=1]{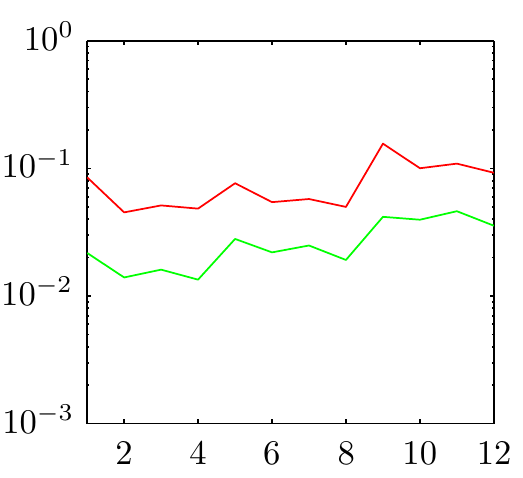}
	};
	
	\node[inner sep=0em,outer sep=0em,below=0.0em of Tensionb] (Tensionc) {
		\includegraphics[scale=1]{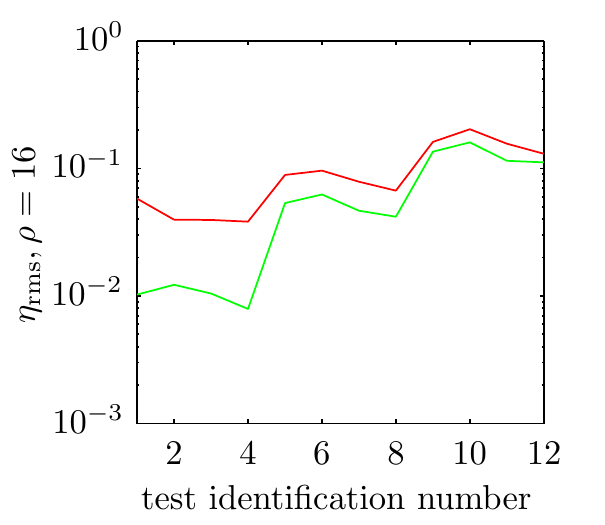}
		};
	\node[inner sep=0em,outer sep=0em,below=0.0em of Shearb] (Shearc) {
		\includegraphics[scale=1]{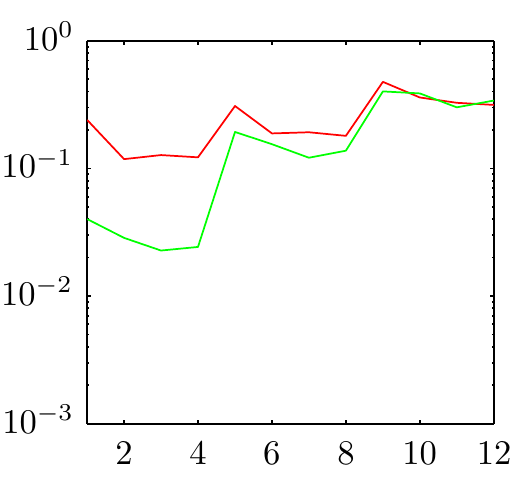}
		};
	\node[inner sep=0em,outer sep=0em,below=0.0em of Bendingb] (Bendingc) {
		\includegraphics[scale=1]{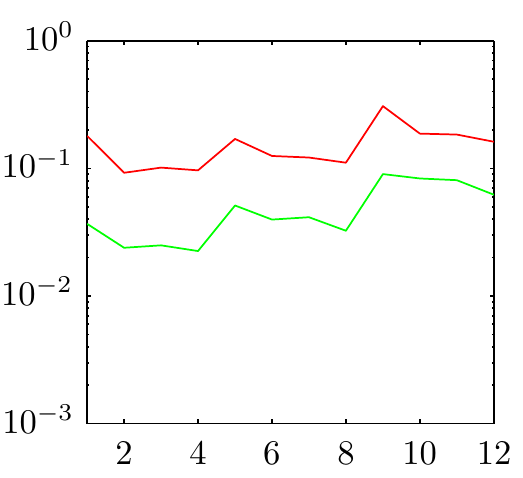}				
	};	

 	\node[mynode] at ([shift={(82.5:3.5)}]Tension) (TMVEmedium) {\begin{tabular}{c}MVE\\ medium mesh
 	\end{tabular}};
 	\node[mynode,left=0.5em of TMVEmedium] (TMVEfine) {\begin{tabular}{c}MVE\\ fine mesh
 	\end{tabular}};
 	\node[mynode,right=0.5em of TMVEmedium] (TMVEcoarse) {\begin{tabular}{c}MVE\\ coarse mesh
 	\end{tabular}};	

 	\node[mynode] at ([shift={(87:3.5)}]Shear) (SMVEmedium) {\begin{tabular}{c}MVE\\ medium mesh
 	\end{tabular}};
 	\node[mynode,left=0.5em of SMVEmedium] (SMVEfine) {\begin{tabular}{c}MVE\\ fine mesh
 	\end{tabular}};
 	\node[mynode,right=0.5em of SMVEmedium] (SMVEcoarse) {\begin{tabular}{c}MVE\\ coarse mesh
 	\end{tabular}};	 	
 	
 	\node[mynode] at ([shift={(87:3.5)}]Bending) (BMVEmedium) {\begin{tabular}{c}MVE\\ medium mesh
 	\end{tabular}};
 	\node[mynode,left=0.5em of BMVEmedium] (BMVEfine) {\begin{tabular}{c}MVE\\ fine mesh
 	\end{tabular}};
 	\node[mynode,right=0.5em of BMVEmedium] (BMVEcoarse) {\begin{tabular}{c}MVE\\ coarse mesh
 	\end{tabular}};	  	

	\node[mynode,above=1em of TMVEmedium] (Ltension) {Tension};
	\node[mynode,above=1em of SMVEmedium] (Lshear) {Shear};
	\node[mynode,above=1em of BMVEmedium] (Lbending) {Bending};
 		
	\node[mynode,above=1em of Lshear] (head) {\begin{tabular}{c}Computed\\ results
		\end{tabular}};

	\node[] at (-1.515,2) (Tmat1) {};
	\node[right=0.207em of Tmat1] (Tmat2) {};
	\node[right=0.207em of Tmat2] (Tmat3) {};
	\node[right=0.207em of Tmat3] (Tmat4) {};
	\node[right=0.207em of Tmat4] (Tmat5) {};
	\node[right=0.207em of Tmat5] (Tmat6) {};
	\node[right=0.207em of Tmat6] (Tmat7) {};
	\node[right=0.207em of Tmat7] (Tmat8) {};
	\node[right=0.207em of Tmat8] (Tmat9) {};				
	\node[right=0.207em of Tmat9] (Tmat10) {};
	\node[right=0.207em of Tmat10] (Tmat11) {};	
	\node[right=0.207em of Tmat11] (Tmat12) {};	

	\node[] at (3.8,2) (Tmat1a) {};
	\node[right=0.207em of Tmat1a] (Tmat2a) {};
	\node[right=0.207em of Tmat2a] (Tmat3a) {};
	\node[right=0.207em of Tmat3a] (Tmat4a) {};
	\node[right=0.207em of Tmat4a] (Tmat5a) {};
	\node[right=0.207em of Tmat5a] (Tmat6a) {};
	\node[right=0.207em of Tmat6a] (Tmat7a) {};
	\node[right=0.207em of Tmat7a] (Tmat8a) {};
	\node[right=0.207em of Tmat8a] (Tmat9a) {};				
	\node[right=0.207em of Tmat9a] (Tmat10a) {};
	\node[right=0.207em of Tmat10a] (Tmat11a) {};	
	\node[right=0.207em of Tmat11a] (Tmat12a) {};
	
	\node[] at (9.11,2) (Tmat1b) {};
	\node[right=0.207em of Tmat1b] (Tmat2b) {};
	\node[right=0.207em of Tmat2b] (Tmat3b) {};
	\node[right=0.207em of Tmat3b] (Tmat4b) {};
	\node[right=0.207em of Tmat4b] (Tmat5b) {};
	\node[right=0.207em of Tmat5b] (Tmat6b) {};
	\node[right=0.207em of Tmat6b] (Tmat7b) {};
	\node[right=0.207em of Tmat7b] (Tmat8b) {};
	\node[right=0.207em of Tmat8b] (Tmat9b) {};				
	\node[right=0.207em of Tmat9b] (Tmat10b) {};
	\node[right=0.207em of Tmat10b] (Tmat11b) {};	
	\node[right=0.207em of Tmat11b] (Tmat12b) {};		

 	\draw[myarrow] (head.south) -- ++(0,-0.10) -| (Ltension.north);	
 	\draw[myarrow] (head.south) -- ++(0,-0.10) -| (Lshear.north);
 	\draw[myarrow] (head.south) -- ++(0,-0.10) -| (Lbending.north);

 	\draw[myarrow] (Ltension.south) -- ++(0,-0.10) -| (TMVEfine.north);	
 	\draw[myarrow] (Ltension.south) -- ++(0,-0.10) -| (TMVEmedium.north);
 	\draw[myarrow] (Ltension.south) -- ++(0,-0.10) -| (TMVEcoarse.north);	

 	\draw[myarrow] (Lshear.south) -- ++(0,-0.10) -| (SMVEfine.north);	
 	\draw[myarrow] (Lshear.south) -- ++(0,-0.10) -| (SMVEmedium.north);
 	\draw[myarrow] (Lshear.south) -- ++(0,-0.10) -| (SMVEcoarse.north);	

 	\draw[myarrow] (Lbending.south) -- ++(0,-0.10) -| (BMVEfine.north);	
 	\draw[myarrow] (Lbending.south) -- ++(0,-0.10) -| (BMVEmedium.north);
 	\draw[myarrow] (Lbending.south) -- ++(0,-0.10) -| (BMVEcoarse.north);	

 	\draw[] (TMVEfine.south) -- ++(0,-0.1) -| (Tmat1.center);
 	\node [right=0em of Tmat1.center,scale=0.45,rotate=90,anchor=north west,inner sep=0.25em,outer sep=0em] {$G_1, K_1, G_2$};
 	\draw[] (TMVEfine.south) -- ++(0,-0.1) -| (Tmat2.center);	 	
 	\node [right=0em of Tmat2.center,scale=0.45,rotate=90,anchor=north west,inner sep=0.25em,outer sep=0em] {$G_1, K_1, K_2$};
 	\draw[] (TMVEfine.south) -- ++(0,-0.1) -| (Tmat3.center);	 	
 	\node [right=0em of Tmat3.center,scale=0.45,rotate=90,anchor=north west,inner sep=0.25em,outer sep=0em] {$G_1, G_2, K_2$};
 	\draw[] (TMVEfine.south) -- ++(0,-0.1) -| (Tmat4.center);	
 	\node [right=0em of Tmat4.center,scale=0.45,rotate=90,anchor=north west,inner sep=0.25em,outer sep=0em] {$K_1, G_2, K_2$}; 	
 	\draw[] (TMVEmedium.south) -- ++(0,-0.1) -| (Tmat5.center);
 	\node [right=0em of Tmat5.center,scale=0.45,rotate=90,anchor=north west,inner sep=0.25em,outer sep=0em] {$G_1, K_1, G_2$};
	\draw[] (TMVEmedium.south) -- ++(0,-0.1) -| (Tmat6.center);
 	\node [right=0em of Tmat6.center,scale=0.45,rotate=90,anchor=north west,inner sep=0.25em,outer sep=0em] {$G_1, K_1, K_2$};
 	\draw[] (TMVEmedium.south) -- ++(0,-0.1) -| (Tmat7.center);
 	\node [right=0em of Tmat7.center,scale=0.45,rotate=90,anchor=north west,inner sep=0.25em,outer sep=0em] {$G_1, G_2, K_2$};
 	\draw[] (TMVEmedium.south) -- ++(0,-0.1) -| (Tmat8.center);
 	\node [right=0em of Tmat8.center,scale=0.45,rotate=90,anchor=north west,inner sep=0.25em,outer sep=0em] {$K_1, G_2, K_2$}; 	
	\draw[] (TMVEcoarse.south) -- ++(0,-0.1) -| (Tmat9.center);
 	\node [right=0em of Tmat9.center,scale=0.45,rotate=90,anchor=north west,inner sep=0.25em,outer sep=0em] {$G_1, K_1, G_2$}; 	
 	\draw[] (TMVEcoarse.south) -- ++(0,-0.1) -| (Tmat10.center);
 	\node [right=0em of Tmat10.center,scale=0.45,rotate=90,anchor=north west,inner sep=0.25em,outer sep=0em] {$G_1, K_1, K_2$}; 	
 	\draw[] (TMVEcoarse.south) -- ++(0,-0.1) -| (Tmat11.center);
 	\node [right=0em of Tmat11.center,scale=0.45,rotate=90,anchor=north west,inner sep=0.25em,outer sep=0em] {$G_1, G_2, K_2$}; 	
 	\draw[] (TMVEcoarse.south) -- ++(0,-0.1) -| (Tmat12.center); 	
 	\node [right=0em of Tmat12.center,scale=0.45,rotate=90,anchor=north west,inner sep=0.25em,outer sep=0em] {$K_1, G_2, K_2$}; 	 	
 	\draw[] (SMVEfine.south) -- ++(0,-0.1) -| (Tmat1a.center);	
 	\node [right=0em of Tmat1a.center,scale=0.45,rotate=90,anchor=north west,inner sep=0.25em,outer sep=0em] {$G_1, K_1, G_2$}; 	 	 	
 	\draw[] (SMVEfine.south) -- ++(0,-0.1) -| (Tmat2a.center);	 	
 	\node [right=0em of Tmat2a.center,scale=0.45,rotate=90,anchor=north west,inner sep=0.25em,outer sep=0em] {$G_1, K_1, K_2$}; 	 	
 	\draw[] (SMVEfine.south) -- ++(0,-0.1) -| (Tmat3a.center);	 	
 	\node [right=0em of Tmat3a.center,scale=0.45,rotate=90,anchor=north west,inner sep=0.25em,outer sep=0em] {$G_1, G_2, K_2$}; 	 	
 	\draw[] (SMVEfine.south) -- ++(0,-0.1) -| (Tmat4a.center);	
 	\node [right=0em of Tmat4a.center,scale=0.45,rotate=90,anchor=north west,inner sep=0.25em,outer sep=0em] {$K_1, G_2, K_2$}; 	 	 	
 	\draw[] (SMVEmedium.south) -- ++(0,-0.1) -| (Tmat5a.center);
 	\node [right=0em of Tmat5a.center,scale=0.45,rotate=90,anchor=north west,inner sep=0.25em,outer sep=0em] {$G_1, K_1, G_2$}; 	 	 	
 	\draw[] (SMVEmedium.south) -- ++(0,-0.1) -| (Tmat6a.center);
 	\node [right=0em of Tmat6a.center,scale=0.45,rotate=90,anchor=north west,inner sep=0.25em,outer sep=0em] {$G_1, K_1, K_2$}; 	 	
 	\draw[] (SMVEmedium.south) -- ++(0,-0.1) -| (Tmat7a.center);
 	\node [right=0em of Tmat7a.center,scale=0.45,rotate=90,anchor=north west,inner sep=0.25em,outer sep=0em] {$G_1, G_2, K_2$}; 	 	
 	\draw[] (SMVEmedium.south) -- ++(0,-0.1) -| (Tmat8a.center);
 	\node [right=0em of Tmat8a.center,scale=0.45,rotate=90,anchor=north west,inner sep=0.25em,outer sep=0em] {$K_1, G_2, K_2$}; 	 	
 	\draw[] (SMVEcoarse.south) -- ++(0,-0.1) -| (Tmat9a.center);
 	\node [right=0em of Tmat9a.center,scale=0.45,rotate=90,anchor=north west,inner sep=0.25em,outer sep=0em] {$G_1, K_1, G_2$}; 	
 	\draw[] (SMVEcoarse.south) -- ++(0,-0.1) -| (Tmat10a.center);
 	\node [right=0em of Tmat10a.center,scale=0.45,rotate=90,anchor=north west,inner sep=0.25em,outer sep=0em] {$G_1, K_1, K_2$}; 	 	 	
 	\draw[] (SMVEcoarse.south) -- ++(0,-0.1) -| (Tmat11a.center);
 	\node [right=0em of Tmat11a.center,scale=0.45,rotate=90,anchor=north west,inner sep=0.25em,outer sep=0em] {$G_1, G_2, K_2$}; 	 	 	
 	\draw[] (SMVEcoarse.south) -- ++(0,-0.1) -| (Tmat12a.center); 	 
 	\node [right=0em of Tmat12a.center,scale=0.45,rotate=90,anchor=north west,inner sep=0.25em,outer sep=0em] {$K_1, G_2, K_2$}; 	 	 	
 	\draw[] (BMVEfine.south) -- ++(0,-0.1) -| (Tmat1b.center);	 	
 	\node [right=0em of Tmat1b.center,scale=0.45,rotate=90,anchor=north west,inner sep=0.25em,outer sep=0em] {$G_1, K_1, G_2$}; 	 	
 	\draw[] (BMVEfine.south) -- ++(0,-0.1) -| (Tmat2b.center);	 	
 	\node [right=0em of Tmat2b.center,scale=0.45,rotate=90,anchor=north west,inner sep=0.25em,outer sep=0em] {$G_1, K_1, K_2$}; 
 	\draw[] (BMVEfine.south) -- ++(0,-0.1) -| (Tmat3b.center);	 	
 	\node [right=0em of Tmat3b.center,scale=0.45,rotate=90,anchor=north west,inner sep=0.25em,outer sep=0em] {$G_1, G_2, K_2$};  	
 	\draw[] (BMVEfine.south) -- ++(0,-0.1) -| (Tmat4b.center);	
 	\node [right=0em of Tmat4b.center,scale=0.45,rotate=90,anchor=north west,inner sep=0.25em,outer sep=0em] {$K_1, G_2, K_2$};  	
 	\draw[] (BMVEmedium.south) -- ++(0,-0.1) -| (Tmat5b.center);
 	\node [right=0em of Tmat5b.center,scale=0.45,rotate=90,anchor=north west,inner sep=0.25em,outer sep=0em] {$G_1, K_1, G_2$}; 	 	
 	\draw[] (BMVEmedium.south) -- ++(0,-0.1) -| (Tmat6b.center);
 	\node [right=0em of Tmat6b.center,scale=0.45,rotate=90,anchor=north west,inner sep=0.25em,outer sep=0em] {$G_1, K_1, K_2$}; 
 	\draw[] (BMVEmedium.south) -- ++(0,-0.1) -| (Tmat7b.center);
 	\node [right=0em of Tmat7b.center,scale=0.45,rotate=90,anchor=north west,inner sep=0.25em,outer sep=0em] {$G_1, G_2, K_2$};  	
 	\draw[] (BMVEmedium.south) -- ++(0,-0.1) -| (Tmat8b.center);
 	\node [right=0em of Tmat8b.center,scale=0.45,rotate=90,anchor=north west,inner sep=0.25em,outer sep=0em] {$K_1, G_2, K_2$};  	
 	\draw[] (BMVEcoarse.south) -- ++(0,-0.1) -| (Tmat9b.center);
 	\node [right=0em of Tmat9b.center,scale=0.45,rotate=90,anchor=north west,inner sep=0.25em,outer sep=0em] {$G_1, K_1, G_2$}; 	 	
 	\draw[] (BMVEcoarse.south) -- ++(0,-0.1) -| (Tmat10b.center);
 	\node [right=0em of Tmat10b.center,scale=0.45,rotate=90,anchor=north west,inner sep=0.25em,outer sep=0em] {$G_1, K_1, K_2$}; 
 	\draw[] (BMVEcoarse.south) -- ++(0,-0.1) -| (Tmat11b.center);
 	\node [right=0em of Tmat11b.center,scale=0.45,rotate=90,anchor=north west,inner sep=0.25em,outer sep=0em] {$G_1, G_2, K_2$};  	
 	\draw[] (BMVEcoarse.south) -- ++(0,-0.1) -| (Tmat12b.center); 	 	
 	\node [right=0em of Tmat12b.center,scale=0.45,rotate=90,anchor=north west,inner sep=0.25em,outer sep=0em] {$K_1, G_2, K_2$};  	
	 	 	 	
 	
	\end{tikzpicture}

	\caption{RMS values, defined in Eq.~\eqref{SubSect:Ex1:Eq1}, corresponding to all~$108$ test cases. For the GDIC-IDIC approach, the best configuration is presented, i.e.~$\eta_\mathrm{rms}(h_\mathrm{opt})$, whereas for the BE-IDIC approach~$\eta_\mathrm{rms}$ does not depend on~$\widehat{h}$. Identification carried out for zero-noise images.}
	\label{SubSect:Ex1:Fig3}
\end{figure}
\begin{figure}
	\flushleft
	\mbox{}\hspace{4em}\includegraphics[scale=1]{BargraphLegend.pdf}\vspace{-1.0em}\\ 	
 	\centering
	\subfloat[$\eta_\mathrm{rms}$]{\includegraphics[scale=1]{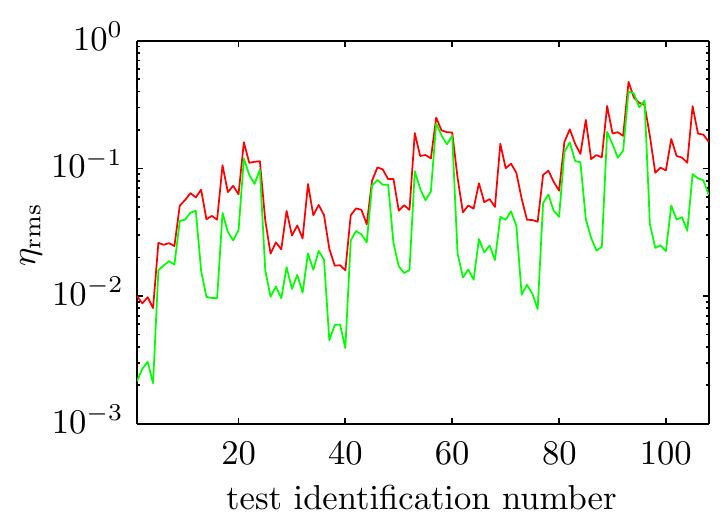}\label{SubSect:Ex1:Fig4a}}\hspace{0.5em}
	\subfloat[$h_\mathrm{opt}$ for GDIC]{\includegraphics[scale=1]{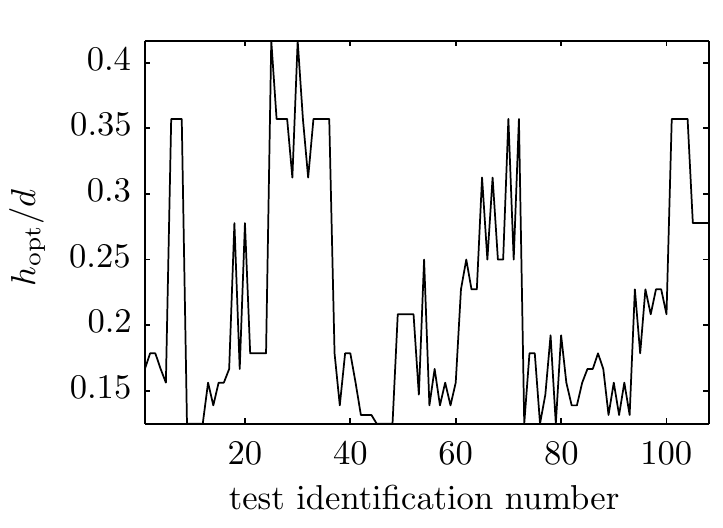}\label{SubSect:Ex1:Fig4b}}
	\caption{RMS values and corresponding~$h_\mathrm{opt}$ of the GDIC mesh as functions of individual tests. (a)~RMS values for both methods, and~(b) optimal GDIC mesh element sizes~$h_\mathrm{opt}$ relative to the inclusions' diameter~$d = 1$. In both cases, zero-noise images were used.}
	\label{SubSect:Ex1:Fig4}
\end{figure}

In order to evaluate the performance for all~$108$ test cases (recall Fig.~\ref{SubSect:BCGDIC:Fig0}), Fig.~\ref{SubSect:Ex1:Fig3} reveals the corresponding RMS values~$\eta_\mathrm{rms}$. The curves clearly show that the BE-IDIC method is in practically all cases more accurate compared to the best results of the GDIC-IDIC approach. The only exception is the shear test for coarse MVE meshes and material contrast ratio~$\rho = 16$. For this particular configuration one can argue, based on the general trends emerging in all figures, that the kinematic freedom provided by coarse MVE meshes is insufficient. For fine MVE meshes, on the other hand, the differences between the two methods approach one order of magnitude.

Fig.~\ref{SubSect:Ex1:Fig4} finally presents the RMS values for all test cases stacked together along with corresponding optimal GDIC mesh element sizes. Interestingly, the optimal value~$h_\mathrm{opt}$ for the GDIC-IDIC approach could hardly be guessed a priori, nor a posteriori (without knowledge of~$\bs{\lambda}_\mathrm{ex}$), as it varies from test to test. This means that the accuracy and precision of the GDIC-IDIC will always be less when applied in practice to real tests. The mean of~$h_\mathrm{opt}$ over all tests equals~$0.2162\, d$, which is a rather low value relative to inclusions' diameter~$d$. This result shows once again that a high level of detail should be captured by the MVE BCs.

In terms of CPU time, identification of one material parameter combination took approximately $10$~times more ($66.9$ versus $5.8$~s) for the BE-IDIC approach (initialized through GDIC) compared to the GDIC-IDIC approach. Corresponding memory footprint was approximately~$130$ times more ($975.9$ versus $7.2$~MB), mainly due to the fact that the sparse data storage of kinematic sensitivity fields has not been used. Note that computing times are based on a Matlab implementation where computationally intensive parts were coded in C++ and linked to the main code through mex files. Due to this heterogeneity, reported computing times and their ratios may not be representative.
%
%
\subsection{Image Noise Study}
\label{SubSect:Noise}
In order to examine the effect of image noise, random white Gaussian noise is superimposed on both the reference and deformed images, i.e.
\begin{equation}
\begin{aligned}
\widetilde{\bs{f}} &= \bs{f}+\zeta \, 2^8 \, \bs{\mathcal{N}}, \\
\widetilde{\bs{g}} &= \bs{g}+\zeta \, 2^8 \, \bs{\mathcal{N}},
\end{aligned}
\label{SubSect:Noise:Eq1}
\end{equation}
where~$\zeta \in \frac{1}{100}\{1, \dots, 5\}$ reflects the intensity of the image noise, $\bs{f}$ and~$\bs{g}$ are matrices storing the evaluations of the images~$f$ and~$g$ at pixel positions, and~$\bs{\mathcal{N}}$ denotes a matrix of the same dimensions as~$\bs{f}$ and~$\bs{g}$ filled with iid Gaussian random variables having zero mean and unit variance. In Eq.~\eqref{SubSect:Noise:Eq1}, the value~$2^8$ has been used because the full dynamic range of 8-bit digitization was exploited, recall Fig.~\ref{SubSect:Speckle:Fig1b}. 

\begin{figure}
 	\centering
	\includegraphics[scale=1]{BCNoiseLsigLegend.pdf}\vspace{-0.5em}\\
	\subfloat[BE-IDIC: tension]{\includegraphics[scale=1]{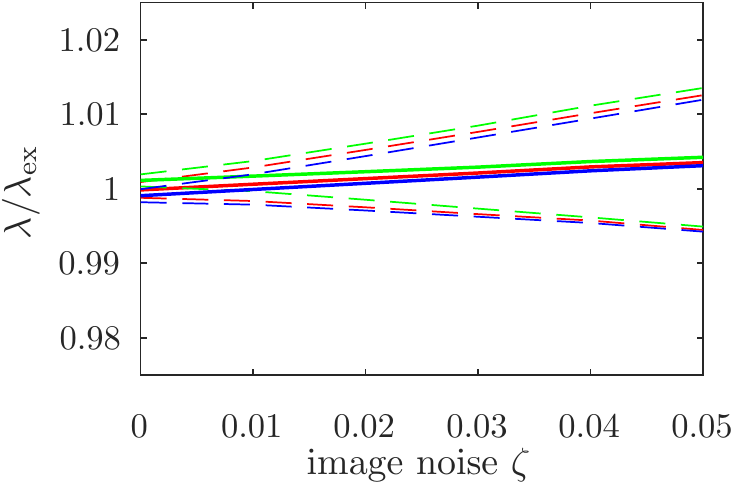}\label{SubSect:Noise:Fig1a}}\hspace{1.0em}
	\subfloat[GDIC-IDIC: tension]{\includegraphics[scale=1]{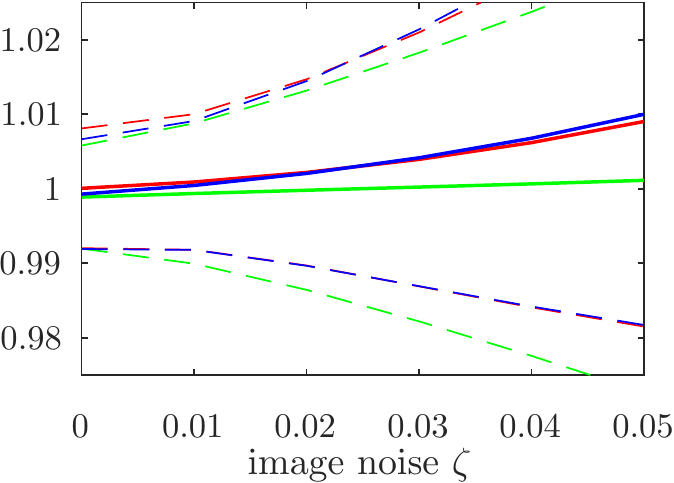}\label{SubSect:Noise:Fig1b}}\\
	\subfloat[BE-IDIC: shear]{\includegraphics[scale=1]{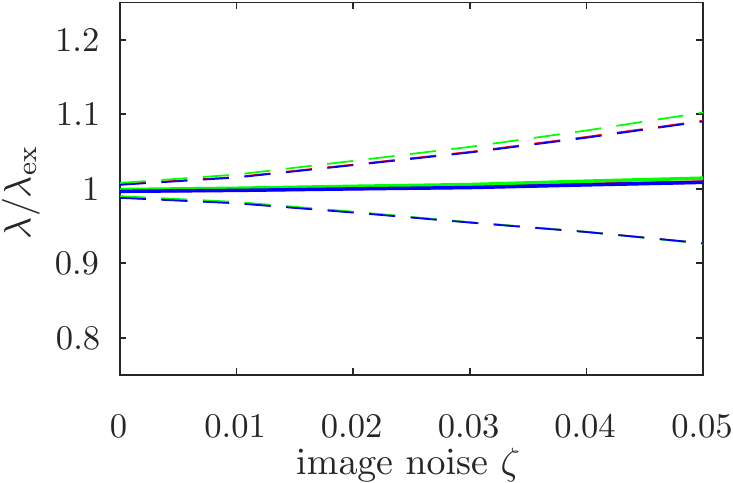}\label{SubSect:Noise:Fig1c}}\hspace{0.5em}
	\subfloat[GDIC-IDIC: shear]{\includegraphics[scale=1]{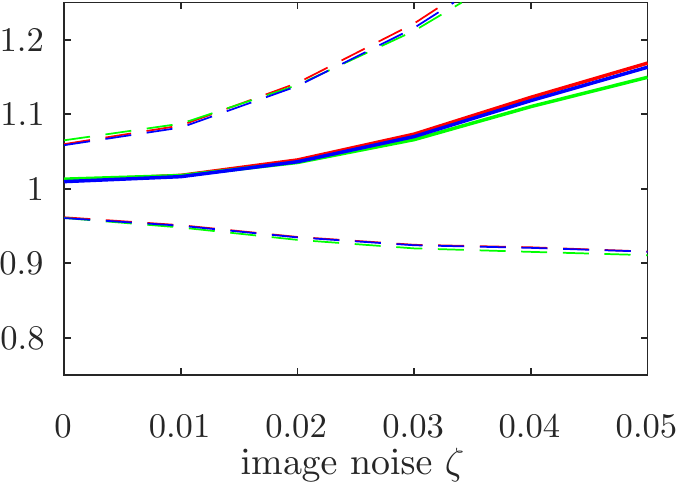}\label{SubSect:Noise:Fig1d}}\\
	\subfloat[BE-IDIC: bending]{\includegraphics[scale=1]{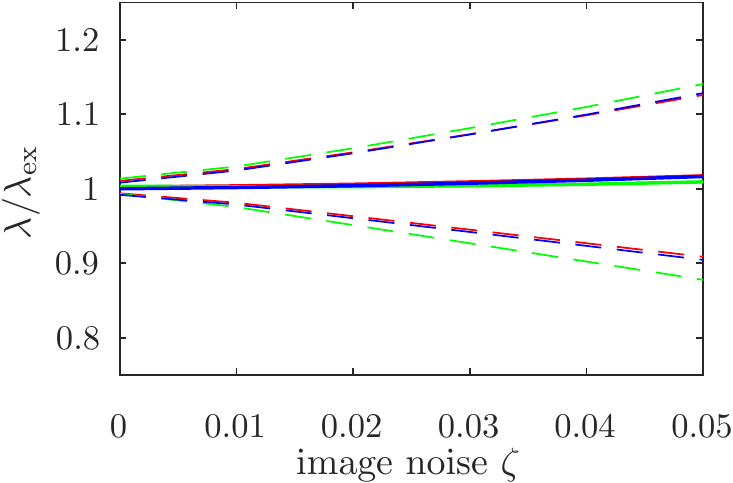}\label{SubSect:Noise:Fig1e}}\hspace{0.5em}
	\subfloat[GDIC-IDIC: bending]{\includegraphics[scale=1]{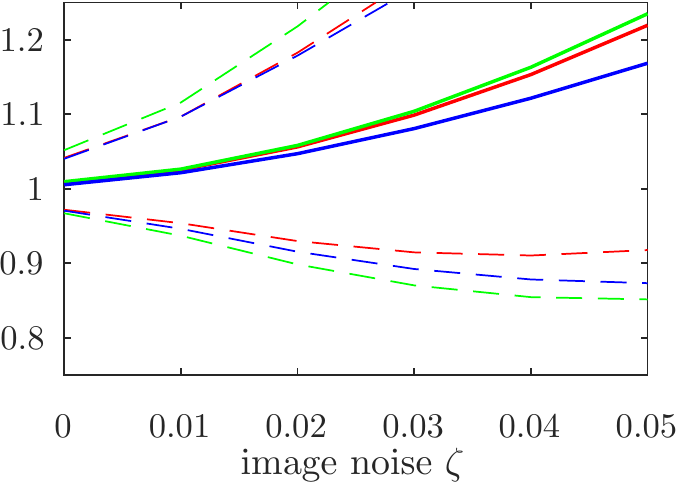}\label{SubSect:Noise:Fig1f}}	
	\caption{Identified material parameters for noisy images for~(a) and~(b) tension, (c) and~(d) shear, and~(e) and~(f) bending test. The GDIC-IDIC approach uses the best GDIC mesh in the mean. In all cases, $\rho = 4$ and fine MVE meshes used.}
	\label{SubSect:Noise:Fig1}
\end{figure}

The mean and standard deviations of the identified parameters obtained from correlations of the noisy images~$\widetilde{f}$ and~$\widetilde{g}$ are shown in Fig.~\ref{SubSect:Noise:Fig1} as functions of~$\zeta$. In order to separate the influence of noise as much as possible, the presented results correspond to fine MVE meshes only. As the optimal element size~$h_\mathrm{opt}$ in the GDIC-IDIC approach is unknown, the presented results correspond to the GDIC mesh element size that is closest to the mean optimal element size computed for fine MVE meshes and all tests. The figures clearly show that the BE-IDIC approach achieves significantly less biased results in terms of the mean values (important when numerous measurements are carried out), and also a significantly smaller standard deviation (important when only a limited number of tests is performed).
%
%
\section{Summary and Conclusions}
\label{Sect:Conclusion}
In this contribution, a systematic study has been presented revealing the significant effects induced by inaccuracy in Boundary Conditions~(BCs) prescribed to a Microstructural Volume Element~(MVE) used in micromechanical parameter identification carried out by Integrated Digital Image Correlation~(IDIC). To this end, heterogeneous specimens with simple random microstructures have been subjected to three virtual mechanical tests under plane strain conditions. The main results can be summarized as follows:
\begin{enumerate}
	\item A high accuracy in BCs prescribed to the MVE model is essential, as even a small degree of error may strongly deteriorate the systematic and statistical accuracy of the identified parameters.
	
	\item The intrinsic phenomenon of error locking in BCs (in GDIC based methods) has been discussed and its effects on micromechanical parameter identification have been demonstrated in the case of random noise and smoothing of kinematic boundary data.

	\item Effects of errors in BCs obtained directly from Global Digital Image Correlation~(GDIC) have been investigated and proven to be significant. Typically, a balance between the random error and inaccuracy due to smoothing needs to be reached, which can hardly be guessed a priori.
	
	\item In order to remove the adverse effects of GDIC errors locked in the MVE boundary, it is important to treat kinematic Degrees Of Freedom~(DOFs) associated with nodes located on the MVE boundary as unknowns in the IDIC procedure, as also pointed by~\cite{Fedele:2015} for FEMU. The improved accuracy, however, goes along with higher computational and memory requirements (approximately $10$~times more computational time and $130$~times more memory compared to the GDIC-IDIC approach).
	
	\item Adaptivity in the MVE boundary of the BE-IDIC approach has been shown to automatically guarantee a required level of detail captured by boundary conditions, not known a priori and yet needed for accurate microstructural parameter identification. Other kinds of regularization in boundary displacements need to be approached carefully due to the inherent local fluctuations.
	
	\item Image noise analyses have revealed that noise further decreases the accuracy of the identified results, especially when the BCs are extracted from GDIC. When the DOFs of the nodes at the MVE boundary are used as DOFs in the IDIC procedure, overall more accurate results are obtained than for the GDIC-IDIC approach.

	\item Boundary sensitivity functions at the MVE boundary have indicated that under the given circumstances, the tension test is approximately one order of magnitude less sensitive to errors in the prescribed BCs than the shear and bending tests. As this test is also the least complex micro-mechanical test to perform under in-situ microscopic observation, this simple test is most appropriate for identification of microstructural parameters.

\end{enumerate}

Finally, note that the presented results were obtained for the exact constitutive model, which is a rather unlikely situation in real experiments, and that also other significant sources of errors exist. For accurate identification it is desirable, nevertheless, to eliminate as many sources of potential error as possible, which may be best accomplished by enriching the IDIC DOFs with displacements at the boundary of the employed microstructural model. Sensitivity analyses to various other sources of errors and tests on real experiments are further required, but lie outside the scope of the current contribution.
%
%
%
%
%
%
%
%
\section*{Acknowledgements}
The research leading to these results has received funding from the European Research Council under the European Union's Seventh Framework Programme (FP7/2007-2013)/ERC grant agreement \textnumero~[339392].
%
%
%


\end{document}